\documentclass[english,12pt,aps,prd,a4paper,preprintnumbers,floatfix,nofootinbib,showpacs,superscriptaddress, notitlepage]{revtex4-1} 
 \pdfoutput=1
\usepackage[usenames,dvipsnames]{color}  
\usepackage{graphicx}
\usepackage{setspace}
\usepackage{braket}
\usepackage{caption}
\usepackage{subcaption}
\usepackage{amsmath}
\usepackage{amssymb}
\usepackage[colorlinks=true,citecolor=darkred,urlcolor=darkred, pdfborder={0 0 0}]{hyperref}
\usepackage[normalem]{ulem}
\usepackage{xcolor}

\makeatletter
\def\p@subsection{}
\makeatother

\definecolor{darkred}{rgb}{0.6,0,0}

\definecolor{linkcolor}{rgb}{0,0,0.5}

\def\gsim{\raise0.3ex\hbox{$\;>$\kern-0.75em\raise-1.1ex\hbox{$\sim\;$}}}
\def\lsim{\raise0.3ex\hbox{$\;<$\kern-0.75em\raise-1.1ex\hbox{$\sim\;$}}}

\def\beqn#1{\begin{equation}\label{#1}}
\def\eeqn{\end{equation}}

\def\beqa#1{\begin{eqnarray}\label{#1}}
\def\eeqa{\end{eqnarray}}

\def\Z2{$\mathcal{Z_2}$}

\def\vb#1{\vbox to #1 pt{}}

\newcommand {\ignore}[1]{}

\newcommand{\sm}{{SM }}

\def\SM{$\mathrm{SU(3)_c \otimes SU(2)_L \otimes U(1)_Y}$ }

\def\321{$\mathrm{SU(3) \otimes SU(2) \otimes U(1)}$ }

\newcommand{\AddrAHEP}{
  AHEP Group, Institut de F\'{i}sica Corpuscular --
  CSIC/Universitat de Val\`{e}ncia, Parc Cient\'ific de Paterna.\\
 C/ Catedr\'atico Jos\'e Beltr\'an, 2 E-46980 Paterna (Valencia) - SPAIN}

\newcommand{\AddrIST}{Departamento de F\'{\i}sica and CFTP,
Instituto Superior T\'{e}cnico, Universidade de Lisboa, \\
Avenida Rovisco Pais 1, 1049-001 Lisboa, Portugal}
 
\newcommand{\AddrIISERB}{Department of Physics, Indian Institute of Science Education and Research - Bhopal, \\ 
Bhopal Bypass Road, Bhauri, Bhopal 462066, India}

\begin{document}
  
\title{\color{BrickRed} Dynamical inverse seesaw mechanism as a simple benchmark for electroweak breaking and Higgs boson studies  }
\author{Sanjoy Mandal}\email{smandal@ific.uv.es}
\affiliation{\AddrAHEP}
\author{Jorge C. Rom\~ao}\email{jorge.romao@tecnico.ulisboa.pt}
\affiliation{\AddrIST}
\author{Rahul Srivastava}\email{rahul@iiserb.ac.in}
\affiliation{\AddrIISERB}
\author{Jos\'{e} W. F. Valle}\email{valle@ific.uv.es}
\affiliation{\AddrAHEP}
 \begin{abstract}
  \vspace{1cm} 
  The Standard Model~(SM) vacuum is unstable for the measured values of the top Yukawa coupling and Higgs mass.
  Here we study the issue of vacuum stability when neutrino masses are generated through spontaneous low-scale lepton number violation.
  In the simplest dynamical inverse seesaw, the \sm Higgs has two siblings: a massive $CP$-even scalar plus a massless Nambu-Goldstone boson, called majoron.
  For TeV scale breaking of lepton number, Higgs bosons can have a sizeable decay into the invisible majorons.
  We examine the interplay and complementarity of vacuum stability and perturbativity restrictions, with collider constraints on visible and invisible Higgs boson decay channels.
  This simple framework may help guiding further studies, for example, at the proposed FCC facility.
\end{abstract}
\maketitle
\section{Introduction}
\label{sec:introduction}
The main \emph{leitmotiv} of the LHC has been to elucidate the mechanism of spontaneous symmetry breaking in the SM.
With the discovery of a scalar particle with properties closely resembling those of the \sm Higgs, the ATLAS~\cite{Aad:2012tfa} and CMS~\cite{Chatrchyan:2012ufa} experiments
have achieved this goal, although only partially.
The Higgs discovery provides motivation for further studies, for example, at the upcoming FCC facility~\cite{Abada:2019ono,Abada:2019lih} and complementary lepton machines. 
If this scalar particle is indeed the \sm Higgs boson, its mass measurement allows us to study the stability of the vacuum up to high energies through the renormalization group equations (RGEs). 
Given the measured values of the top quark and Higgs boson masses, the \sm Higgs quartic coupling remains perturbative all the way up to the Planck scale ($M_P$),
but goes negative well below, Fig.~\ref{fig:RG-SM-Running}.
Thus, the Higgs vacuum in the \sm is not stable.  

The discovery of neutrino masses~\cite{Kajita:2016cak,McDonald:2016ixn,deSalas:2020pgw} provides us with another important milestone in particle physics.
It brings to surface one of the most important \sm shortcomings, i.e. the absence of neutrino masses, which stands out as a key problem.
Therefore, despite its outstanding achievements, it is now widely expected that the \sm cannot be the final theory of nature up to the Planck scale.
It is therefore important to analyze the problem of vacuum instability within neutrino mass extensions of the SM.
 
The purpose of this paper is to re-examine the consistency, i.e. the stability-perturbativity of the electroweak vacuum within neutrino mass extensions of the SM
and also to confront the resulting restrictions with information available from collider experiments LEP and LHC. 
We adopt the seesaw paradigm realized within the minimal \SM gauge structure. There are two versions, ``explicit''~\cite{Schechter:1980gr} and ``dynamical''~\cite{Chikashige:1980ui,Schechter:1981cv}.
In the former case lepton number is broken explicitly, while in the second, the breaking occurs via the vacuum expectation value (vev) of a \SM singlet scalar $\sigma$.
This ``dynamical'' variant harbors a physical Nambu-Goldstone boson, called majoron $J$~\cite{Chikashige:1980ui,Schechter:1981cv}.
The most obvious way to account for the small neutrino masses is to assume very heavy right-handed neutrinos as mediators. 
As an interesting alternative to such high-scale type-I seesaw, we have the low-scale extensions such as the inverse seesaw mechanism~\cite{Mohapatra:1986bd}. %
For sizeable Dirac-type Yukawa couplings one finds that the Higgs vacuum stability problem can become worse than in the SM
~\cite{Khan:2012zw,Rodejohann:2012px,Bonilla:2015kna,Rose:2015fua,Lindner:2015qva,Ng:2015eia,Bambhaniya:2016rbb,Garg:2017iva,Mandal:2019ndp}.

One of the attractive features of low-scale seesaw models is that we can have large Dirac-type Yukawa coupling even with light mediators, e.g. at the $\mathcal{O}(\text{TeV})$ scale.
In this case, the Yukawa couplings will evolve for a much longer range, compared to the high scale type-I seesaw~\cite{Mandal:2020lhl}. 
As a result the Higgs quartic coupling can become negative much sooner than in SM. Thus, it will have larger negative effect upon vacuum stability. 
However, we will see how in dynamical low-scale seesaw scenarios~\cite{GonzalezGarcia:1988rw} electroweak vacuum stability can be substantially improved~\cite{Bonilla:2015kna,Mandal:2020lhl}.

The prototype model is characterized by a very simple set of scalar bosons: in addition to the SM-like Higgs boson $H_{125}$ found at the LHC, there is another $CP$-even scalar $H'$.
The mixing angle $\sin\theta$ between the CP-even scalars plays a key role for our study.
Moreover, there is a massless CP-odd boson, the majoron $J$, the physical Nambu-Goldstone associated to spontaneous breaking of lepton number symmetry. 
In such low-scale seesaw the majoron can couple substantially to the Higgs boson~\cite{Joshipura:1992hp}, leading to potentially large invisible decays, e.g. $H_{125}\to JJ$ and $H' \to JJ$.
A sizeable mixing between the two $CP$-even scalars can have important phenomenological consequences, particularly for collider experiments like the LHC.
As a result of this mixing, the couplings of the SM-like Higgs scalar $H_{125}$ can deviate appreciably from the SM values.
These can modify the so-called signal ``strength parameter'' $\mu_f$ associated to a given ``visible'' final-state $f$, which can be tested at the LHC~\cite{TheATLASandCMSCollaborations:2015bln,Aad:2019mbh}.
The modified couplings and the existence of invisible decays are constrained by Higgs measurements at the LEP and LHC experiments~\cite{Bonilla:2015uwa,Bonilla:2015jdf,Fontes:2019uld}.
Here we adopt the conservative range,  
\begin{equation}
  \label{eq:muf}
0.8\leq \mu_f\leq 1, 
\end{equation}
while for the invisible Higgs boson decays we take bound coming from the CMS experiment~\cite{Sirunyan:2018owy}\footnote{The present bound from ATLAS for invisible Higgs decays is $\text{BR}(H_{125}\to\text{Inv})\leq 26\%$~\cite{Aaboud:2019rtt}.},
\begin{equation}
  \label{eq:inv}
 \text{BR}(H_{125}\to\text{Inv})\leq 19\%~. 
\end{equation}

Collider bounds and vacuum stability conditions lead to complementary constraints on the allowed range of the mixing angle $\sin\theta$.
The goal of this work is to exploit this complementarity to test the simplest dynamical inverse seesaw scenario.
We confront the collider limits with the consistency restrictions arising from the stability-perturbativity of the electroweak vacuum. 

The work is organized as follows: in Sec.~\ref{sec:vacuum stability in SM} we briefly summarize the issue of vacuum stability in the SM.
In Secs.~\ref{sec:inverse seesaw and vacuum stability} and \ref{sec:majoron completion and vacuum stability}, we discuss vacuum stability in the inverse seesaw mechanism
with explicit lepton number breaking as well as in the dynamical inverse seesaw mechanism.
We use the two-loop RGEs given in Ref.~\cite{Mandal:2020lhl} to evolve all the SM parameters as well as the new ones.
We derive the full two-loop RGEs of the relevant parameters of the dynamical inverse seesaw, and list them in Appendix~\ref{app:inverse-seesaw-majoron}.  
In Sec.~\ref{sec:invisible decay widths}, we discuss the production and decays of the two CP-even scalars $H_{125}$ and $H'$, including both visible as well as invisible decay modes.
In Sec.~\ref{sec:Constraints for Case I} and \ref{sec:Constraints for Case II}, we study in detail the sensitivities of Higgs boson searches at LEP and LHC.
In Sec.~\ref{sec:Vacuum Stability}, we address the issue of vacuum stability taking into account the collider constraints. Finally in Sec.~\ref{Conclusion} we conclude.
\section{Vacuum stability in the Standard Model}
\label{sec:vacuum stability in SM}
\begin{figure}[h]
\centering
\includegraphics[width=0.75\textwidth]{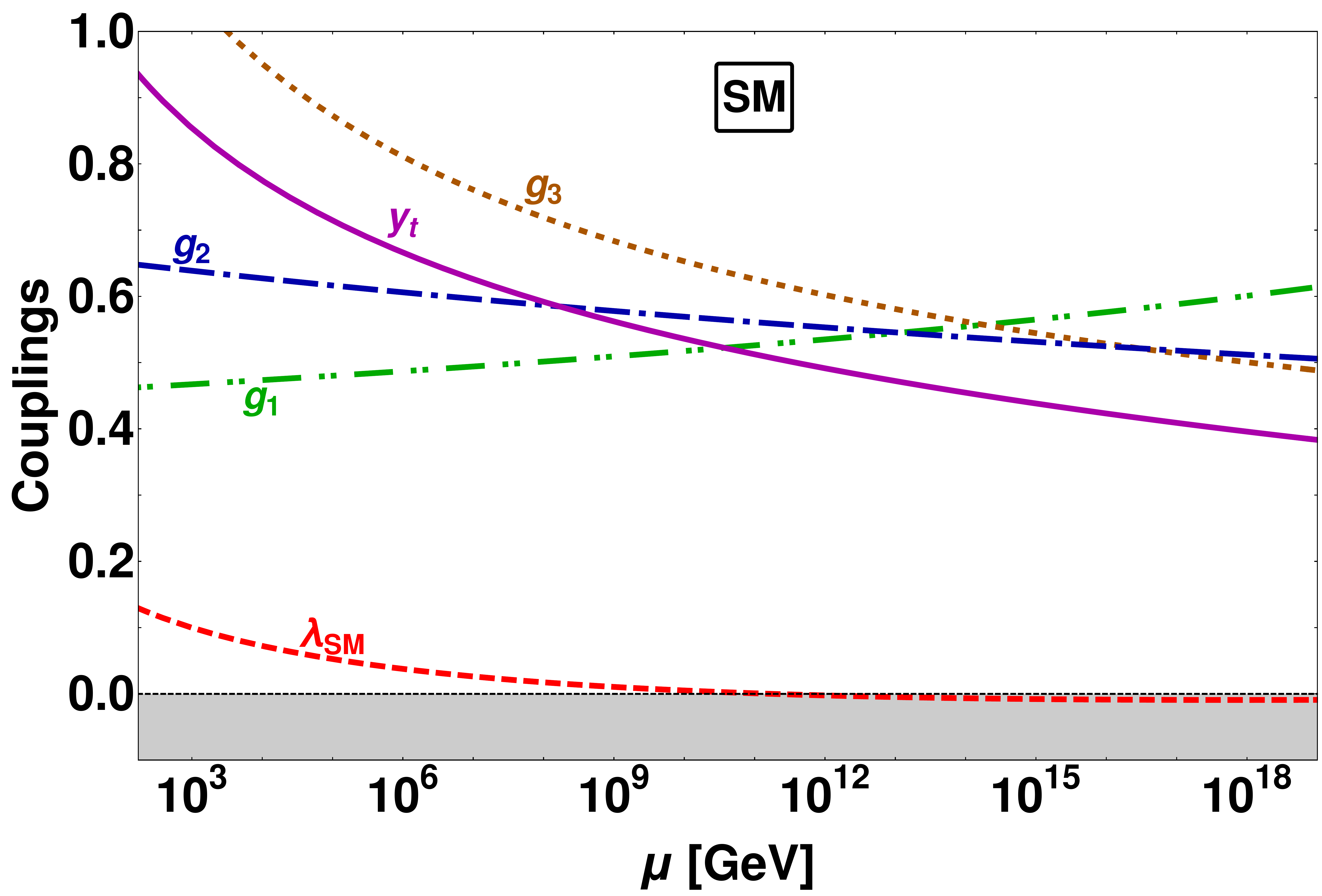}
\caption{\footnotesize{The renormalization group evolution of the \sm gauge couplings $g_{1}$, $g_{2}$, $g_{3}$, the top quark Yukawa coupling $y_{t}$ and the quartic Higgs boson self-coupling
    $\lambda_{\text{SM}}$.}}
\label{fig:RG-SM-Running}
\end{figure}

Before starting in earnest it is useful to briefly sum up the lessons from previous vacuum stability studies in the SM.
If the 125 GeV scalar discovered at LHC is indeed the \sm Higgs boson then we can determine its quartic coupling at the electroweak scale.
This measurement can subsequently be used to study the stability of the fundamental vacuum at high energies, all the way up to Planck scale. 
In Fig.~\ref{fig:RG-SM-Running}, we summarize the status of the electroweak vacuum within the SM, following the discussion of Ref.~\cite{Mandal:2019ndp,Mandal:2020lhl}.

Here we adopt the $\overline{\text{MS}}$ scheme, taking the parameter values at low scale as input, see \cite{Mandal:2019ndp} for details. The top quark mass scale is set as $m_t=173\pm 0.4$.
We have used two-loop renormalization group equations~(RGEs) for the quartic coupling $\lambda_{\text{SM}}$, the Yukawa coupling $Y_\nu$, as well as for \SM gauge couplings $g_1$, $g_2$ and $g_3$.
Fig.~\ref{fig:RG-SM-Running} clearly shows that \sm Higgs quartic coupling $\lambda_{\text{SM}}$ goes negative around $\mu\sim 10^{10}$ GeV.
As a result, the potential is not bounded from below, indicating an unstable vacuum.
Note that the \sm vacuum stability is very sensitive to the input value of top-quark mass.
A dedicated analysis shows that SM Higgs vacuum is not absolutely stable, but rather metastable with very long lifetime~\cite{Buttazzo:2013uya,Degrassi:2012ry,Alekhin:2012py}.
In what follows we will examine the implications of vacuum stability requirements within seesaw models of neutrino mass generation.
\section{Inverse seesaw and vacuum stability} 
\label{sec:inverse seesaw and vacuum stability}
The seesaw mechanism based on the \SM gauge group can be realized either in ``high-scale'' or in ``low-scale'' regimes.
The vacuum stability issue has been examined in the high-scale type-I seesaw mechanism with large Yukawa couplings in Ref.~\cite{Mandal:2019ndp}.
It has been found that majoron extensions of these schemes can restore the vacuum stability all the way up to Planck scale once one takes into account the scalar threshold corrections.   
However, such high scale seesaw schemes typically involves mediator masses much larger than the electroweak scale, unaccessible at collider experiments.
Here we revisit the issue, but in the context of type-I ``low-scale" seesaw mechanism, in which mediators would be accessible to high energy colliders.
The simplest prototype is the inverse seesaw mechanism~\cite{Mohapatra:1986bd,GonzalezGarcia:1988rw,CentellesChulia:2020dfh}. 
In inverse seesaw, lepton number is violated by introducing extra \sm gauge singlet fermions $S_i$ with small Majorana mass terms, associated to the conventional ``right-handed'' neutrinos $\nu_i^c$.
The relevant part of the Lagrangian is given by
\begin{align} 
 -\mathcal{L}=\sum_{ij} Y_{\nu}^{ij} L_{i}\tilde{\Phi} \nu^c_{j} + M^{ij} \nu^c_{i} S_{j}  + \frac{1}{2}\mu^{ij}_S S_{i} S_{j} + \text{H.c.} 
 \label{lag-inv-seesaw}
\end{align}
where $L_{i} = \left(\nu , \ell \right)^{T};\,$ $i = 1,2,3$ are the lepton doublets, $\Phi$ is the Higgs doublet and $\nu_i^c, S_i$ are \sm gauge singlet fermions. 
The $\nu_i^c, S_i$ transform under the lepton number symmetry $U(1)_L$ as $\nu_i^c \sim -1$ and $S_i \sim +1$, respectively.
The smallness of light neutrino masses is controlled by the lepton number violating Majorana mass parameter $\mu_S$.  
This allows the Yukawa couplings $Y_\nu$ to be sizeable, even when the messenger mass scale $M$ lies in the TeV scale, without conflicting with the observed smallness of the neutrino masses.

\begin{figure}[h]
\centering
\includegraphics[width=0.59\textwidth]{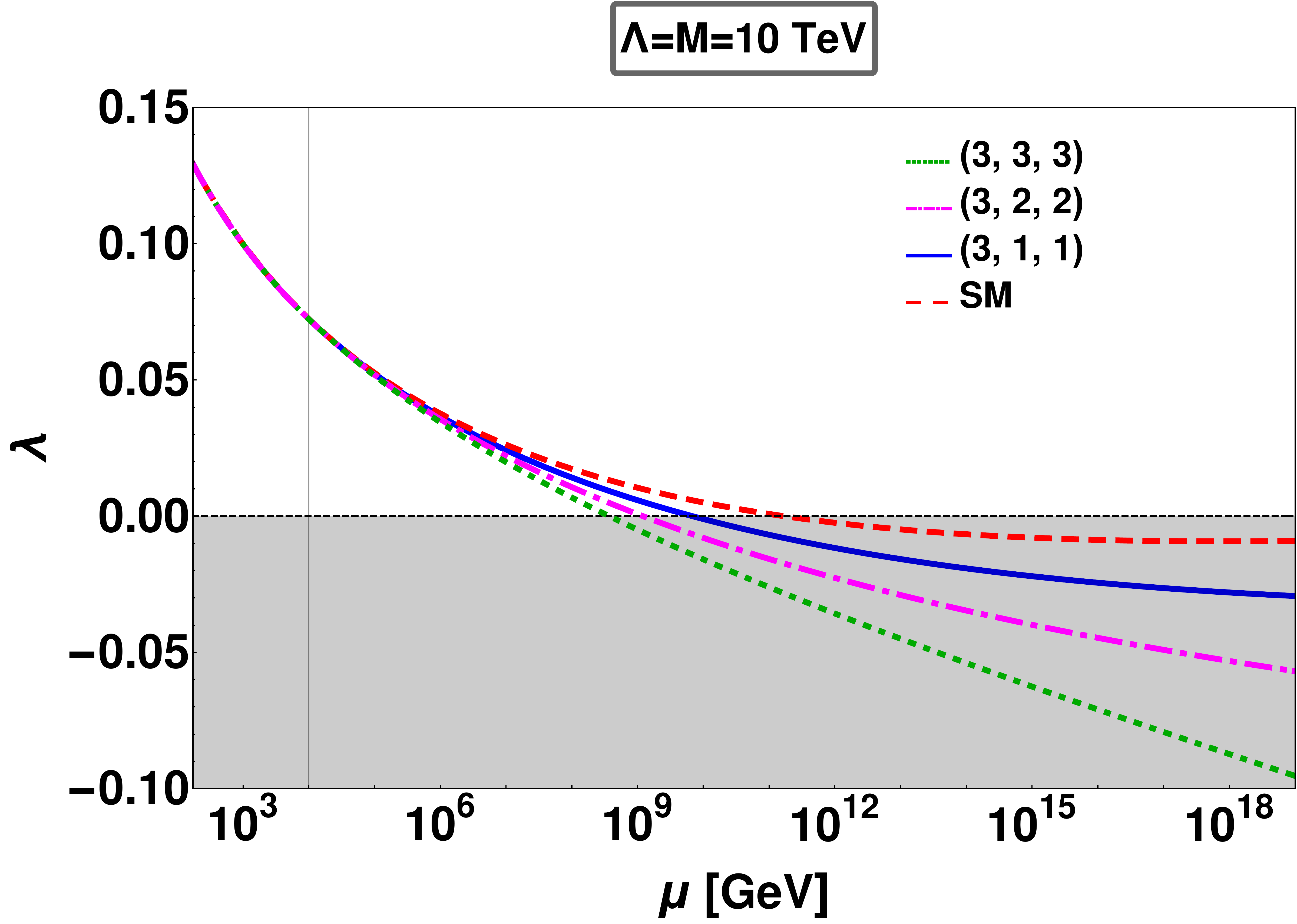}
\caption{\footnotesize{
    Comparing the evolution of the quartic Higgs self-coupling $\lambda$ in the SM~(dashed, red) with various inverse-seesaw extensions with explicit lepton number violation:
    (3,1,1) denoted in solid (blue), (3,2,2) dot-dashed (magenta) and (3,3,3) dotted (green), see text for details.
}}
\label{fig:Three-species}
\end{figure}

Thanks to the potentially large Dirac neutrino Yukawa coupling $Y_\nu$ required to generate adequate neutrino masses in such schemes the vacuum stability problem aggravates. 
To examine the effect of the new fermions $\nu^c$ and $S$ upon the stability of the electroweak Higgs vacuum we need to take into account the effect of the threshold corrections.
To begin with, below the threshold scale $\Lambda\approx M$, we need to integrate out the new fermions, so the theory is just the SM plus an effective dimension-five Weinberg operator. 
This affects the running of Higgs quartic coupling $\lambda_\kappa$ below the scale $\Lambda$, though the correction is negligibly small. 
As a result, in the effective theory, the running of $\lambda_\kappa$ below the scale $\Lambda$ is almost same as in the SM. 
Above the threshold scale $\Lambda$ we have the full Ultra-Violet (UV) complete theory.
Now the Yukawa coupling $Y_\nu$ will affect the running of the Higgs quartic coupling which we now call $\lambda$ so as to distinguish it from the quartic coupling below the threshold scale. 
The two-loop system of RGEs governing the evolution of $\lambda$, $Y_\nu$ and the \sm couplings are listed in Ref.~\cite{Mandal:2020lhl}.
Integrating out the heavy neutrinos also introduces threshold corrections to the SM Higgs quartic coupling $\lambda$ at the scale $\Lambda$~\cite{Casas:1999cd,Mandal:2020lhl,Mandal:2019ndp}.
As a result we also need to consider the shift due to threshold corrections in $\lambda$ at $\Lambda$ when solving RGEs. The threshold corrections imply that 
\begin{align}
\lambda(\Lambda)\to\lambda(\Lambda)-\frac{5n^2}{32\pi^4}\text{Tr}(Y_\nu^\dagger Y_\nu)^2,
\label{shift}
\end{align}
where $n$ is the number of singlet fermions $\nu^c$. 
Having set up our basic scheme, let us now look at the impact of the new Yukawa coupling $Y_\nu$ on the stability of the Higgs vacuum.
As shown in Fig.~\ref{fig:Three-species}, above the threshold scale $\Lambda$, Yukawa coupling $Y_\nu$ can completely dominate the RGEs behaviour of quartic coupling $\lambda$.

In Fig.~\ref{fig:Three-species} we compare the evolution of the Higgs quartic coupling $\lambda$ within the SM~(dashed, red) with the $(3, n, n)$ inverse seesaw completion.
Here $n$ denotes the number of $\nu^c$ and $S$ species.
We show the results for $n = 1$ (solid, blue), $n = 2$ (dot-dash, magenta) and $n = 3$ (dot, green).
For this comparison, we have fixed the Yukawa coupling $|Y_\nu|$ = 0.4 for the (3,1,1) case.
For $(3,n,n)$ with $n\geq 2$, we took the diagonal entries of $Y_\nu$ as $Y_\nu^{ii} = 0.4$, while all the off-diagonal ones are neglected.  
We have fixed the threshold scale, which also sets the mass scale of the singlet neutrinos, as $\Lambda = 10$ TeV. 
One sees how, the larger the value of $n$, the more strongly the $(3, n, n)$ inverse seesaw scenarios aggravate the Higgs vacuum stability problem.  
This destabilizing effect of the neutrino Yukawas can be potentially cured in the presence of other particles that can revert the trend found above.
\section{Majoron completion and vacuum stability} 
\label{sec:majoron completion and vacuum stability}
As a well-motivated completion of the above scheme, we now turn to the dynamical version of the inverse seesaw mechanism~\cite{GonzalezGarcia:1988rw}.
Lepton number is now promoted to a spontaneously broken symmetry within the \SM gauge framework. 
To do this, in addition to the \sm singlets $\nu_i^c$ and $S_i$, we add a complex scalar singlet $\sigma$ carrying two units of lepton number.
This symmetry is then broken by the vev of this complex singlet $\sigma$. The relevant Lagrangian is given by 
 \begin{align}
-\mathcal{L} = \sum_{i,j}^{3} Y_{\nu}^{ij} L_{i} \tilde{\Phi} \nu^c_{j} + M^{ij} \nu^c_{i}  S_{j} + Y_{S}^{ij} \sigma S_{i} S_{j} + \text{H.c.} 
\end{align}
The neutral component of the doublet $\Phi$ and the singlet $\sigma$ acquire vevs $\frac{v_\Phi}{\sqrt{2}}$ and $\frac{v_\sigma}{\sqrt{2}}$, respectively leading to the light neutrino masses given by 
\begin{align}
 m_{\nu}\simeq \frac{v_\Phi^{2}}{\sqrt{2}}Y_{\nu}M^{-1}Y_{S}v_\sigma M^{-1T}Y_{\nu}^{T}
\end{align}
For $m_\nu \sim 0.1 \,\rm{eV}$, we can have Yukawa couplings $Y_{\nu}$ of order one, for TeV scale $v_\sigma$ and $M$.  
\subsection{Scalar Potential} 
\label{sec:scalar potential}
The \SM as well as the global lepton number symmetry invariant scalar potential is  given by
\begin{align}
V=\mu_\Phi^2 \Phi^{\dagger}\Phi + \mu_\sigma^2 \sigma^{\dagger}\sigma + \lambda_\Phi (\Phi^{\dagger}\Phi)^2  + \lambda_\sigma (\sigma^{\dagger}\sigma)^2  + \lambda_{\Phi \sigma} (\Phi^{\dagger}\Phi) (\sigma^{\dagger}\sigma)
\label{potential}
\end{align}
\subsection*{Consistency conditions: boundedness and perturbativity} 
The above scalar potential must be bounded from below. This implies that at any given energy scale $\mu$, the quartic couplings should satisfy
\begin{align}
\lambda_\Phi(\mu) > 0,\,\,\lambda_\sigma(\mu) > 0,\,\,\lambda_{\Phi \sigma}(\mu)+2\sqrt{\lambda_\Phi(\mu)\lambda_\sigma(\mu)} > 0,
\label{stability condition}
\end{align}
where $\lambda_i(\mu)$ are the values of the quartic couplings at the running scale $\mu$.
To have an absolutely stable vacuum, one needs to satisfy the condition given in Eq.~\ref{stability condition} at each and every energy scale. 

To ensure perturbativity, we take a conservative approach of simply requiring that 
\begin{align}
\lambda_{\Phi}(\mu)\leq \sqrt{4\pi},\,\,\lambda_\sigma(\mu)\leq \sqrt{4\pi}\,\,\text{and}\,\,|\lambda_{\Phi \sigma}(\mu)|\leq \sqrt{4\pi}.
\label{perturbativity limit}
\end{align}
There will be additional constraints from unitarity or from electroweak precision data through the $S$, $T$ and $U$ parameters.
However, for our parameter range of interest they lead to rather loose constraints~\cite{Bonilla:2015uwa} compared to the LHC bounds which we will shortly discuss in detail.
\subsection*{Mass spectrum}

In order to obtain the mass spectrum for the scalars after \SM and lepton-number symmetry breaking, we expand the scalar fields as 
\begin{align}
\phi^0=\frac{1}{\sqrt{2}}(v_\Phi+R_1+i I_1),\\
\sigma=\frac{1}{\sqrt{2}}(v_\sigma+R_2+i I_2)
\end{align}
Using this expansion, the potential in \eqref{potential} leads to a physical massless Goldstone boson, namely the majoron $J=\text{Im}\,\sigma$ plus two massive neutral CP-even scalars $H_i(i=1,2)$.
The mass matrix of  CP-even Higgs  scalars in the basis $(R_1,R_2)$ reads as 
\begin{align}
M_R^2=
\begin{bmatrix}
2\lambda_\Phi v_\Phi^2  &  \lambda_{\Phi \sigma}v_\Phi v_\sigma \\
\lambda_{\Phi \sigma} v_\Phi v_\sigma & 2\lambda_\sigma v_\sigma^2 
\end{bmatrix}
\end{align}
with the mass eigenvalues given by 
\begin{align}
m_{H_1}^2 &=\lambda_\Phi v_\Phi^2 +\lambda_\sigma v_\sigma^2 - \sqrt{(\lambda_\Phi v_\Phi^2 - \lambda_\sigma v_\sigma^2)^2+(\lambda_{\Phi \sigma}v_\Phi v_\sigma)^2}\\
m_{H_2}^2 &=\lambda_\Phi v_\Phi^2 +\lambda_\sigma v_\sigma^2 + \sqrt{(\lambda_\Phi v_\Phi^2 - \lambda_\sigma v_\sigma^2)^2+(\lambda_{\Phi \sigma}v_\Phi v_\sigma)^2}
\end{align}
where the scalars $H_1$ and $H_2$ have masses $m_{H_1}$ and $m_{H_2}$ respectively, and by convention $m_{H_1}^2\leq m_{H_2}^2$ throughout this work.
One of these scalars must be identified with $H_{125}$ i.e. the scalar discovered at LHC. We have two possibilities, either $H_1 \equiv H_{125}$ or $H_2 \equiv H_{125}$. Here we consider both.
The two mass eigenstates $H_i$ are related with the $R_1, R_2$ fields through the rotation matrix $O_R$ as,  
\begin{align}
\begin{bmatrix}
H_1\\
H_2\\
\end{bmatrix}
=O_R
\begin{bmatrix}
R_1\\
R_2\\
\end{bmatrix}
=
\begin{bmatrix}
\cos\theta & \sin\theta \\
-\sin\theta & \cos\theta \\
\end{bmatrix}
\begin{bmatrix}
R_1\\
R_2\\
\end{bmatrix},
\label{mixing relation}
\end{align}
where $\theta$ is the mixing angle. The rotation matrix satisfies
\begin{align}
O_R M_R^2 O_R^T=\text{diag}(m_{H_1}^2, m_{H_2}^2)
\label{diagonalisation}
\end{align}
We can use Eq.~(\ref{mixing relation}) and (\ref{diagonalisation}) to solve for the potential parameters $\lambda_\Phi$, $\lambda_\sigma$, $\lambda_{\Phi \sigma}$ in terms of mixing angle $\theta$
and the scalar masses $m_{H_i}$ as
\begin{align}
\lambda_\Phi &=\frac{m_{H_1}^2\cos^2\theta+m_{H_2}^2\sin^2\theta}{2v_\Phi^2}
\label{definition lambda phi}
\end{align}
\begin{align}
\lambda_\sigma &=\frac{m_{H_1}^2\sin^2\theta+m_{H_2}^2\cos^2\theta}{2v_\sigma^2}
\label{definition lambda sigma}
\end{align}\vspace{-0.5cm}
\begin{align}
\lambda_{\Phi \sigma} &=\frac{\sin 2\theta (m_{H_1}^2-m_{H_2}^2)}{2 v_\Phi v_\sigma}
\label{definition lambda phisigma}
\end{align}
\subsection*{Vacuum stability and spontaneous lepton number violation }
\begin{figure}[h]
\centering
\includegraphics[width=0.3\textwidth]{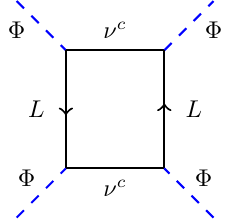}~~~~~~~~
\includegraphics[width=0.3\textwidth]{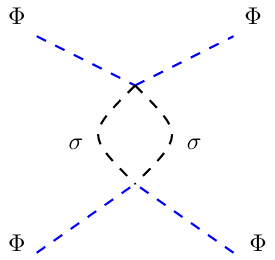}
\caption{\footnotesize{\textbf{Left panel}: the destabilizing effect of right-handed neutrinos on the evolution of the Higgs quartic coupling.
    \textbf{Right panel}: one-loop correction to the $\Phi$ quartic coupling due to its interaction with the singlet $\sigma$ that drives spontaneous lepton number violation in dynamical inverse seesaw models.} }
\label{fig:both-contributions-to-lambda}
\end{figure}

We now look at the stability of the electroweak vacuum in more detail.
To see how the couplings evolve with energy we use the full two-loop RGEs governing the evolution of the Higgs quartic coupling~\cite{Bonilla:2015kna},
which are listed in Appendix~\ref{app:inverse-seesaw-majoron}.  
However, to understand the main features, its enough to look at the one-loop $\beta$ functions for the quartic couplings, which are given as follows  
\begin{align} 
16\pi ^2 \beta_{\lambda_\Phi} & =  
+\lambda_{\Phi\sigma}^{2} + 24 \lambda_\Phi^{2}  
- \frac{9}{5} g_{1}^{2} \lambda_\Phi - 9 g_{2}^{2} \lambda_\Phi  + 12 \lambda_\Phi y_t^2 +4 \lambda_\Phi \mbox{Tr}\Big({Y_\nu  Y_{\nu}^{\dagger}}\Big)     \nonumber \\ 
& - 6  y_t^4  - 2 \mbox{Tr}\Big({Y_\nu  Y_{\nu}^{\dagger}  Y_\nu  Y_{\nu}^{\dagger}}\Big) 
 +\frac{27}{200} g_{1}^{4} +\frac{9}{20} g_{1}^{2} g_{2}^{2} +\frac{9}{8} g_{2}^{4} 
\label{one-loop-lambdaphi}
\end{align}
\vspace{-1cm}
\begin{align}
16\pi^2 \beta_{\lambda_{\Phi\sigma}} & =  
\frac{1}{10} \lambda_{\Phi\sigma} \Big( + 40 \lambda_{\Phi\sigma} + 80 \lambda_{\sigma} + 120 \lambda_\Phi 
+ 40 \mbox{Tr}\Big({Y_S  Y_S^*}\Big) + 60 y_t^2 \nonumber \\
& + 20 \mbox{Tr}\Big({Y_\nu  Y_{\nu}^{\dagger}}\Big) 
 - 9 g_{1}^{2} - 45 g_{2}^{2} \Big) 
\label{lambda phi sigma running}
\end{align}
\vspace{-1cm}
\begin{align}
16\pi^2 \beta_{\lambda_{\sigma}} & =  
2 \Big(10 \lambda_{\sigma}^{2}  + \lambda_{\Phi\sigma}^{2}  
+ 4 \lambda_{\sigma} \mbox{Tr}\Big({Y_S  Y_S^*}\Big)  
- 8 \mbox{Tr}\Big({Y_S  Y_S^*  Y_S  Y_S^*}\Big)  \Big)
\label{lambdasigma-running}
\end{align}
where $\beta_f=\mu \frac{df}{d\mu}$.
Notice that the one-loop contributions of the new fermions $\nu_i^c$ and scalars $\sigma$ to the beta-function of $\lambda_\Phi$ shown in Fig.~\ref{fig:both-contributions-to-lambda} and
Eq.~(\ref{one-loop-lambdaphi}) have the opposite sign.
Indeed, the one-loop diagram in the right panel of Fig.~\ref{fig:both-contributions-to-lambda} leads to a ``positive'' $+\lambda_{\Phi\sigma}^2$ term in the RGE of the quartic coupling $\lambda_\Phi$.
This should be contrasted with the destabilizing effect coming from the left panel of Fig.~\ref{fig:both-contributions-to-lambda} associated to the fermion Yukawa $Y_\nu$. 
We will show shortly that, for appropriate values of the scalar quartic coupling $\lambda_{\Phi \sigma}$, this positive contribution can indeed overcome the destabilizing effect of the fermion Yukawas.

Vacuum stability in this model can be studied in two different regimes namely i) $v_\sigma\gg v_\Phi$
and ii) $v_\sigma\approx \mathcal{O}(v_\Phi)$. Due to its potential testability at LHC, here we focus on the second possibility~\footnote{
  For the first case, in the limit $v_\sigma\gg v_\Phi$ the heavy CP-even scalar $H_2 \equiv H'$ almost decouples. 
  The threshold corrections at $\Lambda=m_{H'}$ induce a shift in the Higgs quartic coupling, $\delta\lambda=\frac{\lambda_{\Phi \sigma}^2}{4\lambda_\sigma}$.
  The net effect is a positive shift to this self-coupling above the threshold scale $\Lambda$, improving the chances of keeping $\lambda_\Phi$ positive~\cite{Mandal:2020lhl}.}
For $v_\sigma\approx \mathcal{O}(v_\Phi)$ the threshold scale $\Lambda = M$ will be at the mass scale of the fermions which we take to be $M \approx  \mathcal{O}(10 \, \rm{TeV})$. 
At the  threshold scale only the heavy fermions are integrated out, while all the scalars remain in the resulting effective theory below the threshold. 
%

\begin{figure}[h]
\centering
\includegraphics[width=0.49\textwidth]{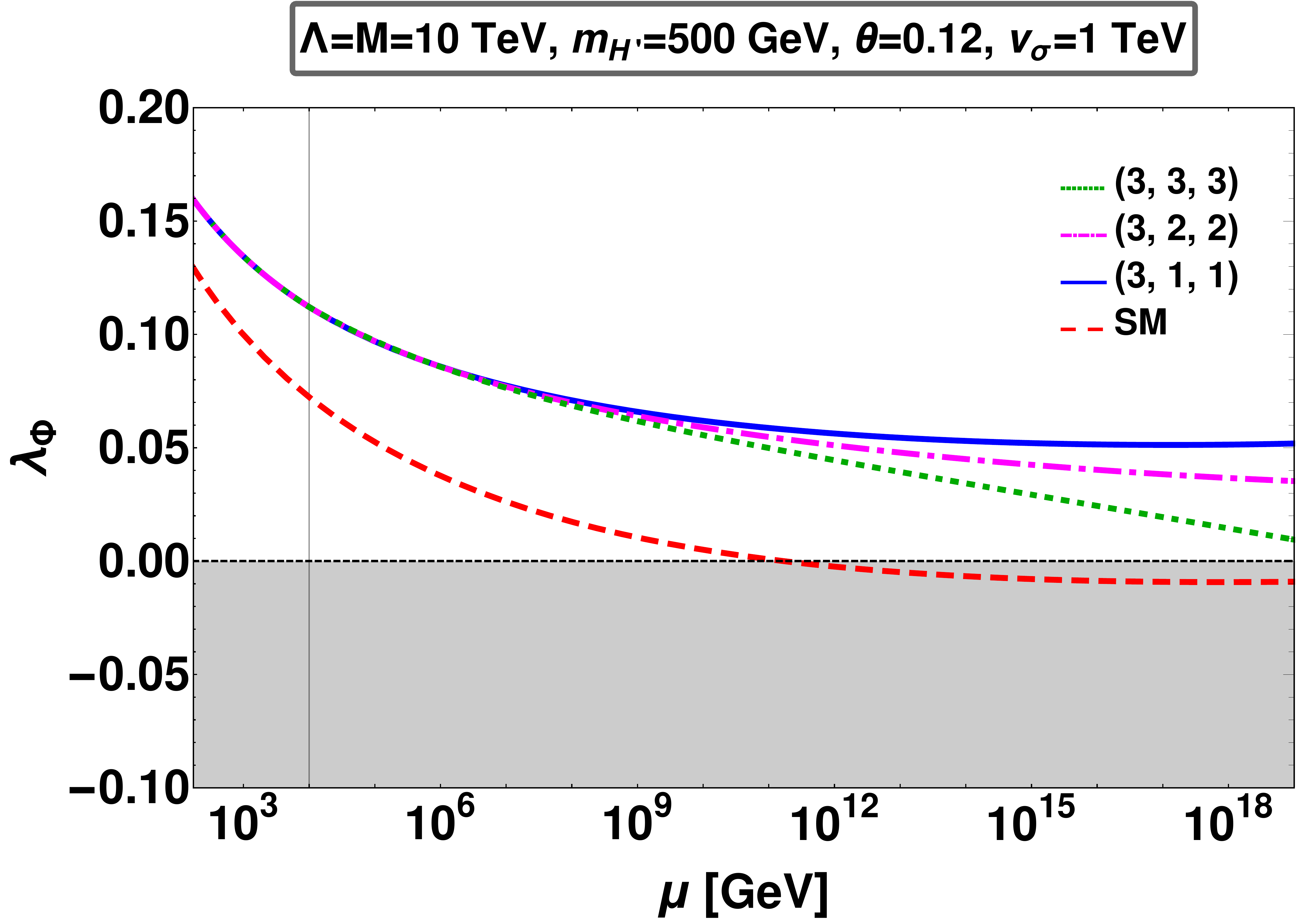}
\includegraphics[width=0.49\textwidth]{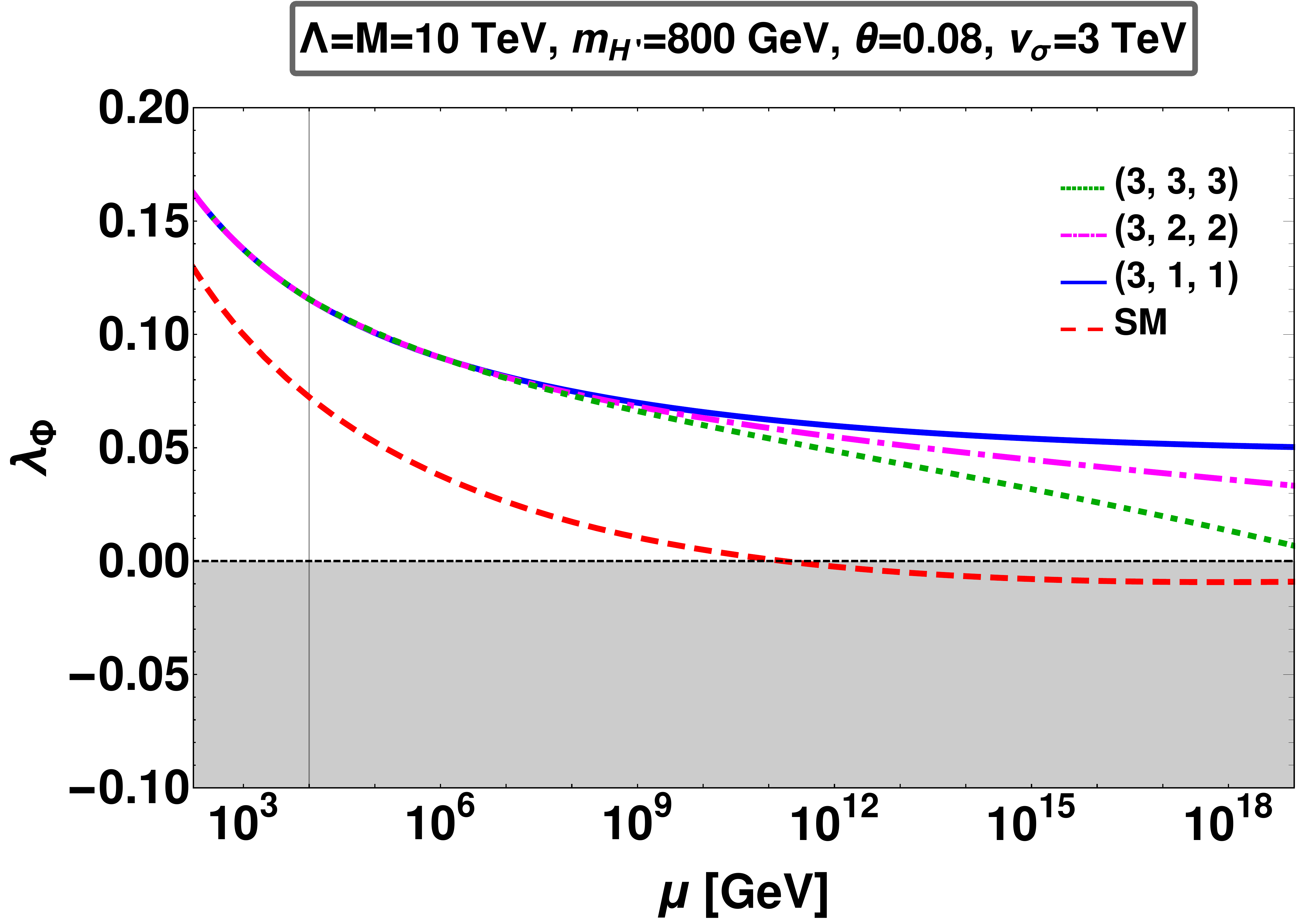}
\caption{\footnotesize{Comparing the evolution of the quartic Higgs self-coupling $\lambda_\Phi$ in the SM~(red dashed) with the majoron inverse seesaw mechanism:
    the minimal (3,1,1) is denoted in solid (blue), (3,2,2) is dot-dashed (magenta) and (3,3,3) is dotted (green).
    Left panel is for $v_\sigma=1$ TeV and right panel is for $v_\sigma=3$ TeV. See text for details. 
}}
\label{fig:Three-species1TeV}
\end{figure}

The RG evolution of the Higgs coupling in the majoron inverse seesaw model is illustrated in Fig.~\ref{fig:Three-species1TeV} for two different benchmarks. 
In this plot, we have fixed the singlet neutrino scale $\Lambda = 10$ TeV. 
Above the threshold scale the full two-loop RGEs (see Appendix~\ref{app:inverse-seesaw-majoron}) are used.
Below the threshold scale the RGEs of the effective theory, where the heavy fermions have been integrated out, is employed. 
 As the mass of the heavy scalar $H_{2} \equiv H'$ close to the electroweak scale, we have neglected the small range between $M_Z$ and $m_{H'}$ and have run all the quartic couplings from the scale $M_Z$.
 The left panel is for the benchmark $m_{H'}=500$ GeV, $\theta=0.12$, $v_\sigma=1$ TeV, whereas the right panel has $m_{H'}=800$ GeV, $\theta=0.08$, $v_\sigma=3$ TeV. 
 We have taken the Yukawa coupling $|Y_\nu| = 0.4$ for (3,1,1) case, while for the $(3,n,n)$ with $n\geq 2$ we took $Y_\nu^{ii} = 0.4$ with zero off-diagonal entries. 
To avoid overcrowding the plot, only the evolution of the quartic scalar coupling $\lambda_\Phi$ has been shown in Fig.~\ref{fig:Three-species1TeV}. 

In Fig.~\ref{fig:Three-species1TeV} we compare the \sm case (dashed, red) with the (3,1,1) (solid, blue), (3,2,2) (dot-dash, magenta) and (3,3,3) (dot, green) Majoron inverse seesaw evolution curves. 
One sees from  Fig.~\ref{fig:Three-species1TeV} that, by adequate choices for the quartic couplings $\lambda_\sigma, \lambda_{\Phi \sigma}$, or equivalently $\theta$ and $m_{H'}$, 
we can have stable vacuum all the way up to the Planck scale, even for sizeable Yukawa couplings. 
Given the stabilizing effect of the quartic couplings, especially  $\lambda_{\Phi\sigma}$,  one might be tempted to always take them sufficiently large i.e. $\lambda_{\Phi\sigma} \approx \mathcal{O}(1)$. 
However, taking the quartic couplings $\approx \mathcal{O}(1)$ can lead to nonperturbative couplings after renormalization group evolution. 
One should start with moderately small values of $\lambda_{\Phi\sigma}$, to prevent its effect in the RGEs making the other quartic couplings, e.g. $\lambda_\sigma$, nonperturbative at scales far below $M_P$.
This can be clearly seen in the approximation $Y_S\approx 0$, where $\beta_{\lambda_\sigma}$ is always positive and hence $\lambda_\sigma$ can only increase. 
For relatively large input values of $\lambda_\sigma$ and small $\lambda_{\Phi\sigma}$, the running of $\lambda_\sigma$ can be approximated as $\beta_{\lambda_\sigma}\propto \lambda_{\sigma}^2$.
Hence, it can encounter a Landau pole at a scale far below the Planck scale.
Similarly, with a very large starting value of $\lambda_\Phi$ and small $\lambda_{\Phi\sigma}$, the running of $\lambda_\Phi$ will be dominated by the term $+24\lambda_\Phi^2$.
As a result one can hit a Landau pole at a scale far below $M_P$.
In Appendix.~\ref{app:few discussions of RGEs} we have discussed in detail each of these scenarios. 
Thus $\lambda_{\Phi\sigma}$ can neither be taken too small, nor too large, only a small optimal parameter range will satisfy both stability and perturbativity constraints.
However, this small optimal range can be probed in an important way by colliders such as the LHC, as we will discuss in the next section.
\section{Collider constraints and invisible Higgs boson Decays}  
\label{sec:invisible decay widths}
In this section we examine how Higgs measurements at LEP and LHC can constrain the parameter space of the Majoron inverse seesaw model. For this we have considered the following scenarios: \\[-.3cm]

\underline{{\bf Case I}}: $m_{H_1} < 125$ GeV with $H_2 \equiv H_{125}$  i.e. $m_{H_2} = 125$~GeV. In this case we have assumed the lighter Higgs scalar $H_1 \equiv H'$
in the mass range $15\,\text{GeV}\leq m_{H_1}\leq 120$~GeV.\\[-.3cm]

\underline{{\bf Case II}}: $m_{H_2} > 125$ GeV with $H_1 \equiv H_{125}$  i.e. $m_{H_1}=125$~GeV. In this case we have taken $m_{H_2} \equiv m_{H'} > 130$~GeV for the heavier Higgs scalar.\\[-.3cm]

We should remind the reader that, by definition, we always take $m_{H_2} > m_{H_1}$.\\[-.3cm]

Rather than discussing collider constraints in terms of quartic couplings, its more convenient to use the mass basis quantities, e.g. scalar masses and mixing angles, since experimental results are
quoted in terms of these quantities.  
In our simple model, the mixing angle $\theta$, the mass $m_{H'}^2$ and the ratio of the two vevs $\tan\beta=\frac{v_\Phi}{v_\sigma}$~(with $v_\Phi=\frac{2m_W}{g}$) can be taken as free parameters,
in terms of which all others can be fixed. \\[-.3cm] 

Before discussing the collider constraints, notice that in the Majoron inverse seesaw extension, the coupling of the Higgs boson to \sm particles gets modified according to the substitution rule 
\begin{align}
h_{\rm{SM}}   \to
\begin{cases}
\sin\theta H' +\cos\theta H_{125}, & \text{for Case I}\\
\cos\theta H_{125} -\sin\theta H', & \text{for Case II}
\end{cases}
\end{align}
The decay widths to \sm states can be obtained from those of the SM with the help of this substitution rule. 

The scalar sector harbors phenomenologically important interactions involving trilinear couplings $H_i JJ$ and $H_i H_j H_j$  given as
\begin{equation}
\mathcal{L}_{H_iJJ} =g_{H_1 JJ} H_1 J^2 + g_{H_2 JJ}H_2 J^2,~~~~~~~~~~
\mathcal{L}_{H_2H_1H_1} =g_{H_2 H_1 H_1} H_2 H_1^2,
\end{equation}
where $J$ denotes the Majoron and 
\begin{align}
g_{H_i JJ}=
\begin{cases}
\frac{\tan\beta}{2v_\Phi} m_{H_i}^2 O_{Ri1},& \text{for Case I where $H_2 = H_{125}$}\\
\frac{\tan\beta}{2v_\Phi} m_{H_i}^2 O_{Ri2},& \text{for Case II where $H_1 = H_{125}$}
\end{cases}
\end{align}
\begin{align}
g_{H_2H_1H_1}^{\text{Case I}}&=
\frac{\tan\beta}{4 v_\Phi}(2 m_{H_1}^2+m_{H_2}^2)\sin 2\theta (\cot\beta\sin\theta-\cos\theta),\,\text{where $H_2 = H_{125}$}\\
g_{H_2H_1H_1}^{\text{Case II}}&=
\frac{\tan\beta}{4 v_\Phi}(2 m_{H_1}^2+m_{H_2}^2)\sin 2\theta (-\cot\beta\cos\theta + \sin\theta),\,\text{where $H_1 = H_{125}$}
\end{align}
The decay widths for $H_i\to JJ$ are given by 
\begin{align}
\Gamma(H_i\to J J)=\frac{g_{H_i JJ}^2}{8\pi m_{H_i}}\sqrt{1-\frac{4 m_{J}^2}{m_{H_i}^2}}
\end{align}
If $m_{H_2}> 2m_{H_1}$, $H_2$ can also decay to $H_1H_1$ with the decay width given by 
\begin{align}
\Gamma(H_2\to H_1 H_1)=\frac{g_{H_2 H_1 H_1}^2}{8\pi m_{H_2}}\sqrt{1-\frac{4 m_{H_1}^2}{m_{H_2}^2}}
\end{align}
These new decay widths will lead to invisible decays as
\begin{align}
\Gamma^{\text{inv}}(H_1)=\Gamma(H_1\to JJ)
\label{H1toJJ}
\end{align}
\begin{align}
\Gamma^{\text{inv}}(H_2)=\Gamma(H_2\to JJ)+\Gamma(H_2\to H_1 H_1\to 4J)
\label{H2toJJ}
\end{align}
Turning to the lighter scalar boson decays to final state $f$ of \sm particles, one finds the branching fractions 
\begin{align}
\text{BR}_f(H_1)=
\begin{cases}
\frac{\sin^2\theta \Gamma_f^\text{SM}(H_1)}{\sin^2\theta \Gamma^\text{SM}(H_1)+\Gamma^{\text{inv}}(H_1)}, & \text{for Case I where $H_2 = H_{125}$}\\
\frac{\cos^2\theta \Gamma_f^\text{SM}(H_1)}{\cos^2\theta \Gamma^\text{SM}(H_1)+\Gamma^{\text{inv}}(H_1)}, & \text{for Case II  where $H_1 = H_{125}$}
\end{cases}
\end{align}
If $\Gamma^{\text{inv}}(H_1)=0$, the branching fraction would be same as that of SM.
The lighter $H_1$ may decay predominantly into pair of majorons depending upon the mass $m_{H_1}$ and mixing angle $\theta$.
The invisible branching ratio for $H_1$ is given by 
\begin{align}
\text{BR}^{\text{inv}}(H_1)=
\begin{cases}
\frac{\Gamma^{\text{inv}}(H_1)}{\sin^2\theta\Gamma^\text{SM}(H_1)+\Gamma^{\text{inv}}(H_1)}, & \text{for Case I where $H_2 = H_{125}$}\\
\frac{\Gamma^{\text{inv}}(H_1)}{\cos^2\theta\Gamma^\text{SM}(H_1)+\Gamma^{\text{inv}}(H_1)}, & \text{for Case II where $H_1 = H_{125}$}
\end{cases}
\end{align}
For the heavier state $H_2$, the branching fraction into SM final state $f$ is 
\begin{align}
\text{BR}_f(H_2)=
\begin{cases}
\frac{\cos^2\theta\Gamma_f^\text{SM}(H_2)}{\cos^2\theta\Gamma^\text{SM}(H_2)+\Gamma(H_2\to JJ)+\Gamma(H_2\to H_1 H_1)}, & \text{for Case I  where $H_2 = H_{125}$}\\
\frac{\sin^2\theta\Gamma_f^\text{SM}(H_2)}{\sin^2\theta\Gamma^\text{SM}(H_2)+\Gamma(H_2\to JJ)+\Gamma(H_2\to H_1 H_1)}, & \text{for Case II  where $H_1 = H_{125}$}
\end{cases}
\end{align}
Similarly, the invisible branching ratio for $H_2$ is given by
\begin{align}
\text{BR}^{\text{inv}}(H_2)=
\begin{cases}
\frac{\Gamma^\text{inv}(H_2)}{\cos^2\theta\Gamma^\text{SM}(H_2)+\Gamma(H_2\to JJ)+\Gamma(H_2\to H_1 H_1)}, & \text{for Case I where $H_2 = H_{125}$}\\
\frac{\Gamma^\text{inv}(H_2)}{\sin^2\theta\Gamma^\text{SM}(H_2)+\Gamma(H_2\to JJ)+\Gamma(H_2\to H_1 H_1)}, & \text{for Case II where $H_1 = H_{125}$}
\end{cases}
\end{align}
The coupling of $H_1$ and $H_2$ to other \sm fermions and gauge bosons are suppressed relative to the standard values by $\sin\theta$~($\cos\theta$) and $\cos\theta$~($\sin\theta$)
for Case I and Case II respectively.  
Hence the single $H_1$ or $H_2$ production cross-sections are given as, 
\begin{align}
\sigma(pp\to H_1)=
\begin{cases}
\sin^2\theta\sigma^{\text{SM}}(pp\to H_1), & \text{for Case I where $H_2 = H_{125}$}\\
\cos^2\theta\sigma^{\text{SM}}(pp\to H_1), & \text{for Case II where $H_1 = H_{125}$}
\end{cases}
\end{align}
\begin{align}
\sigma(pp\to H_2)=
\begin{cases}
\cos^2\theta\sigma^{\text{SM}}(pp\to H_2 ), & \text{for Case I where $H_2 = H_{125}$}\\
\sin^2\theta\sigma^{\text{SM}}(pp\to H_2 ), & \text{for Case II where $H_1 = H_{125}$}
\end{cases}
\end{align}
where $\sigma^{\text{SM}}(pp\to H_1)$ and $\sigma^{\text{SM}}(pp\to H_2)$ are the \sm cross-sections for Higgs production at $m_{H_1}$ and $m_{H_2}$.
Note that they are modified by factors $\sin^2\theta$~($\cos^2\theta$) or $\cos^2\theta$~($\sin^2\theta$) with respect to the conventional ones.

In the following sections, we discuss the constraints on the relevant parameter space of Higgs bosons which follow from searches performed at LEP as well as LHC.
In what follows we will discuss both Case I and Case II. 
In our numerical scans, we restricted the range of the singlet vev to $v_\sigma\in [0.1 \,\text{TeV}, 1\, \text{TeV}]$ and $v_\sigma\in [1 \,\text{TeV}, 3\, \text{TeV}]$ for Cases I and II, respectively.

\section{Case I: Lightest CP even scalar below 125 GeV i.e. $H_2 = H_{125}$}
\label{sec:Constraints for Case I}

We now examine the experimental limits coming from the LHC and LEP experiments, starting with the constraints for Case I. 
Note that in this case $H_2 = H_{125}$ is the SM-like Higgs boson, with $m_{H_2}=125$ GeV, and the mass of the lighter scalar boson $H_1 = H'$ lies in the range $15\,\text{GeV}\leq m_{H'}\leq 120$~GeV. 

\subsection{LEP constraints in the presence of invisible Higgs decays}
\label{LEP limit}

\begin{figure}[t]
\centering
\includegraphics[width=0.45\textwidth]{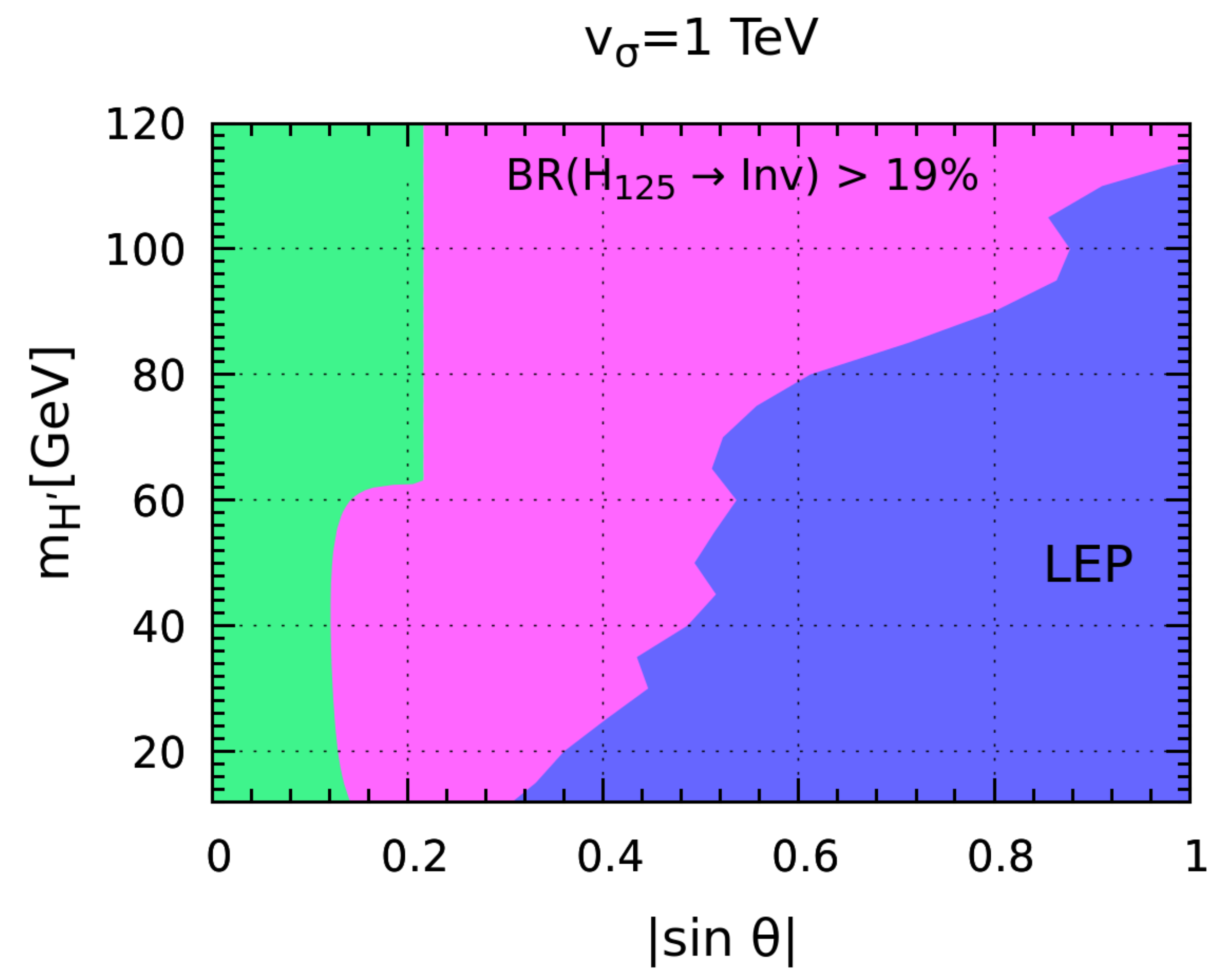}
\includegraphics[width=0.45\textwidth]{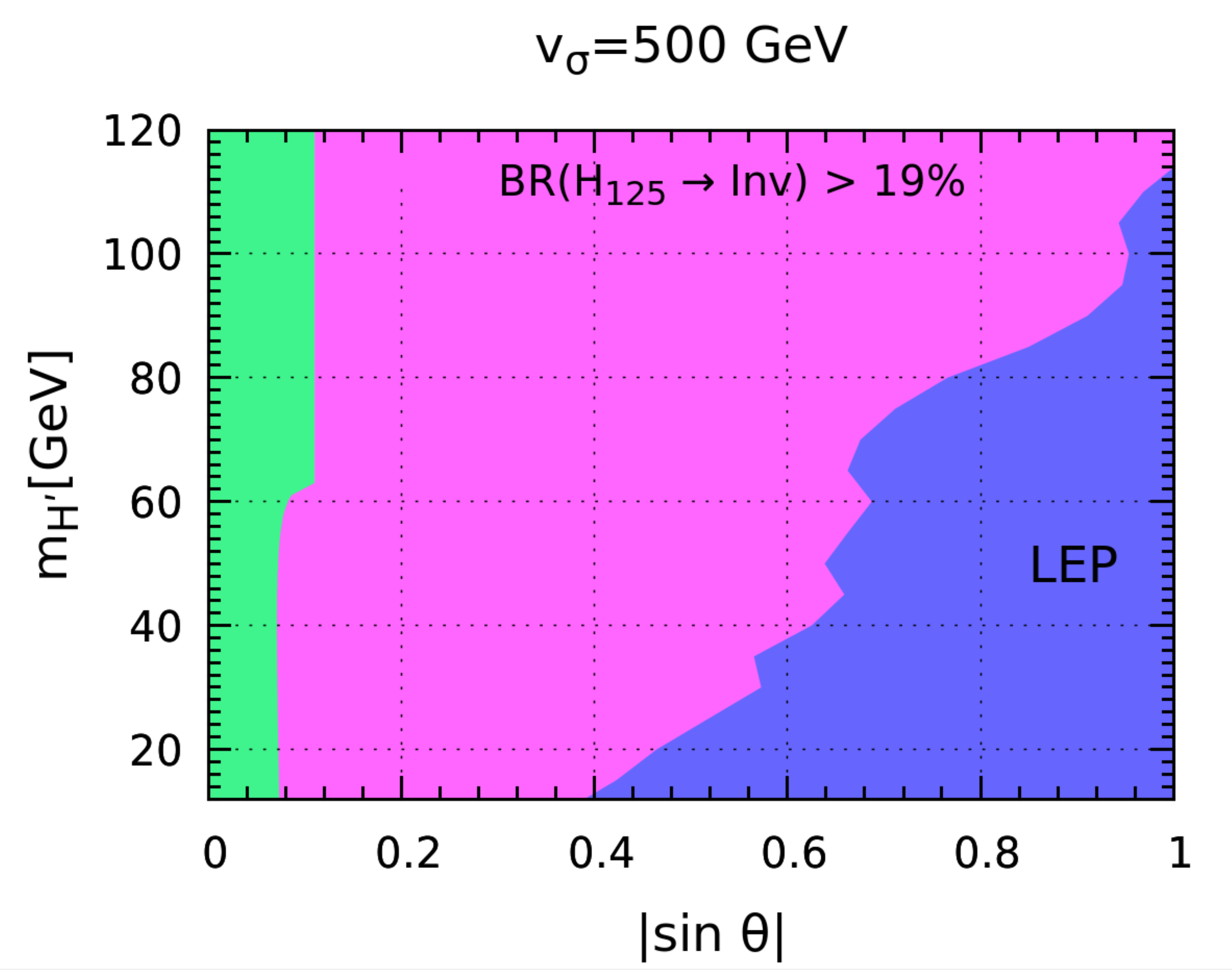}
\includegraphics[width=0.45\textwidth]{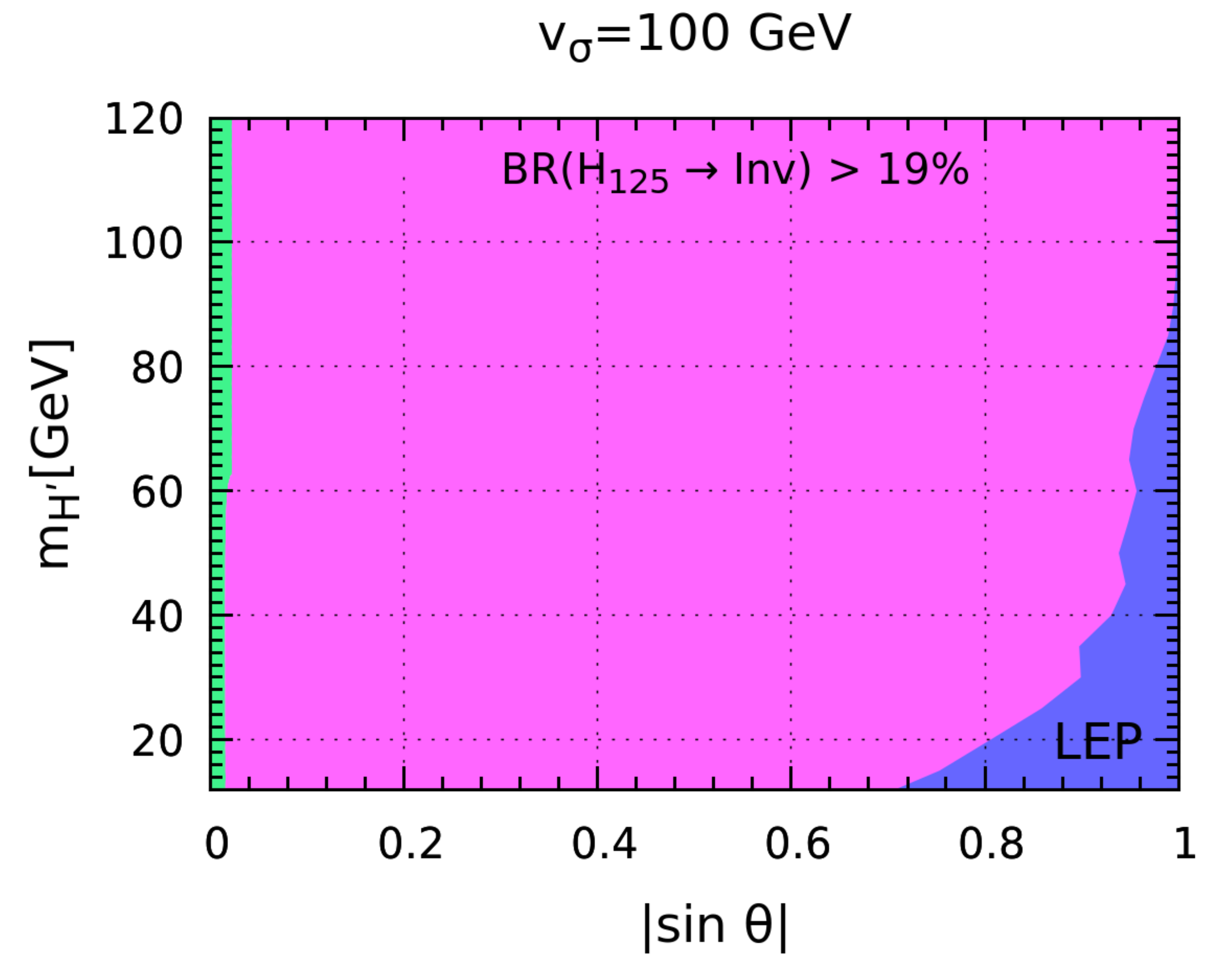}
\caption{\footnotesize{Exclusion region of $m_{H'}$ versus $|\sin\theta|$ for $v_\sigma$=1 TeV~(top left panel), $v_\sigma=0.5$ TeV~(top right panel) and $v_\sigma=0.1$ TeV~(bottom panel).
    The blue regions are excluded by LEP results. The magenta regions correspond to an invisible BR greater than $19\%$ and are therefore excluded by LHC~\cite{Sirunyan:2018koj}.}}
\label{fig:LEPandBR}
\end{figure}

Soon after the start of the LEP experiment, it was realized that in theories with spontaneously broken lepton number the invisible decay of the Higgs boson~\cite{Joshipura:1992hp} had clear impact
on $e^+ e^-$ scattering experiments~\cite{Romao:1992zx,LopezFernandez:1993tk,DeCampos:1994fi,deCampos:1996bg}. 
Let's first start with the constraints that follow from the LEP collider~\cite{Abdallah:2004wy}. For case I these constraints apply to the lighter Higgs boson, $H'$. 
Due to the presence of the invisible decay channel, the visible decay rates get modified. 
For the channel $e^+e^-\to Z H'\to Zb\bar{b}$, the final state is expressed in terms of the SM $h Z$ cross section through 
\begin{align}
\sigma_{H' Z\to b\bar{b}Z}&=\sigma^{\text{SM}}_{h Z}\times R_{H' Z}\times \text{BR}(H' \to b\bar{b})\\ \nonumber
&=\sigma_{hZ}^{\text{SM}}\times C_{Z(H' \to b\bar{b})}^2
\end{align}
where $\sigma_{hZ}^{\text{SM}}$ is the standard cross section, and $R_{H' Z}$ is the suppression factor related to the coupling of the Higgs boson to the gauge boson Z. 
Of course we have $R_{hZ}^{\text{SM}}=1$ in the SM.
In our case, $\text{BR}(H' \to b\bar{b})$ is modified with respect to \sm due to the presence of invisible Higgs decay $H' \to JJ$, see Eq.~(\ref{H1toJJ}).
Thus, in our model $R_{H' Z}=\sin^2\theta$. 
Ref.~\cite{Abdallah:2004wy} gives upper bounds on $C_{Z(H\to b\bar{b})}^2$ for the lightest $CP$-even scalar boson mass in the range from 15 GeV upto 120 GeV. 
From this one can determine the regions of $m_{H'}-|\sin\theta|$ which are currently allowed by the LEP-II searches.
The results are shown in Fig.~\ref{fig:LEPandBR} for three benchmarks values; $v_\sigma=1$ TeV~(upper left panel), $v_\sigma=500$ GeV~(upper right panel) and $v_\sigma=100$ GeV~(bottom panel).
The blue regions are excluded by LEP results. As the $H'$ coupling to $Z$ boson is reduced with respect to that of the SM, lighter masses become allowed. 
%

In Fig.~\ref{fig:LEPandBR}, in addition to the LEP constraints, we have also showed in magenta color the constraints on the invisible decay of $H_{125}$ coming from the LHC.  
As one can see, these constraints supersede those from LEP, and severely restrict the allowed parameter space.
They come from the current upper bound on the branching ratio to invisible decay modes $\text{BR}(H_{125} \to\text{Inv})\leq 19\%$ given by the CMS collaboration~\cite{Sirunyan:2018koj} and a similar one from ATLAS~\cite{Aaboud:2019rtt}.
The results are shown in Fig.~\ref{fig:LEPandBR} with magenta color. 
Notice that the green allowed region has a kink. It is associated with the decay $H_{125}\to H' H'$ for $m_{H'} < \frac{m_{H_{125}}}{2}$. 
Furthermore, as clear from Fig.~\ref{fig:LEPandBR}, smaller $v_\sigma$ values lead to larger invisible decay rate and larger exclusion. 
Also, with decreasing $v_\sigma$, the ratio $\frac{\Gamma(H_{125}\to 4J)}{\Gamma(H_{125}\to JJ)}$ decreases, so the kink gets less prominent. 
This in turn implies larger $\text{BR}(H_{125}\to\text{Inv})$ and smaller $\text{BR}(H'\to b\bar{b})$.
As a result we get stronger exclusion limit from $\text{BR}(H_{125}\to\text{Inv})$ and weaker exclusion limit from LEP results. 
This fact is clearly visible when comparing different panels of Fig.~\ref{fig:LEPandBR}.

\subsection{LHC Constraints in the presence of invisible Higgs decays}  
\label{LHC Constraints}

We now turn to invisible Higgs decays at hadron colliders~\cite{Romao:1992dc}.
Apart from the above LHC limit, we also have the LHC measurements of several visible decay modes of the 125 GeV Higgs boson.
These are given in terms of the so-called signal strength parameters, 
\begin{align}
\mu_f &=\frac{\sigma^{\text{NP}}(pp\to h)}{\sigma^{\text{SM}}(pp\to h)} \frac{\text{BR}^{\text{NP}}(h\to f)}{\text{BR}^{\text{SM}}(h\to f)} \nonumber  \\
&=\frac{\sigma^{\text{NP}}(pp\to h)}{\sigma^{\text{SM}}(pp\to h)} \frac{\Gamma^{\text{NP}}(h\to f)}{\Gamma^{\text{SM}}(h\to f)} \frac{\Gamma^{\text{SM}}(h\to \text{all})}{\Gamma^{\text{NP}}(h\to \text{all})}
\end{align}
where $\sigma$ is the cross-section for Higgs production, NP and SM stand for new physics and SM respectively. 

For the 8 TeV data, we list the results for signal strength parameters from combined ATLAS and CMS analysis~\cite{TheATLASandCMSCollaborations:2015bln} in Table.~\ref{tab:1}. 
\begin{table}[h]
\centering
\begin{tabular}{|c | c | c | c |}
\hline
\hspace{0.25cm} Channel \hspace{0.25cm} &  \hspace{0.25cm} ATLAS  \hspace{0.25cm} & \hspace{0.35cm} CMS  \hspace{0.35cm} & \hspace{0.25cm}ATLAS+CMS \hspace{0.25cm} \\
\hline
$\mu_{\gamma\gamma}$   &
$1.15^{+0.27}_{-0.25}$   &
$1.12 ^{+0.25}_{-0.23}$  &
$1.16^{+0.20}_{-0.18}$
\\*[2mm]
$\mu_{WW}$   &
$1.23^{+0.23}_{-0.21}$  &
$0.91^{+0.24}_{-0.21}$   &
$1.11^{+0.18}_{-0.17}$
\\*[2mm]
$\mu_{ZZ}$   &
$1.51^{+0.39}_{-0.34}$ &
$1.05^{+0.32}_{-0.27}$  &
$1.31^{+0.27}_{-0.24}$
\\*[2mm]
$\mu_{\tau\tau}$  &
$1.41^{+0.40}_{-0.35}$ &
$0.89^{+0.31}_{-0.28}$ &
$1.12^{+0.25}_{-0.23}$
\\*[2mm]
\hline
\end{tabular}
\caption{\label{tab:1} Combined ATLAS and CMS results for the 8 TeV data, Ref.~\cite{TheATLASandCMSCollaborations:2015bln}.} 
\end{table}

For the 13 TeV Run-2, there is no combined final data so far, and the available data is separated by production processes.
Table~\ref{tab:2} compiles the recent results from ATLAS~\cite{Aad:2019mbh}.
\begin{table}[ht]
\begin{center}
\begin{tabular}{|c|c|c|c|c|}
    \hline
 \hspace{0.25cm} Decay \hspace{0.25cm} &\multicolumn{4}{c|}{Production Processes}\\[+2mm]\cline{2-5}
  Mode & \hspace{0.35cm} \texttt{ggF} \hspace{0.35cm} & \hspace{0.35cm} \texttt{VBF} \hspace{0.35cm} & \hspace{0.35cm} \texttt{VH} \hspace{0.35cm} & \hspace{0.35cm} \texttt{ttH} \hspace{0.35cm}  \\[+2mm]
  \hline
  \vb{18}   $H\to \gamma\gamma$
         &$0.96^{+0.14}_{-0.14}$
         &$1.39^{+0.40}_{-0.35}$ 
         &$1.09^{+0.58}_{-0.54}$
         &$1.10^{+0.41}_{-0.35}$ \\[+2mm]
   \vb{18}   $H\to ZZ$
         &$1.04^{+0.16}_{-0.15}$
         &$2.68^{+0.98}_{-0.83}$
         &$0.68^{+1.20}_{-0.78}$
         &$1.50^{+0.59}_{-0.57}$ \\[+2mm]
  \vb{18}  $H\to WW$
         &$1.08^{+0.19}_{-0.19}$
         &$0.59^{+0.36}_{-0.35}$
         &$-$
         &$1.50^{+0.59}_{-0.57}$ \\[+2mm] 
\hline
 \vb{18}   $H\to \tau\tau$
         &$0.96^{+0.59}_{-0.52}$
         &$1.16^{+0.58}_{-0.53}$
         &$-$
         &$1.38^{+1.13}_{-0.96}$ \\[+2mm]
   \vb{18}   $H\to bb$
          &$-$
          &$3.01^{+1.67}_{-1.61}$ 
          &$1.19^{+0.27}_{-0.25}$
          &$0.79^{+0.60}_{-0.59}$ \\[+2mm]
  \hline
\end{tabular}
\end{center}
\caption{ATLAS results for 13 TeV data, taken from Ref.~\cite{Aad:2019mbh}}
\label{tab:2}
\end{table}

\begin{figure}[h]
\centering
\includegraphics[width=0.45\textwidth]{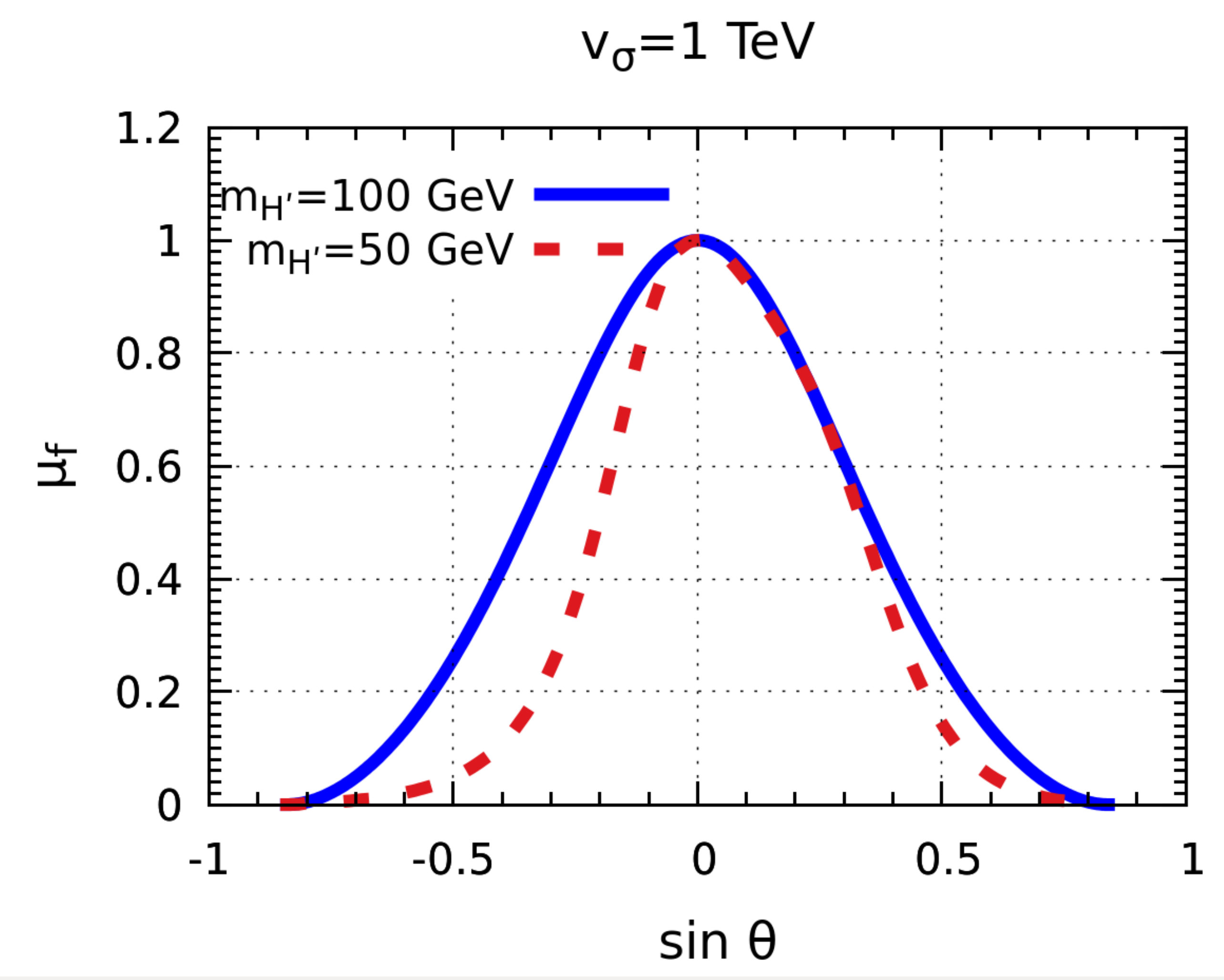}
\includegraphics[width=0.45\textwidth]{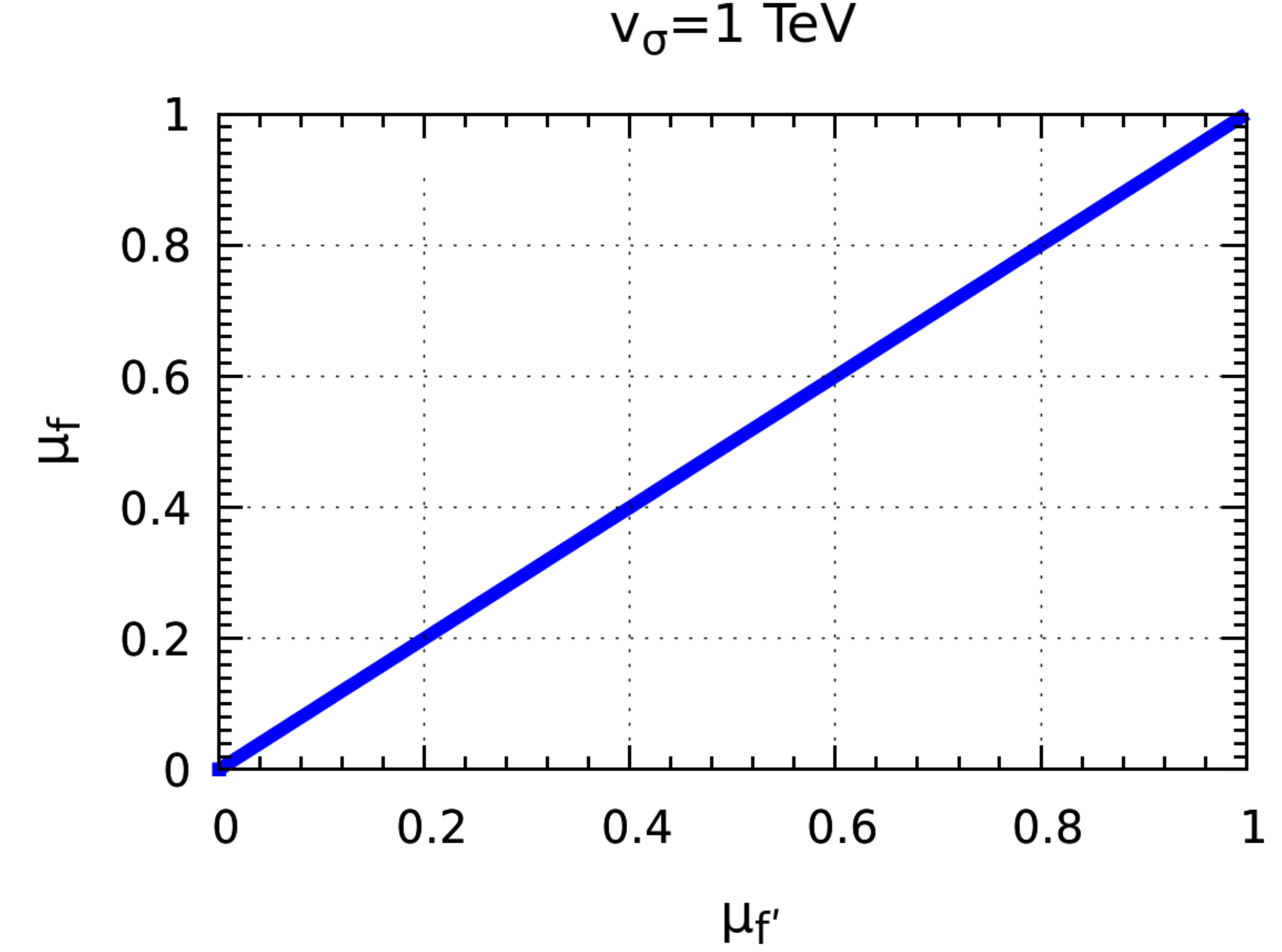}
\caption{\footnotesize{\textbf{Left panel}: The signal strength parameter $\mu_f$ versus $\sin\theta$ for light Higgs masses, $m_{H'}=100$ GeV and 50 GeV. 
    Note that for our model the rate never exceeds the \sm prediction which is 1. \textbf{Right panel}: Correlation between $\mu_f$ and $\mu_{f^\prime}$, where $f\neq f^{\prime}$.
    The straight line reflects the fact that, once  we fix the two Higgs masses, there is essentially only one free parameter left, the mixing angle $\theta$. }}
\label{muf range}
\end{figure}

We note that in our model, the expected signal strength parameter $\mu_f$ for any \sm final state $f$ can only be less than unity, as shown in the left panel of Fig.~\ref{muf range}. 
The left panel of Fig.~\ref{muf range} is plotted for two benchmark values of the light Higgs mass $m_{H'}=100$ GeV~(blue line) and $m_{H'}=50$ GeV~(red dashed line).   
Note that, due to the opening up of the new channel $H_{125} \to H' H'$, the red dashed line is asymmetric in $\theta$, since $g_{H_{125} H' H'}$ is an asymmetric function of mixing angle $\theta$.

In the right panel of Fig.~\ref{muf range}, we show the correlation between $\mu_f$ and $\mu_{f^{\prime}}$ where $f\neq f^{\prime}$.
The straight line reflects the fact that  once we fix two Higgs masses, there is essentially only one free parameter, the mixing angle $\theta$. 
%
Current LHC results indicate that $\mu_f\sim 1$. Taking into account the limits in Tab.~\ref{tab:1} and Tab.~\ref{tab:2}, we assume that the LHC allows for deviations in the range given in
Eq.~\ref{eq:muf}.

\begin{figure}[h]
\centering
\includegraphics[width=0.45\textwidth]{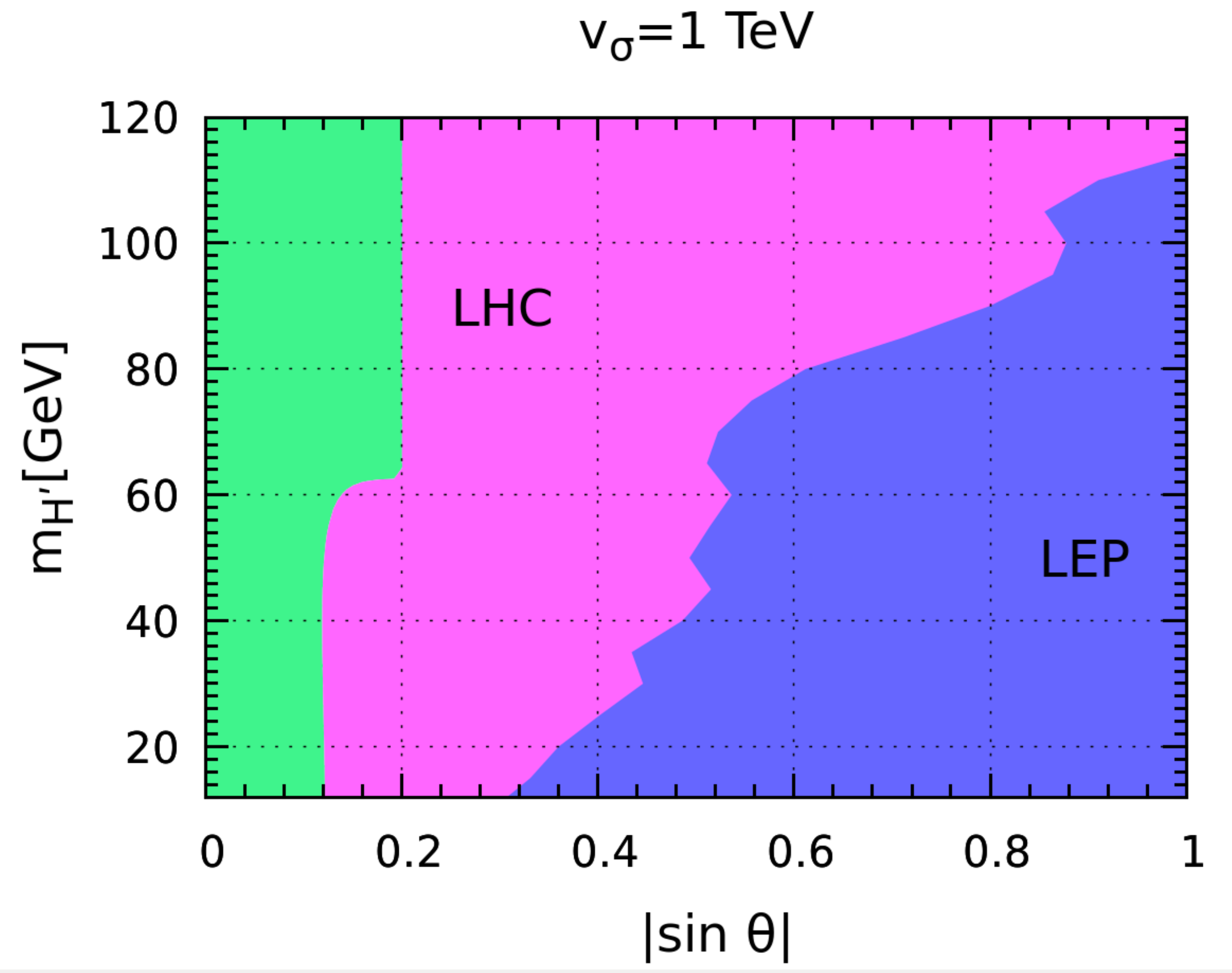}
\includegraphics[width=0.45\textwidth]{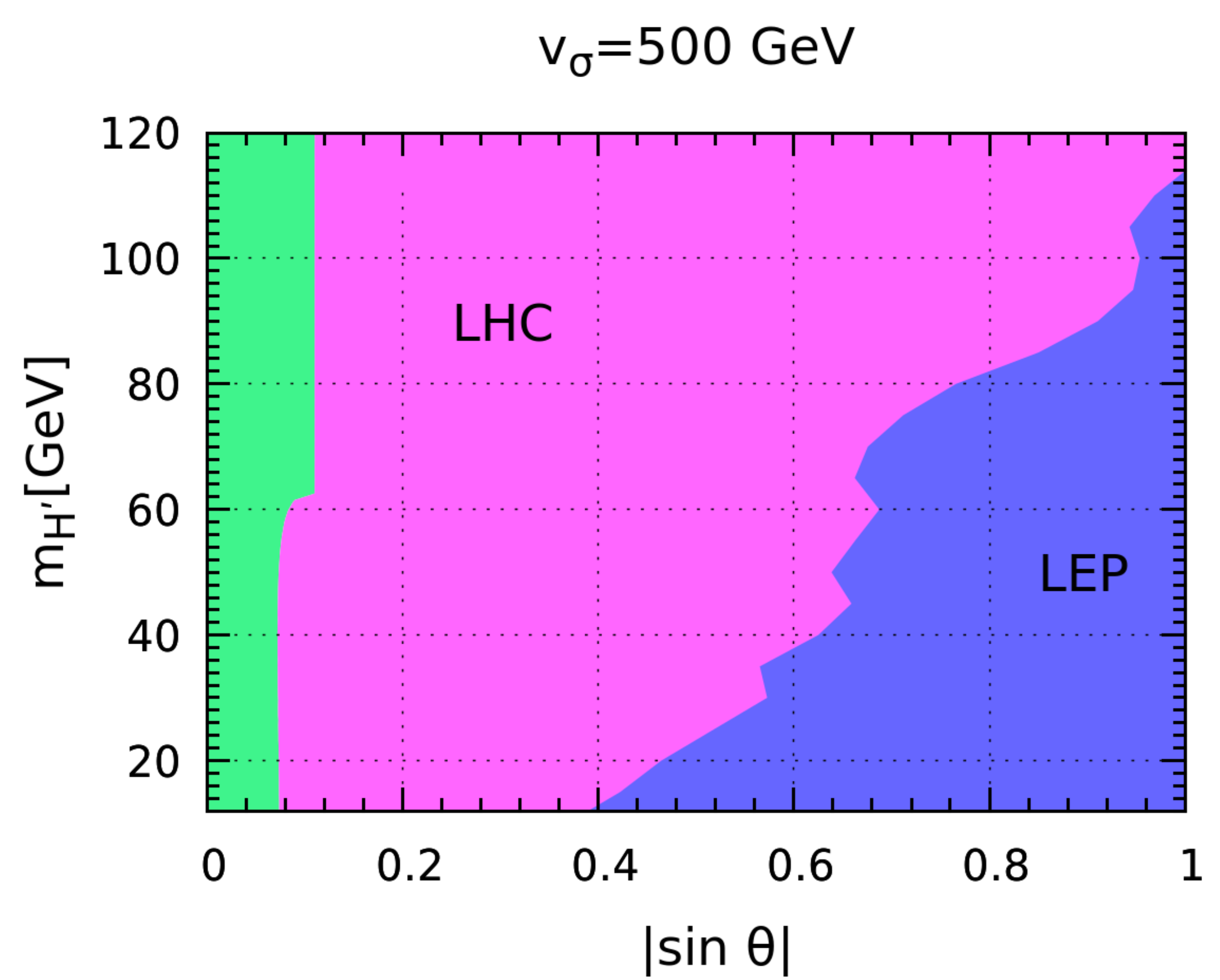}
\includegraphics[width=0.45\textwidth]{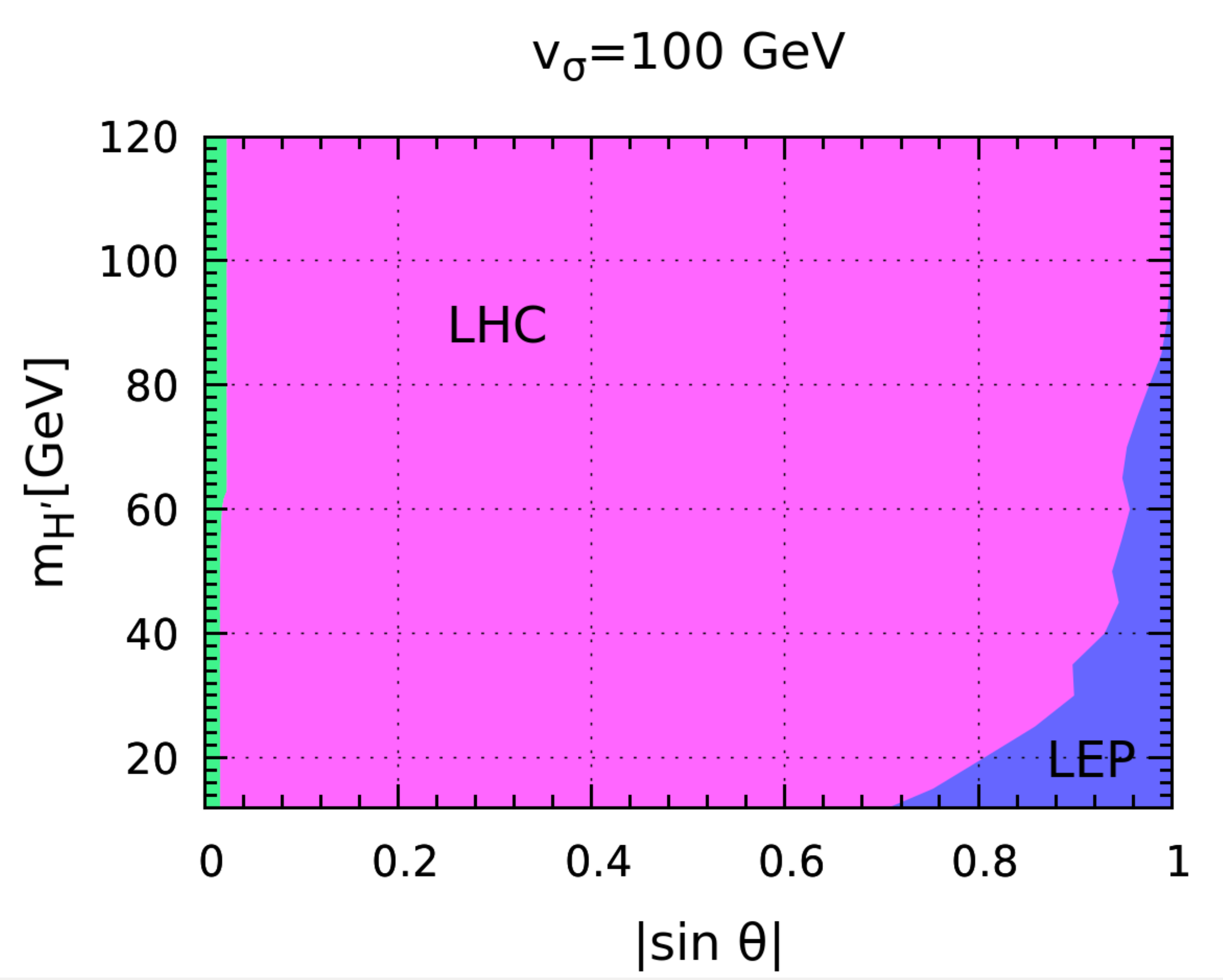}
\caption{\footnotesize{Constraints on $m_{H'}$ versus $\sin\theta$ for $v_\sigma=1$ TeV~(top left panel),  $v_\sigma=0.5$ TeV~(top right) and $v_\sigma=0.1$ TeV~(bottom panel).
    The blue regions are excluded by LEP results and the magenta regions are excluded by the LHC constarint $0.8\leq\mu_f\leq 1$. The green regions pass all the constraints.}}
\label{muWW and LEP}
\end{figure}

Fig.~\ref{muWW and LEP} gives the allowed parameters in the $m_{H'} - |\sin\theta|$ plane obtained by taking the signal strength range of the visible decay channels, $\mu_f$ from LHC
and also the above LEP limits. 
In Fig.~\ref{muWW and LEP} we take $v_\sigma=1$ TeV~(top left panel), 500 GeV~(top right panel) and 100 GeV~(bottom panel). 
As before, the blue region is excluded from LEP, while the magenta region is excluded from the LHC constraint in Eq.~(\ref{eq:muf}). 
The green region is allowed by the LHC limit. As before, the kink is associated with the opening of the decay channel $H_{125}\to H'H'$ for $m_{H'} < \frac{m_{H_{125}}}{2}$. 
%

\begin{figure}[h]
\centering
\includegraphics[width=0.45\textwidth]{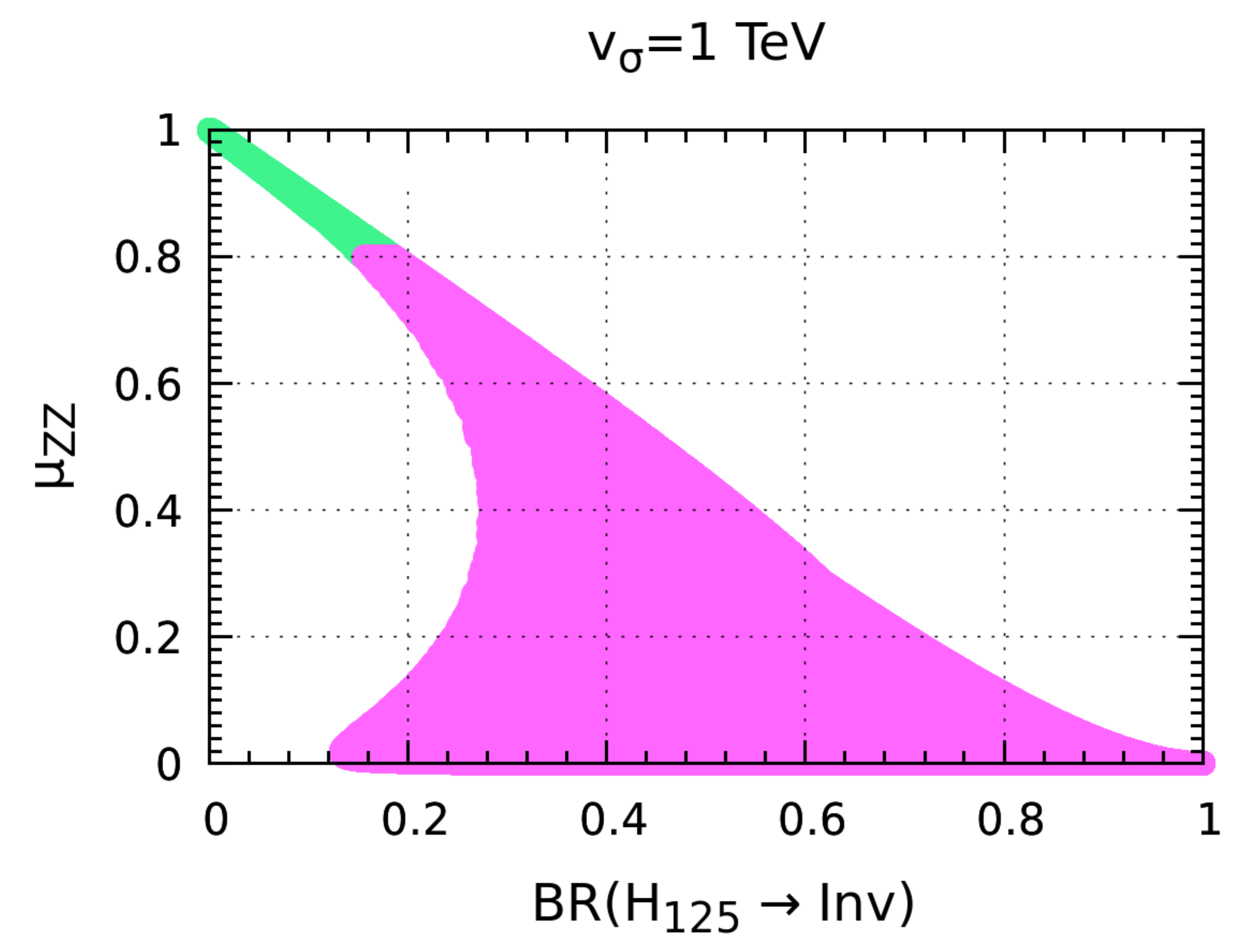}
\includegraphics[width=0.45\textwidth]{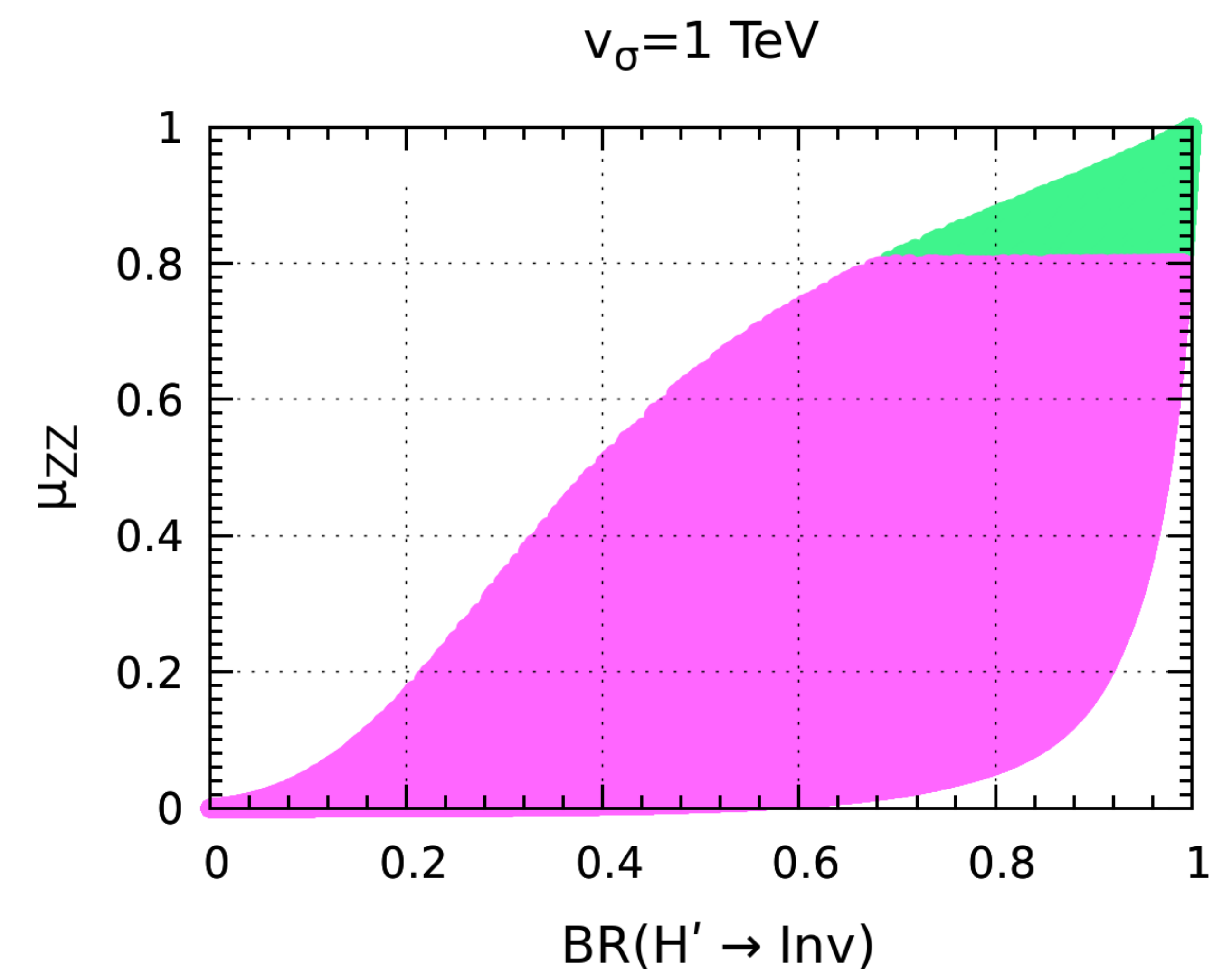}
\caption{\footnotesize{\textbf{Left panel}: $\mu_{VV}$ versus $\text{BR}(H_{125}\to\text{Inv})$.
 \textbf{ Right panel}: $\mu_{VV}$ versus $\text{BR}(H' \to \text{Inv})$. Same color code as used in Fig.~\ref{muWW and LEP}.}}
\label{muZZ vs Invisible}
\end{figure}

\begin{figure}[h]
\centering
\includegraphics[width=0.45\textwidth]{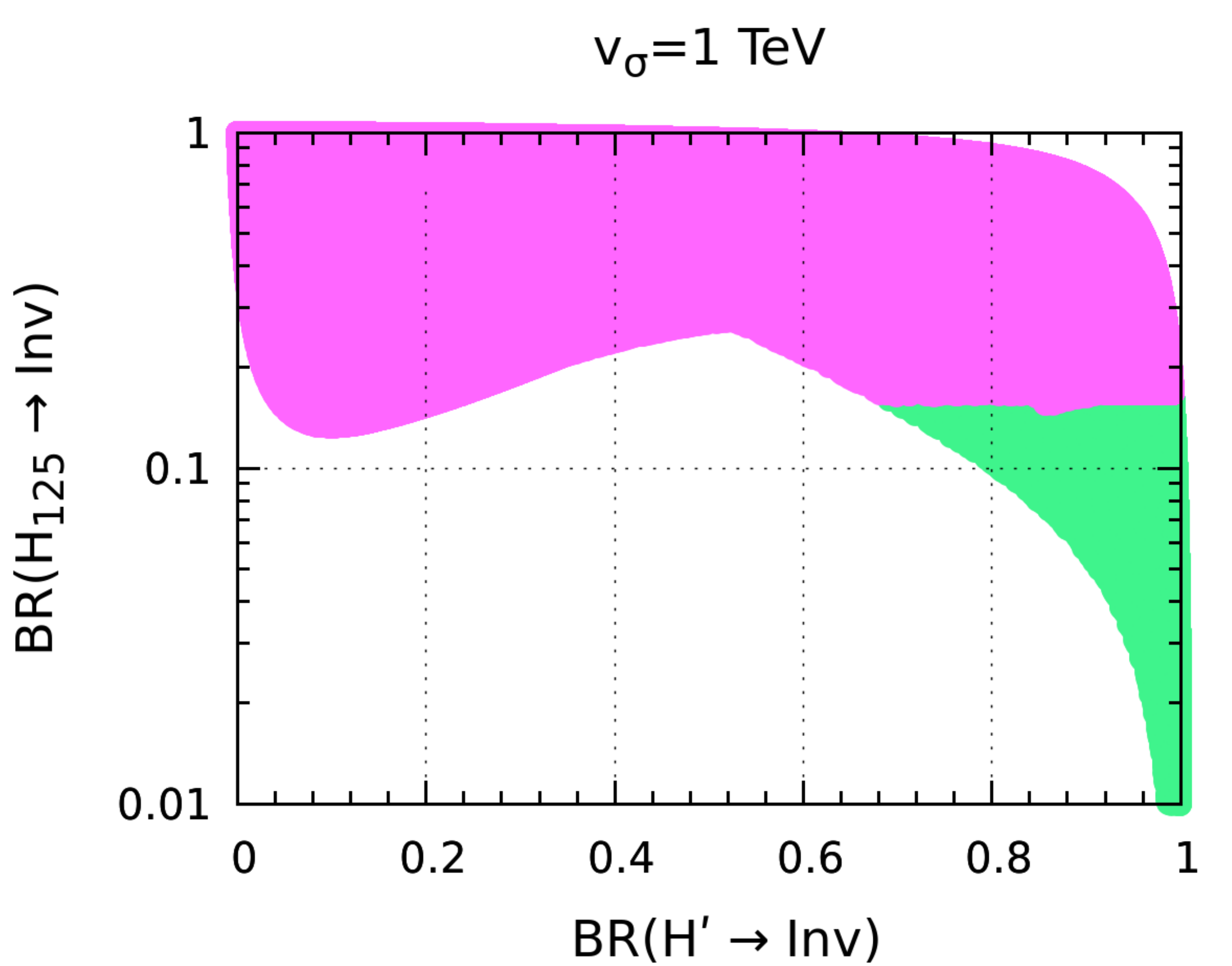}
\includegraphics[width=0.45\textwidth]{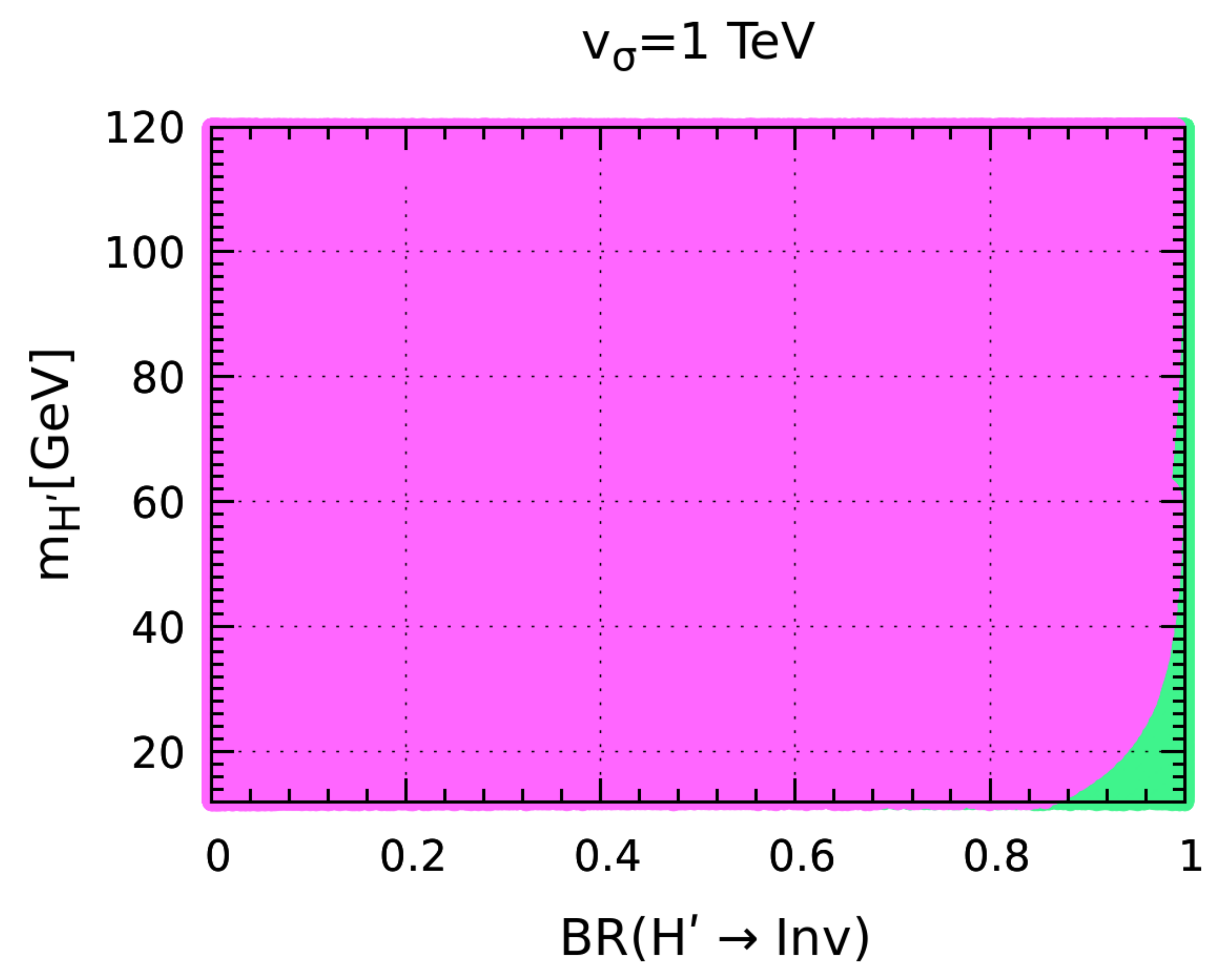}
\caption{\footnotesize{\textbf{Left panel}: $\text{BR}(H_{125} \to \text{Inv})$ as afunction of $\text{BR}(H' \to \text{Inv})$.
    \textbf{ Right panel}: $m_{H'}$ as a function of $\text{BR}(H' \to \text{Inv})$. The color codes are as in Fig.~\ref{muWW and LEP}.}}
\label{Invisible vs Invisible}
\end{figure}

The simplicity of our model implies strong correlations among visible and invisible scalar boson decays. 
In Figs.~\ref{muZZ vs Invisible} and \ref{Invisible vs Invisible} we plot the correlations between $\mu_f$, $\text{BR}(H_{125} \to \text{Inv})$ and $\text{BR}(H' \to \text{Inv})$ for $v_\sigma=1$ TeV.
The color code is the same as in Fig.~\ref{muWW and LEP}. 
The left panel of Fig.~\ref{muZZ vs Invisible} shows that in our model, due to the LHC limit Eq.~(\ref{eq:muf}), the maximum invisible branching ratio of $H_{125}$ is about $20\%$. 
Whereas for the lighter $H'$, the LHC limit Eq.~(\ref{eq:muf}) imples that it should decay mainly via the invisible mode, as shown in the right panel of Fig.~\ref{muZZ vs Invisible}.
In fact, the invisible branching ratio of $H' \to \rm{Inv}$ cannot be less than $\sim 70\%$. 
%
%

The invisible branching ratio of the two scalar bosons $H_{125}$ and $H'$ are shown in the left panel of Fig.~\ref{Invisible vs Invisible} with same color code as in Fig.~\ref{muWW and LEP}. 
This plot again confirms that, while for $H_{125}$ the invisible branching ratio cannot exceed $20\%$, the lighter $H'$ primarily decays in the invisible mode with $\rm{BR}(H' \to \rm{Inv}) \gsim 70\%$.
Finally, in the right panel of Fig.~\ref{Invisible vs Invisible}, we plot $m_{H'}$ versus $\text{BR}(H'\to \text{Inv})$, with the same conventions. 
One sees that, for $m_{H'} \gtrsim 40$ GeV the $H'$ decays almost exclusively in the invisible mode. 
%


\section{Case II: lightest CP-even scalar $H_1 = H_{125}$ is the 125~GeV Higgs}
\label{sec:Constraints for Case II}

In this section we describe the constraints for Case II, in which the lightest CP even scalar is the Standard-Model-like Higgs boson with $m_{H_1}=125$~GeV,
while the heavier one is $H_2 = H'$ with $m_{H'} \geq 130$ GeV. 
As in Case I, we can use the LHC upper limit on the Higgs boson invisible decay~Eq.~(\ref{eq:inv}) and the constraints on Higgs signal strength parameters~in Eq.~(\ref{eq:muf}).
These bounds can then be translated in terms of restrictions on  $|\sin\theta|$ for $m_{H'}>130$ GeV.  

\begin{table}[h]
\centering
\begin{tabular}{|c|cc|cc|}
\hline
         &  & Upper limit on $|\sin\theta|$  & & Upper limit on $|\sin\theta|$       \\
$v_\sigma$ & & from $\mu_f$  & & from $\text{BR}_{H_{125}}^{\text{Inv}}\leq 19\%$ \\
\hline
$700$~GeV  & &
$0.150$ & &
$0.154$
\\*[2mm]
$1$ TeV & & 
$0.201$ & &
$0.218$
\\*[2mm]
$2$ TeV & &
$0.317$ & &
$0.417$
\\*[2mm]
$3$ TeV & &
$0.375$  & &
$0.586$\\
\hline
\end{tabular} 
\caption{\label{tab:3} Maximum allowed values of $|\sin\theta|$ from~Eqs.~(\ref{eq:muf}) and (\ref{eq:inv}).}
  \end{table}
%
%
\begin{figure}[h]
\centering
\includegraphics[width=0.55\textwidth]{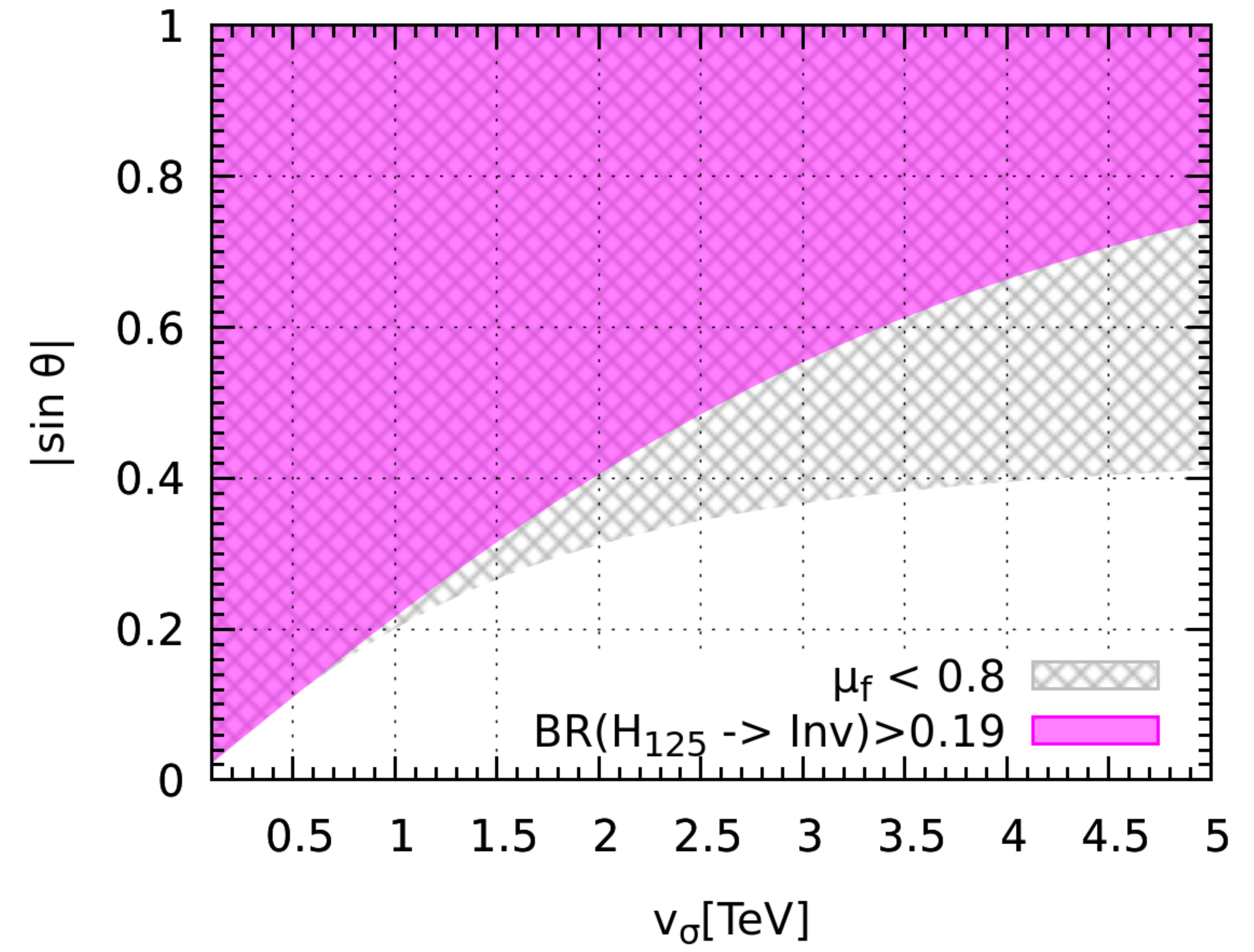}
\caption{\footnotesize{The shaded areas on $|\sin\theta|$ versus $v_\sigma$ are ruled out by the present limit on the invisible Higgs decay in~Eq.~(\ref{eq:inv}) (magenta) and 
    the constraints on the signal strength parameter $\mu_f$ in Eq.~(\ref{eq:muf})~(gray).}}
\label{present limit for case 2}
\end{figure}
%
In Table~\ref{tab:3} and Fig.~\ref{present limit for case 2}, we have give the maximum allowed values of $|\sin\theta|$ from the LHC constraints in~Eqs.~(\ref{eq:inv}) and Eq.~(\ref{eq:muf}). 
As can be seen from Fig.~\ref{present limit for case 2}, for low values of $v_\sigma$ up to around 1 TeV, both~Eq.~(\ref{eq:inv}) and Eq.~(\ref{eq:muf}) lead to similar limits on $|\sin\theta|$. 
However, for $v_\sigma > 1$ TeV the limit from Eq.~(\ref{eq:inv}) gets relaxed since, the larger the $v_\sigma$, the smaller the invisible decay mode $H_{125}\to JJ$.
As a result, for larger $v_\sigma$ values the Higgs invisible decay gives a weaker exclusion limit on $|\sin\theta|$ than that coming from $\mu_f$. 

\begin{figure}[h]
\centering
\includegraphics[height=5.5cm,width=0.48\textwidth]{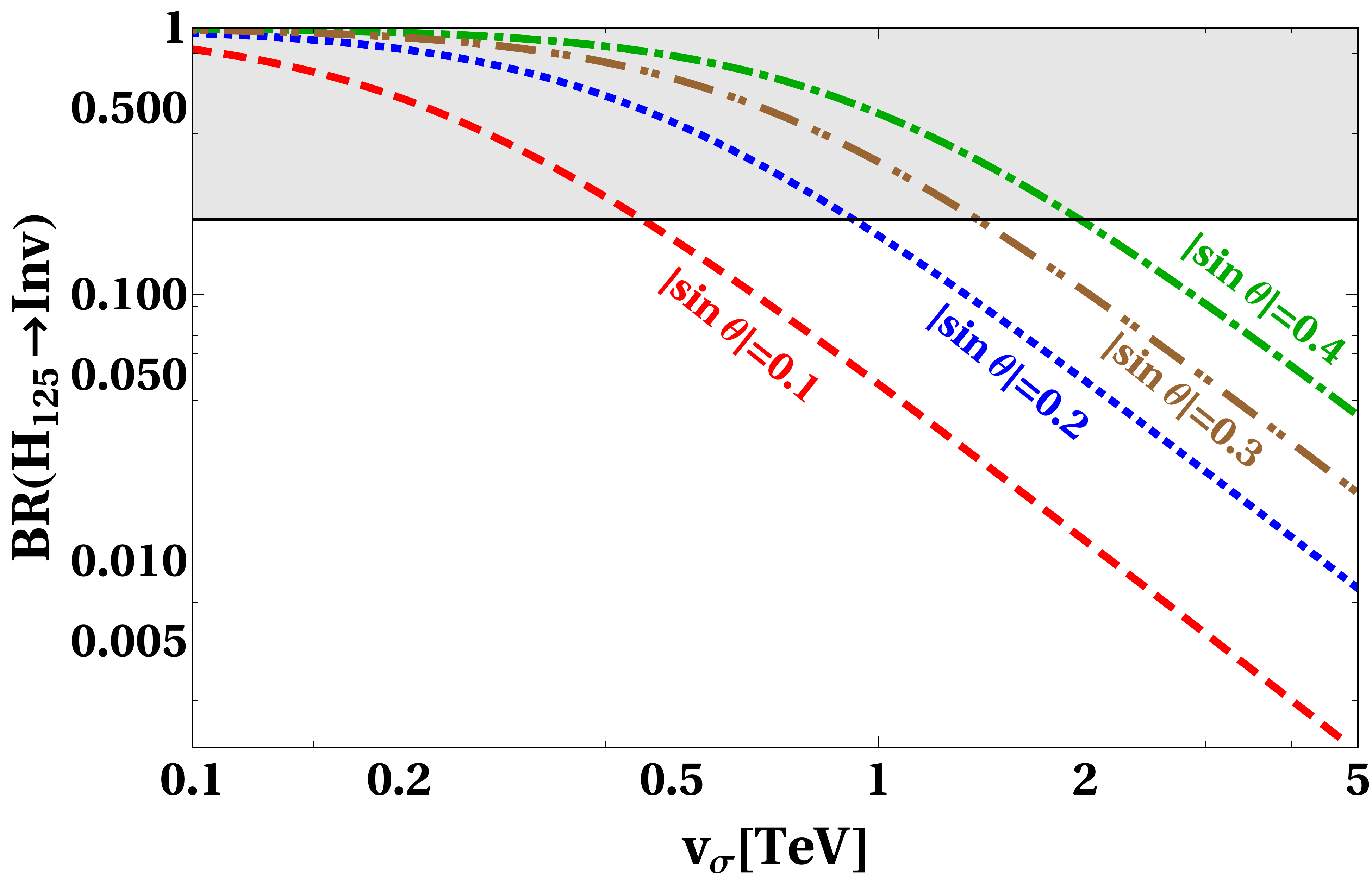}
\includegraphics[height=5.5cm,width=0.48\textwidth]{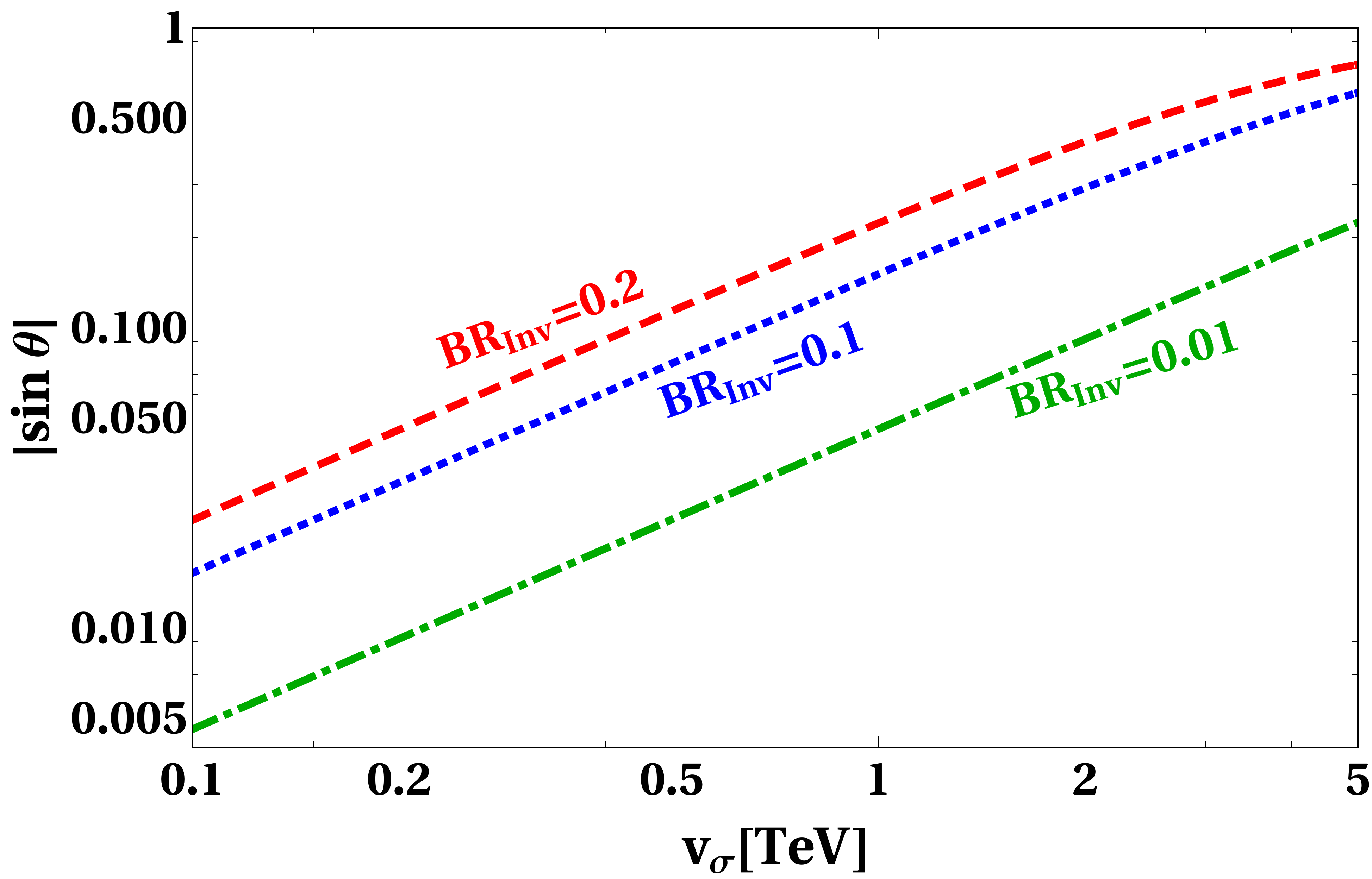}
\caption{\footnotesize{
    \textbf{Left panel}: $\text{BR}(H_{125}\to\text{Inv})$ vs $|\sin\theta|$ for different choices of $|\sin\theta|=0.1$~(red dashed), 0.2~(blue dotted), 0.3~(brown double dot-dashed) and 0.4~(green dot-dashed). The black shaded region is excluded by the LHC limit in Eq.~(\ref{eq:inv}).
    \textbf{Right panel}: Exclusion contours in the plane $|\sin\theta|$ vs $v_{\sigma}$ corresponding to an invisible Higgs branching ratio $\text{BR}(H_{125}\to\text{Inv})$ excluded at $20\%$~(red dashed), $10\%$~(blue dotted) and $1\%$~(green dot-dashed), respectively. The regions above each line are excluded.}}
\label{fig:invisible-vs-vev1}
\end{figure}

Notice that the Higgs invisible branching ratio changes with the scale of dynamical breaking of lepton number, i.e. value of $v_\sigma$ that triggers neutrino mass generation.
Fig.~\ref{fig:invisible-vs-vev1} shows how one can get information on this fundamental scale by Higgs boson measurements. 
These plots show indeed that the Higgs invisible branching ratio can be used to probe the scale of spontaneous lepton number breaking $v_\sigma$. 
As can be seen from Fig.~\ref{fig:invisible-vs-vev1} for a fixed value of the mixing angle $\sin \theta$ the Higgs invisible branching ratio varies in a monotonic fashion with $v_\sigma$. 
Thus, one can use other LHC results i.e. signal strength measurements to obtain limits on $\sin \theta$ and then use Fig.~\ref{fig:invisible-vs-vev1} to constrain the scale of dynamical lepton number breaking. 
For example, for $|\sin\theta| = 0.1$ we find that $v_\sigma$ cannot be less than $500$ GeV while for $|\sin\theta| = 0.2$  $v_\sigma$ cannot be less than $900$ GeV.
Future improvement on the Higgs invisible branching ratio measurement can be used to further constrain the scale of dynamical lepton number breaking, as depicted in Fig.~\ref{fig:invisible-vs-vev1}. 

Finally, as in Case I, one can use~Eqs.~(\ref{eq:muf}) and (\ref{eq:inv}) to obtain correlations between different observables, analogous to the Fig.~\ref{muZZ vs Invisible} and Fig.~\ref{Invisible vs Invisible}. 
However, we will not show these plots explicitly.
Instead, we will make use of such constraints and correlations in conjunction with the vacuum stability constraints in order to obtain complementary limits in Section.~\ref{sec:Vacuum Stability}.

\section{perturbativity and vacuum stability} 
\label{sec:Vacuum Stability}

We now examine the combined implications of collider constraints in conjunction with the restrictions that follow from vacuum stability and perturbativity of the theory. 
As we will see, these two sets of constraints give complementary information on the Majoron inverse seesaw model. 
In most cases, vacuum stability is threatened by the violation of the condition $\lambda_\Phi>0$ and $\lambda_\sigma>0$.
In order to have a stable vacuum one also needs relatively large values of $\lambda_{\Phi \sigma}$, which means non-negligible mixing parameter $|\sin\theta|$ between the two CP even scalars. 
On the other hand the LEP and LHC provide stringent bounds on the mixing between the two CP even scalars i.e. they require small values of $|\sin\theta|$.
We now dicuss this interplay in more detail for both Case I and Case II. 
In subsequent sections we show stable, unstable and non-perturbative $m_{H'}-\sin\theta$ regions, associated to green, red and brown colors, respectively.  
We have categorized these regions using the following criteria: 
\begin{itemize}
 
\item \underline{\bf{Green Region:}}
This is the region where we can have stable vacuum all the way up to the Planck scale, and all the couplings are within their perturbative regime. 
In our numerical scans these conditions are implemented by requiring:
$0<\lambda_\Phi(\mu)<\sqrt{4\pi}$, $0<\lambda_\sigma(\mu)<\sqrt{4\pi}$, $\lambda_{\Phi\sigma}(\mu)+2\sqrt{\lambda_\Phi(\mu)\lambda_\sigma(\mu)}>0$ and $|\lambda_{\Phi\sigma}(\mu)|<\sqrt{4\pi}$ where
$\mu$ is the running scale. 
All other couplings e.g. the gauge and Yukawa couplings are also required to be perturbative till the Planck scale.

\item \underline{\bf{Red Region:}} 
In this region the vacuum is unstable, as the potential becomes unbounded from below at some high energy scale before Planck scale.  
This means that any one or more than one of these conditions are realised: $\lambda_\Phi(\mu)\leq 0$, $\lambda_\sigma(\mu)\leq 0$, $\lambda_{\Phi\sigma}(\mu)+2\sqrt{\lambda_\Phi(\mu)\lambda_\sigma(\mu)}\leq 0$. 
Note that inside the red region there can be parameters for which the potential is unbounded, and also some of the quartic couplings are non-perturbative as well, although we are excluding Landau poles. 

\item \underline{\bf{Brown Region:}} 
This region implies the existence of non-perturbative couplings at some energy scale before the Planck scale. 
This happens if any one of the following conditions holds: $|\lambda_\Phi(\mu)| \geq 4\pi$, $|\lambda_{\sigma}(\mu)| \geq 4\pi$, $|\lambda_{\Phi\sigma}(\mu)| \geq 4\pi$, $|Y_\nu(\mu)| \geq 4\pi$. 
Note that the gauge coupling running here is similar to the SM running and hence they always remain perturbative.
We are also including Landau poles inside the non-perturbative regions, since as one approaches the Landau pole the perturbative approach is no longer reliable.
In Appendix.~\ref{app:few discussions of RGEs}, we discuss how Landau poles can arise in our the RGEs.
There we have also discussed other scenarios leading to non-perturbative couplings, such as the continuous growth of a coupling or the saturation of some coupling with respect to the energy scale $\mu$. 

\end{itemize}

Let us now look at the combined results of the collider constraints and stability-perturbativity constraints of our model. We start with Case I first.  

\subsection{Case I where the heaviest scalar $H_2 = H_{125}$ is the Higgs boson} 
\label{sec:case1}

\begin{figure}[h]
\centering
\includegraphics[width=0.49\textwidth]{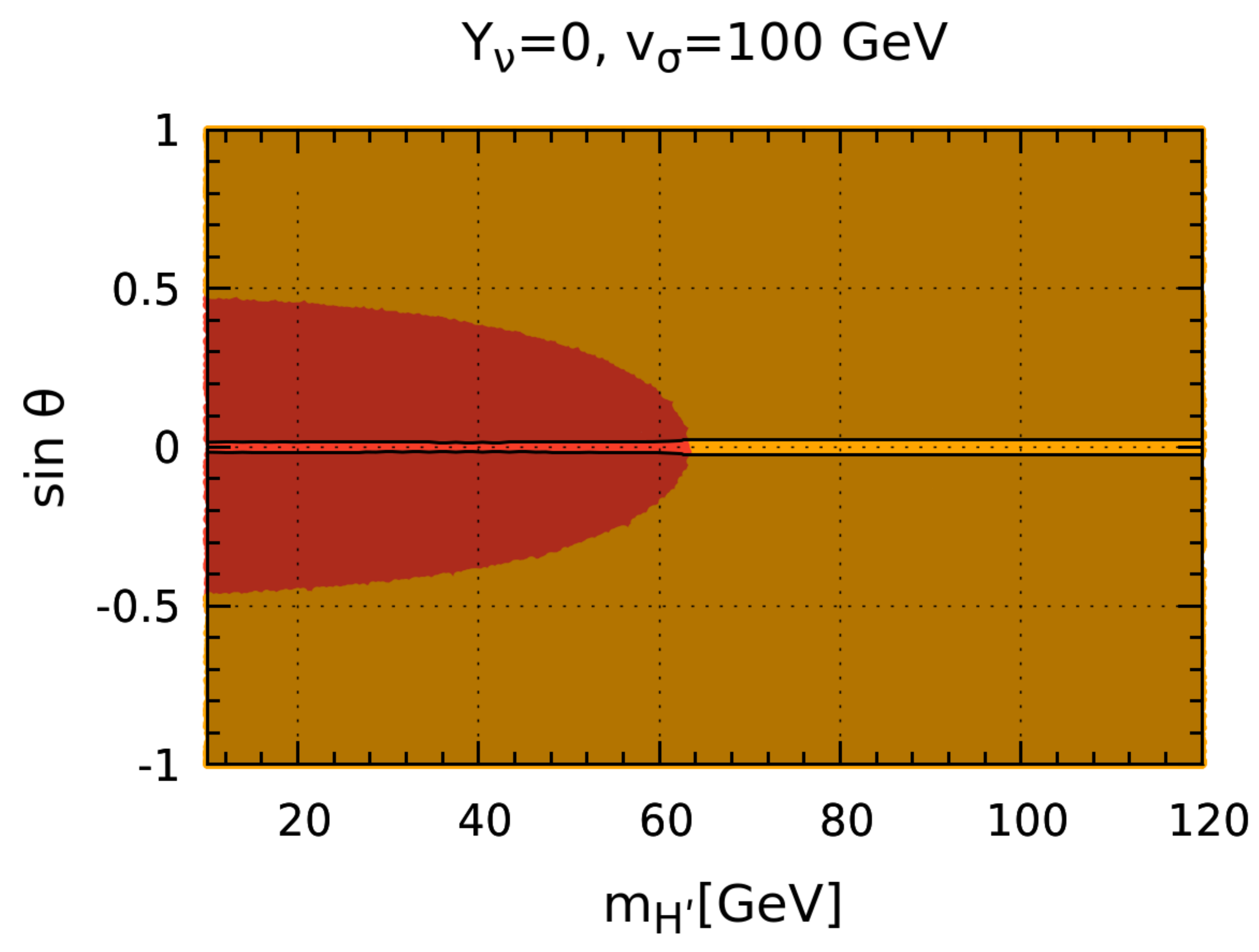}
\includegraphics[width=0.49\textwidth]{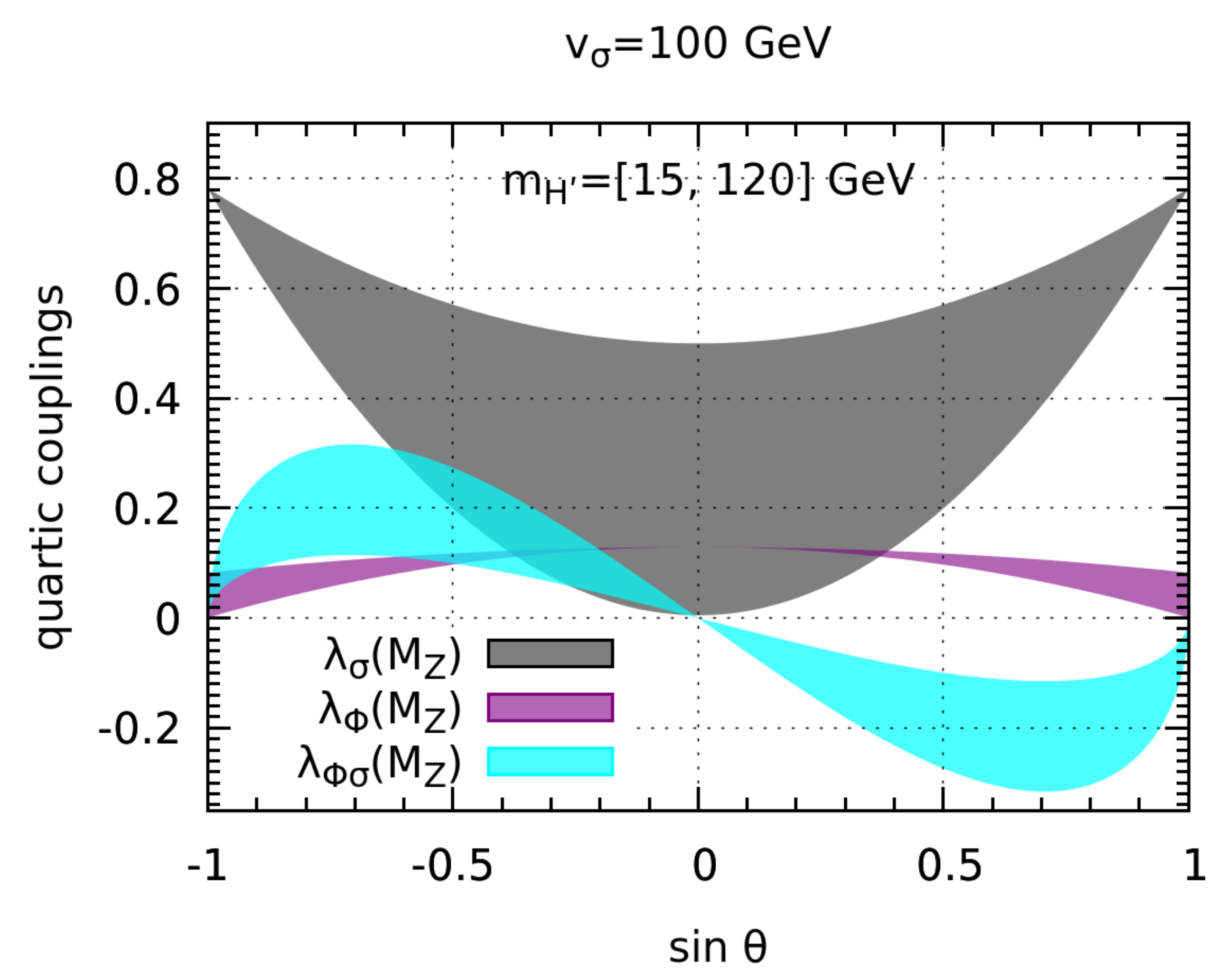}
\caption{\footnotesize{\textbf{Left panel:} Values of $m_{H'}$ and mixing angle $\theta$ leading to an unstable vacuum (in red), to a Landau pole or non-perturbative quartic couplings (in brown),
    at some energy scale below the Planck scale. We have fixed $Y_\nu=0$, $v_\sigma=100$ GeV and the heavy neutrino mass scale $\Lambda=10$ TeV. \\
    \textbf{Right panel:} Initial electroweak values of $\lambda_\Phi$, $\lambda_\sigma$ and $\lambda_{\Phi \sigma}$ as a function of $\sin\theta$. The band is due to the variation of the lighter
    scalar boson mass $m_{H'}\in [15,120]$ GeV range.}}
\label{stability}
\end{figure}

In the left panel of Fig.~\ref{stability}, for the $(3,1,1)$ Majoron inverse seesaw case, we have shown the values of $m_{H'}$ and $\theta$ for $v_\sigma=100$ GeV which lead to either bounded or
unbounded potential, non-perturbative dynamics or Landau Poles. We neglect the Yukawa coupling $Y_\nu$ and take the heavy neutrino mass scale at $M = \Lambda = 10$ TeV. 
We find that even in this extreme case where we have removed the destabilizing contribution of the neutrino Yukawa coupling, there is no parameter space to have a viable vacuum. 
Note that the whole parameter space is ruled out just from vacuum stability and perturbativity considerations.
The region allowed by collider constraints is the one in between the black contours in the left panel of Fig.~\ref{stability}. 
One sees that in this case the collider constraints are very stringent and rule out almost all the parameter space, except for a very thin small region very close to $|\sin\theta| \approx 0$.
However, even this small allowed region is ruled out by the combination of vacuum stability and perturbativity constraints. 

It is easy to understand the different regions in the left panel of Fig.~\ref{stability} with the help of the right panel of Fig.~\ref{stability}.  
There we have shown the corresponding initial values of quartic couplings at the electroweak scale.
The band for the quartic couplings $\lambda_\sigma(M_Z)$~(gray band), $\lambda_\Phi(M_Z)$~(purple) and $\lambda_{\Phi \sigma}(M_Z)$~(cyan) is due to the lighter Higgs mass variation in the range
$15\,\text{GeV}\leq m_{H'}\leq 120\,\text{GeV}$.
We find that for $|\sin\theta|<0.5$ in the low mass regime $m_{H'}<60$~GeV, the value of quartic coupling $\lambda_\Phi(M_Z)$ is $\mathcal{O}(0.1)$ and $0\leq |\lambda_{\Phi\sigma}(M_Z)|\leq 0.1$.
Hence the positive contribution from $\lambda_{\Phi\sigma}$ is not enough to overcome the negative contribution from the top Yukawa coupling. 
As a result, $\lambda_\Phi$ goes negative well below the Planck scale, and the vacuum is unstable. 
For $|\sin\theta|\sim 0$ and $m_{H'}>70$~GeV, we see from the right panel of Fig.~\ref{stability} that $|\lambda_{\Phi\sigma}(M_Z)|\sim 0$ so, from Eq.~\eqref{definition lambda sigma}, we get $\lambda_\sigma(M_Z)>0.3$.
With this small $\lambda_{\Phi\sigma}$ and large $\lambda_\sigma$, the RGEs for $\lambda_\sigma$ at leading order~(Eq.~\ref{lambdasigma-running}) can be approximated as Eq.~\eqref{qua}. 
We find that with this large $\lambda_\sigma(M_Z)>0.3$ the running of $\lambda_\sigma$ will become nonperturbative well before the Planck scale. 
Similarly, for $|\sin\theta| > 0.5$ the $\lambda_\sigma(M_Z)$ is even larger and, again, it quickly becomes non-perturbative below the $M_P$, irrespective of the mass range of $H'$ scalar. 

Finally, we stress that Fig.~\ref{stability} corresponds to the $(3,1,1)$ Majoron inverse seesaw, taking $Y_\nu = 0$. 
A non-zero Yukawa coupling will have an effect on the evolution of the quartic coupling $\lambda_\Phi$, aggravating the vacuum instability problem. 
Moreover, higher $(3,n,n)$ realizations with $n \geq 2$ also will only aggravate the stability problem. 
Thus, we can safely say that at least for $v_\sigma=100$ GeV, there is no viable parameter space within the Majoron inverse seesaw approach to have a stable and perturbative vacuum up to the Planck scale. 
 
Since higher dynamical lepton number breaking scales $v_\sigma$ relax the collider limits, will that help us find a viable parameter space? 
\begin{figure}[h]
\centering
\includegraphics[width=0.49\textwidth]{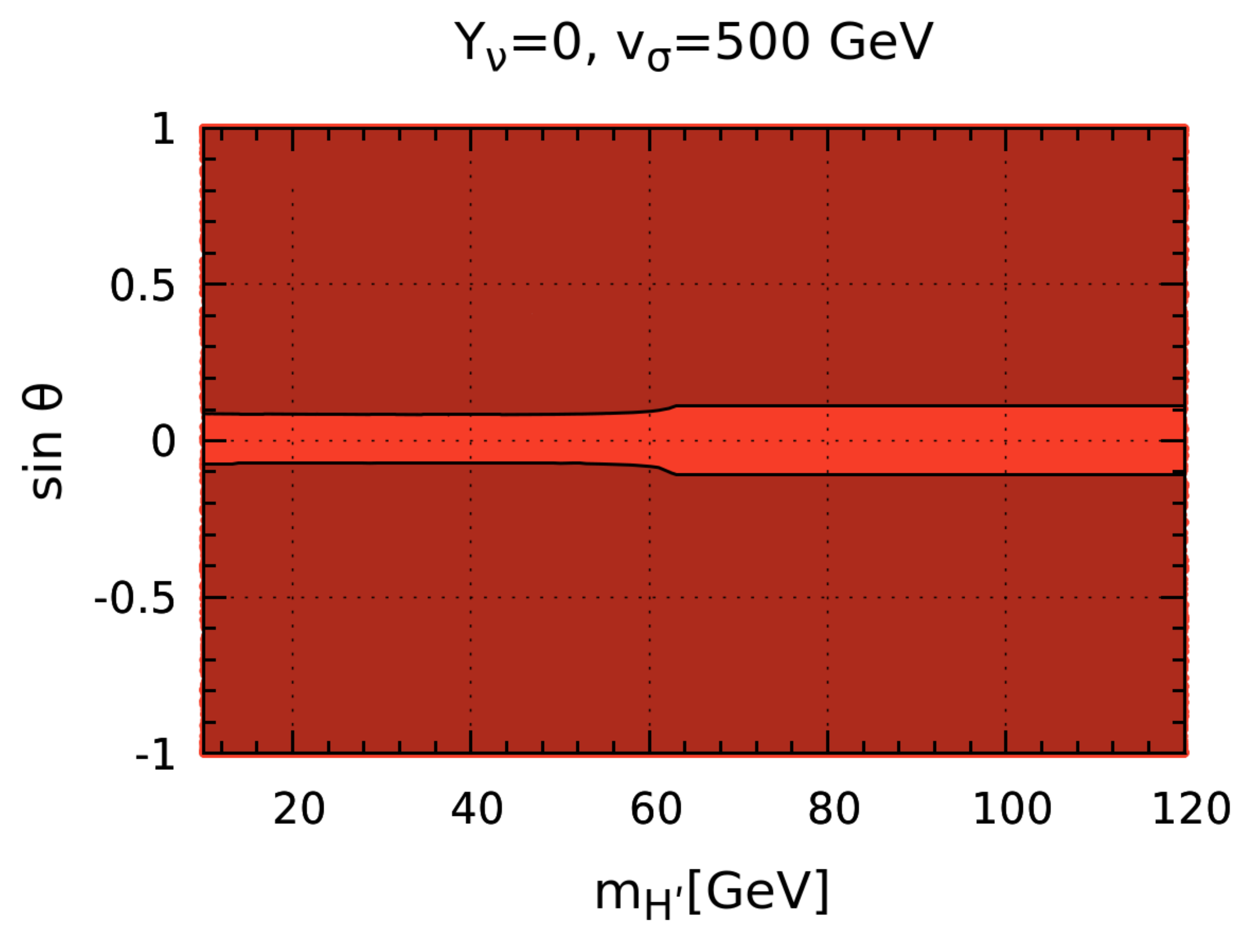}
\includegraphics[width=0.47\textwidth]{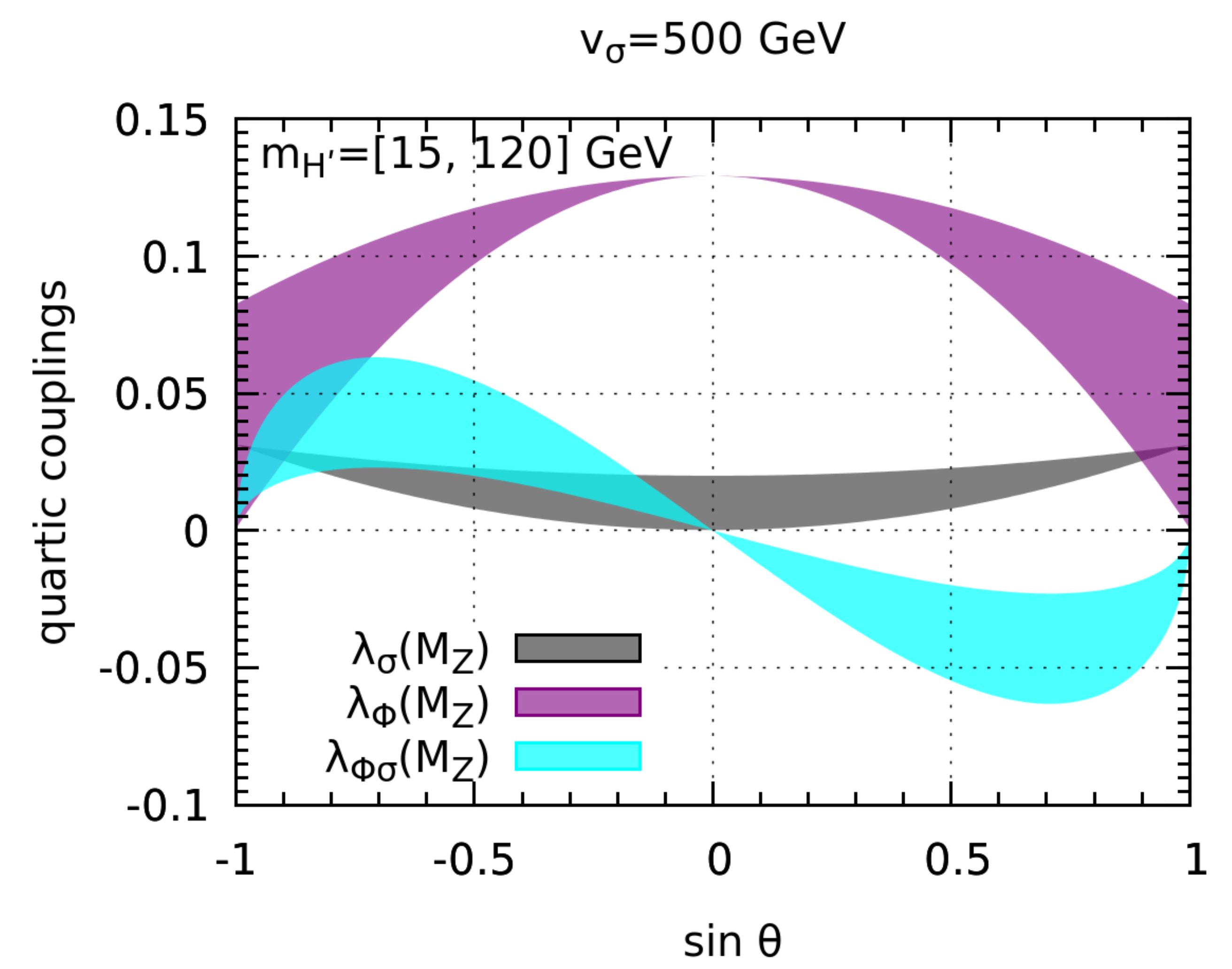}
\includegraphics[width=0.49\textwidth]{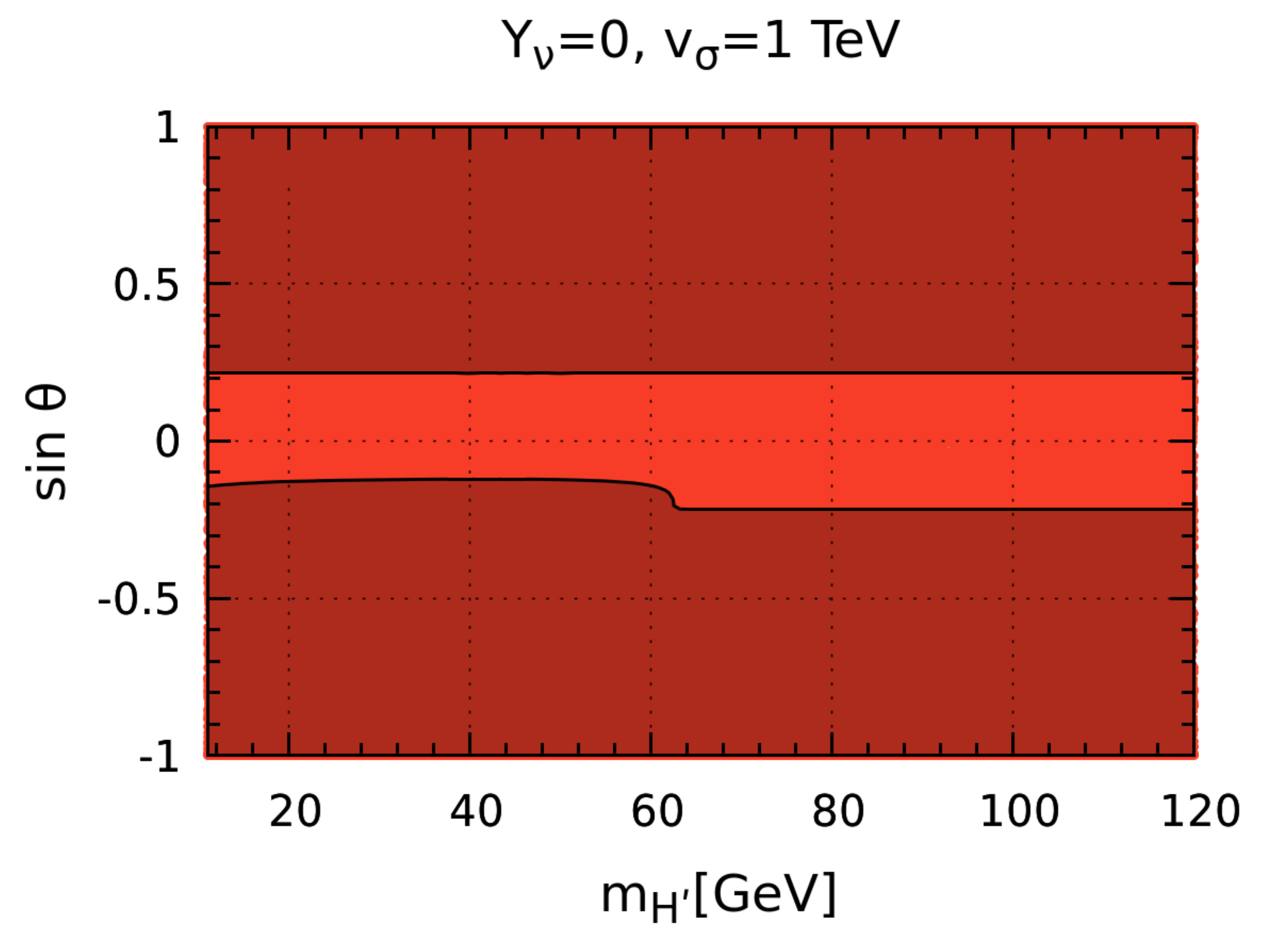}
\includegraphics[width=0.47\textwidth]{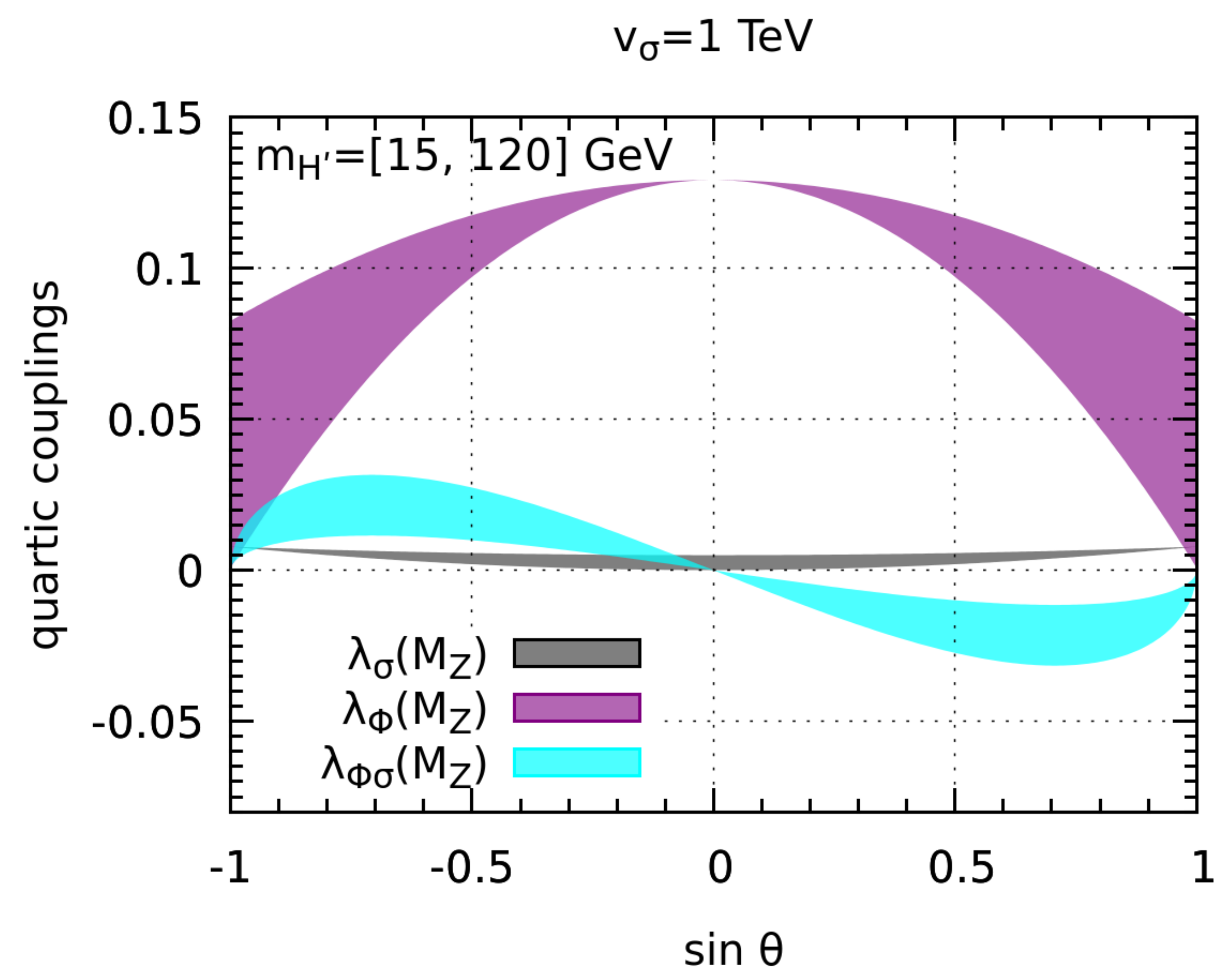}
\caption{\footnotesize{\textbf{Left panels:} The nature of vacuum for the case of $v_\sigma=500$ GeV and $1$ TeV, respectively, with Yukawa coupling $Y_\nu=0$. The color codes are same as in Fig.~\ref{stability}. \\
\textbf{Right panels:} Show the values of $\lambda_\Phi$, $\lambda_\sigma$ and $\lambda_{\Phi \sigma}$ at the electroweak scale as a function of $\sin\theta$. The bands are due to variation of lighter Higgs mass $m_{H'}$ varation 15 GeV to 120 GeV. 
Note that the scale on the Y-axis of right panels is different compared to Fig.~\ref{stability}.}}
\label{lambda plot}
\end{figure}
%
Although the collider restrictions clearly disappear in a genuine high-scale seesaw, as long as we remain within the low-scale seesaw picture, the answer is no!
 This can be seen from the upper left and lower left panel of Fig.~\ref{lambda plot}.
One sees that, for higher values of $v_\sigma$ up to 1TeV, there is no viable region to have a stable vacuum with $m_{H'} < m_{H_{125}}$. 
This figure shows that with $v_\sigma=500$ GeV and $v_\sigma=1$ TeV there is no consistent region left, the vacuum is even more unstable. 
In fact, as can be seen from Fig.~\ref{lambda plot}, the vacuum is now unstable for the entire parameter space!
This can again be seen in terms of the values of the quartic couplings at the electroweak scale, shown in the right panels of Fig.~\ref{lambda plot}. 
One sees that the quartic coupling $\lambda_{\Phi \sigma}$ is small for all values of the mixing angle $\theta$. 
Such small values of $\lambda_{\Phi\sigma}$ are not enough to counter the negative contribution from the top-Yukawa coupling in the evolution of $\lambda_\Phi$. 
On the other hand, now all the quartic couplings are small over the entire range of $m_{H'}$, hence RG running will not hit any Landau pole or non-perturbative quartic couplings. 
Hence we get only unstable vacuum for the entire $m_{H'}-\theta$ plane. 
As before, taking non-zero values of $Y_\nu$ in the $(3,n,n)$ schemes, will only make the problem of vacuum stability worse, due to the new Yukawas and/or extra fermions.
One can try lowering the scale of dynamical lepton number breaking by taking $v_\sigma < 100$ GeV.  
This also doesn't help as in this case, while the parameter space ruled by vacuum instability decreases, the region excluded by non-perturbativity increases.
In the end, no viable region remains.

In short, for Case I with $H_2 = H_{125}$, $15\,\text{GeV}\leq m_{H'}\leq 120\,\text{GeV}$ and $v_\sigma\in [100\,\text{GeV},1\,\text{TeV}]$ we have no regions with consistent vacuum allowed by the LHC
data.\\[-.3cm] 

\begin{figure}[h!]
\centering
\includegraphics[width=0.49\textwidth]{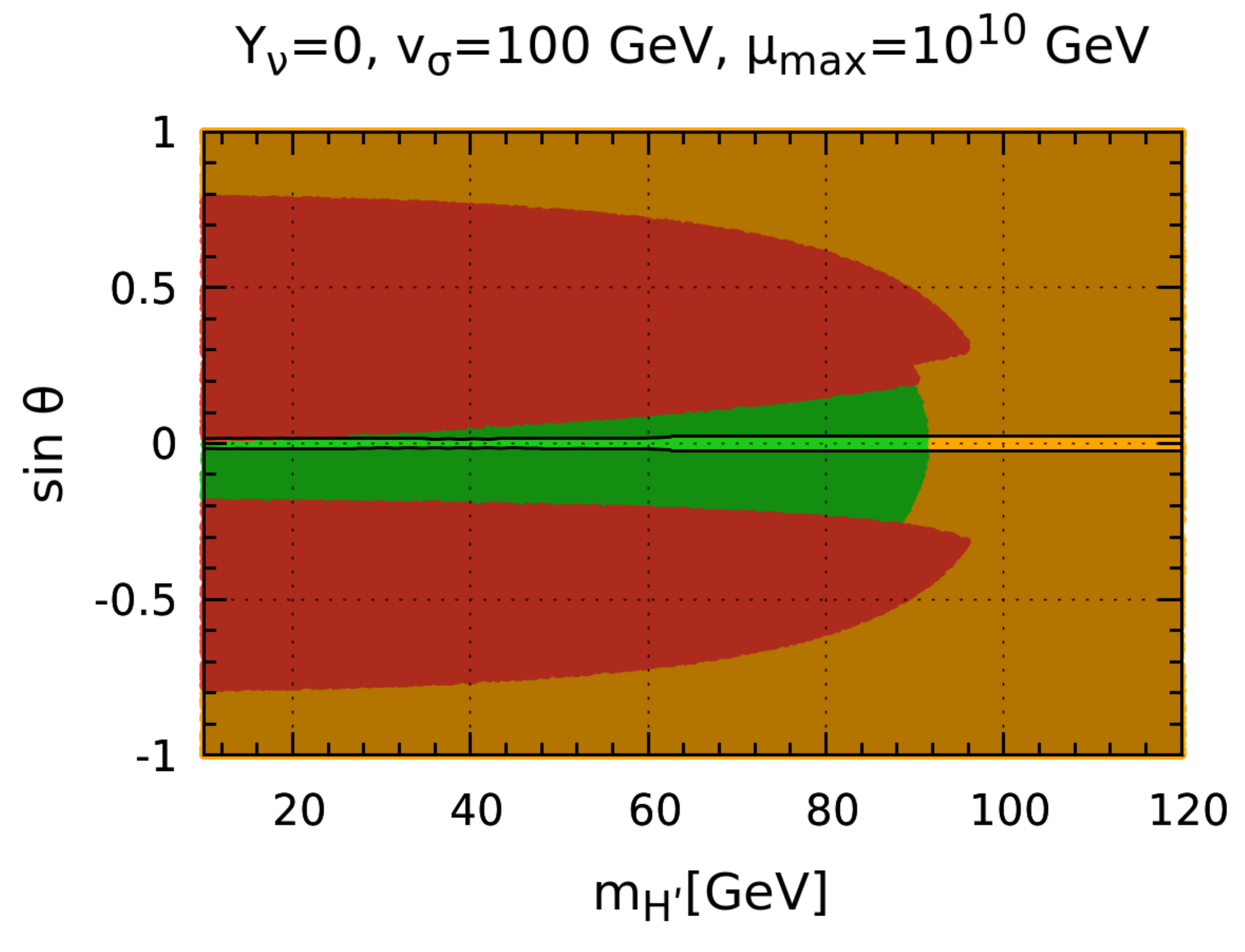}
\includegraphics[width=0.49\textwidth]{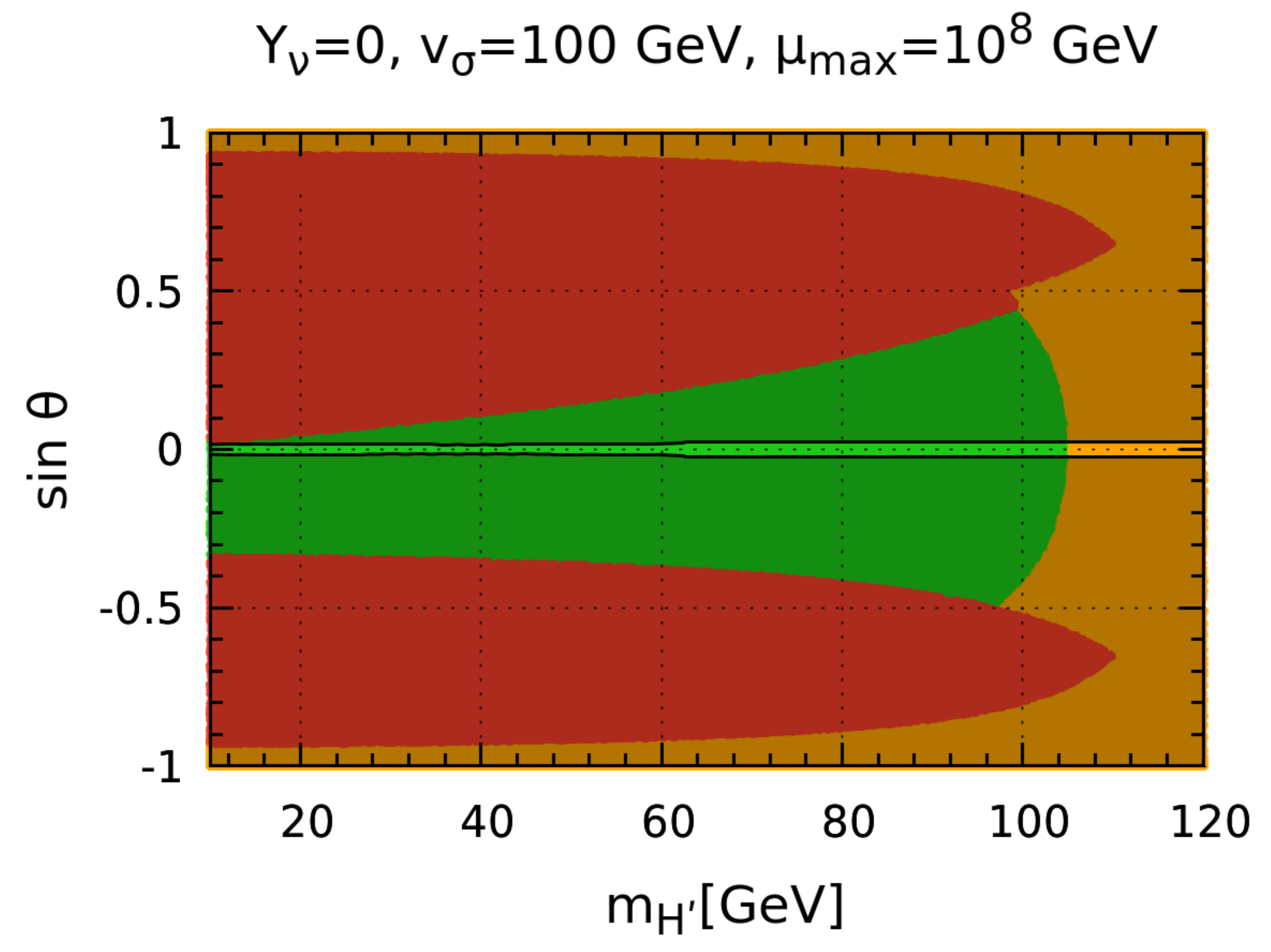}
\includegraphics[width=0.49\textwidth]{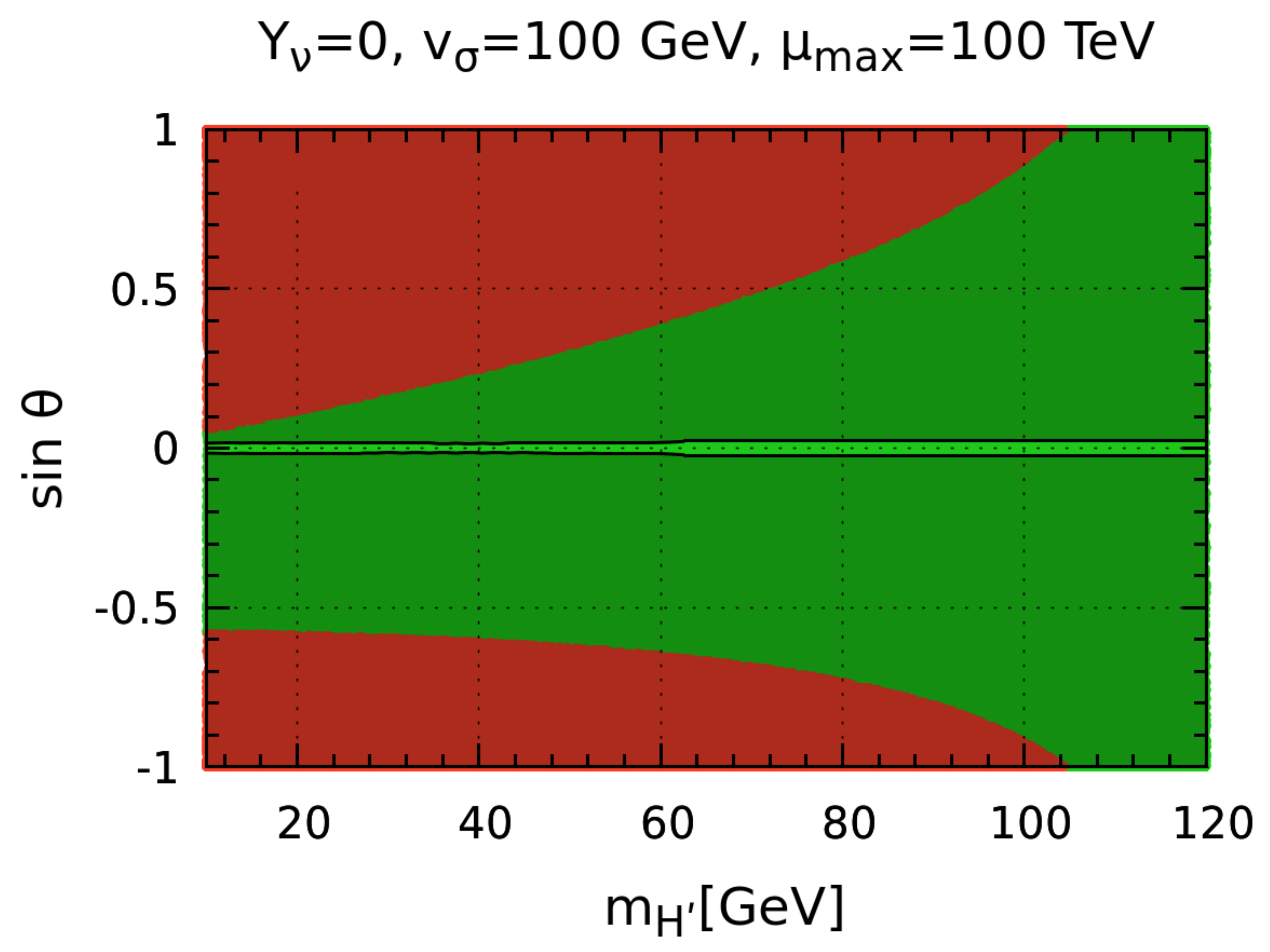}
\caption{\footnotesize{Same as in Fig.~\ref{stability} but now we have considered RGEs evolution only up to $\mu_{\text{max}}=10^{10}$ GeV~(upper left panel),
    $10^{8}$ GeV~(upper right panel) or $100 $ TeV~(lower panel).}}
\label{stability2}
\end{figure}

One should note that in the discussion of Figs.~\ref{stability} and \ref{lambda plot} we have assumed the maximum energy scale to be the Planck scale.
If one demands to have a consistent vacuum only up to some energy scale below $M_P$ the situation changes, as summarised in Fig.~\ref{stability2}.  

In Fig.~\ref{stability2}, we have fixed the maximum energy scale up to which one should have a stable vacuum with perturbative couplings, as $\mu_{\text{max}}=10^{10}$ GeV~(upper left panel),
$10^{8}$ GeV~(upper right panel) and $\mu_{\text{max}}=100$ TeV~(bottom panel)\footnote{
  Note that $\lambda_{\Phi\sigma}(M_Z)$ is an asymmetric function of $\sin\theta$, see Eq.~\eqref{definition lambda phisigma}.
  The evolution of $\lambda_{\Phi\sigma}$ depends on the sign of $\lambda_{\Phi\sigma}(M_Z)$, see~Appendix~\ref{app:inverse-seesaw-majoron}.
  Hence the stability condition $\lambda_{\Phi\sigma}(\mu)+2\sqrt{\lambda_\Phi(\mu)\lambda_\sigma(\mu)}>0$ also depends on the sign of $\sin\theta$, so the green regions are not symmetric in $\sin\theta$. }. 
Unlike the case of $\mu_{\text{max}}=\text{Planck scale}$,  with $\mu_{\text{max}}=10^{10}$ GeV we can indeed have a stable vacuum with perturbative couplings~(green region).
Moreover, this stability region increases with decreasing cut-off scale $\mu_{\text{max}}$ as seen in Fig.~\ref{stability2}.  
However, most of the consistent regions are excluded by the LEP and LHC constraints (these are the regions outside the narrow region delimited by the black contours). 
Thus, even with lower $\mu_{\text{max}}$, only a thin region between the two black lines around $|\sin\theta| \sim 0$ remains, due to the collider constraints. 

Before closing this section, we comment on the possibility that the lepton number symmetry is spontaneously broken at a very low scale, i.e. $v_\sigma\sim \mathcal{O}(\text{KeV})$~\cite{Berezhiani:1992cd}. 
This case unfortunately requires extreme fine tuning to be viable. 
From \eqref{definition lambda phisigma} one sees that it is only allowed if the masses of the CP even scalars is fine-tuned to be nearly degenerate
i.e. $m^2_{H_1} - m^2_{H_2} \sim v_\sigma \sim \mathcal{O}(\text{KeV})$.
This case would require a separate analysis, properly taking into account the threshold corrections coming from integrating out the \sm particles as well as the QCD corrections. 

\subsection{Case II where the lightest scalar is the $H_1 = H_{125}$ Higgs }
\label{sec:case2}

\begin{figure}[h]
\centering
\includegraphics[width=0.49\textwidth]{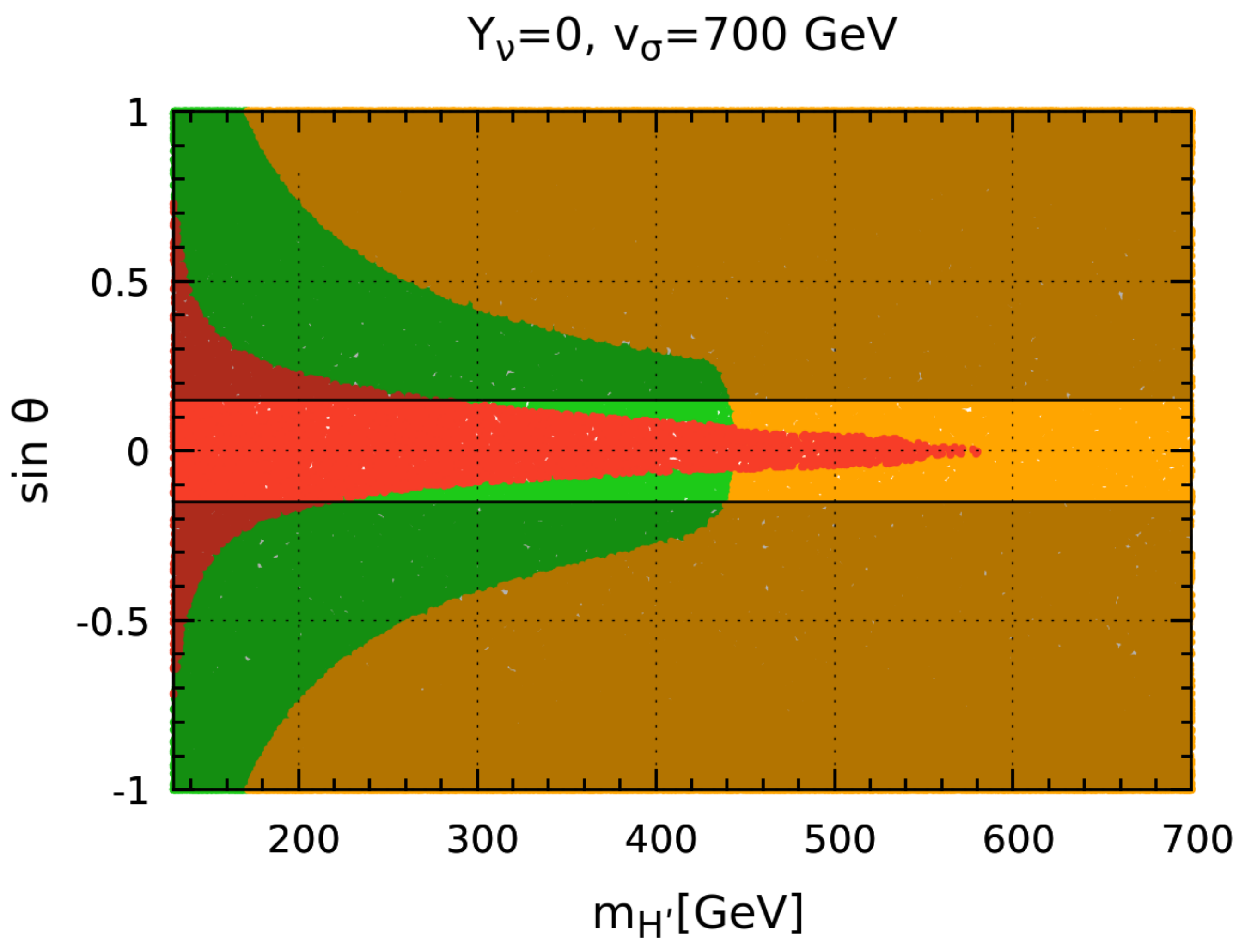}
\includegraphics[width=0.49\textwidth]{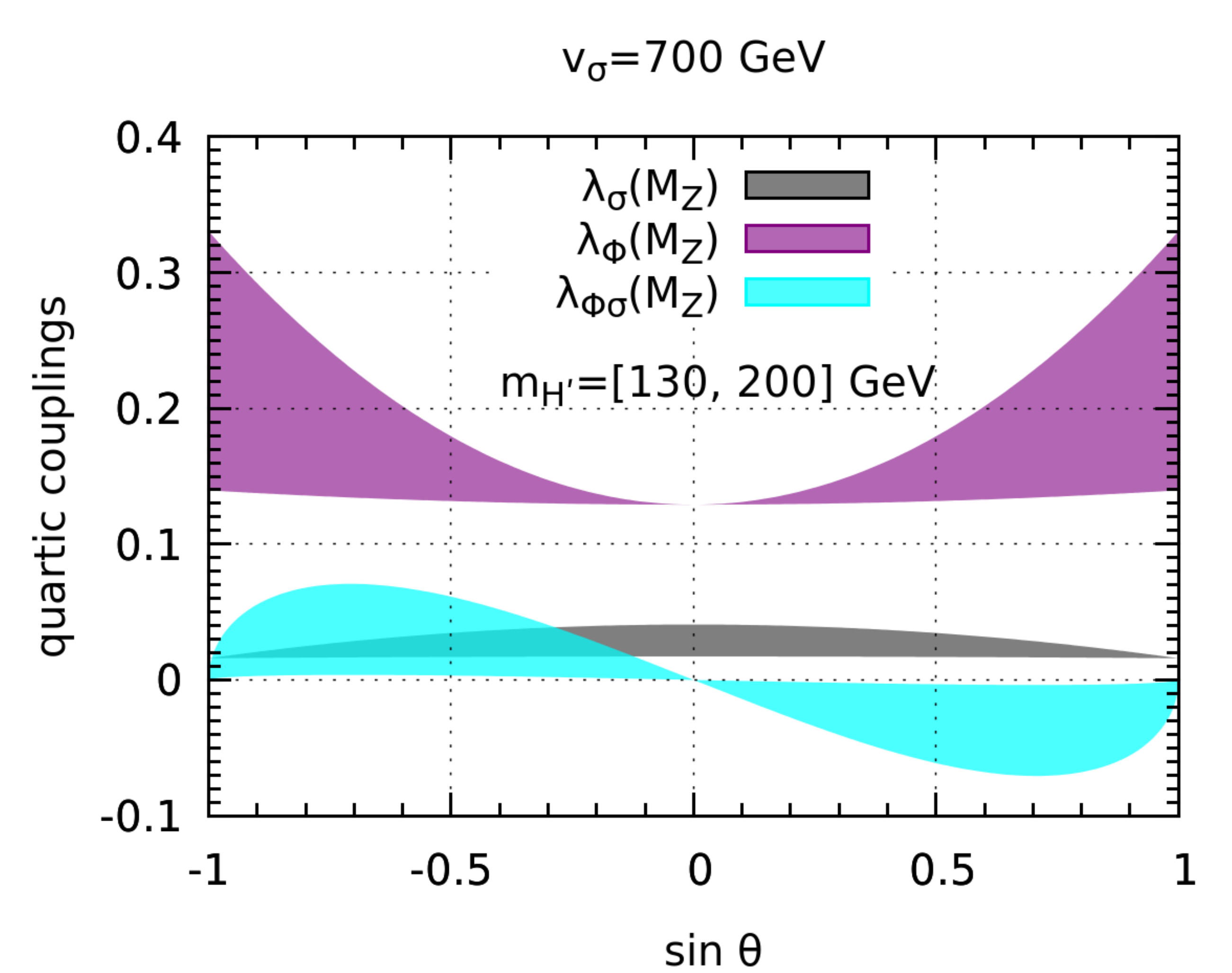}
\includegraphics[width=0.49\textwidth]{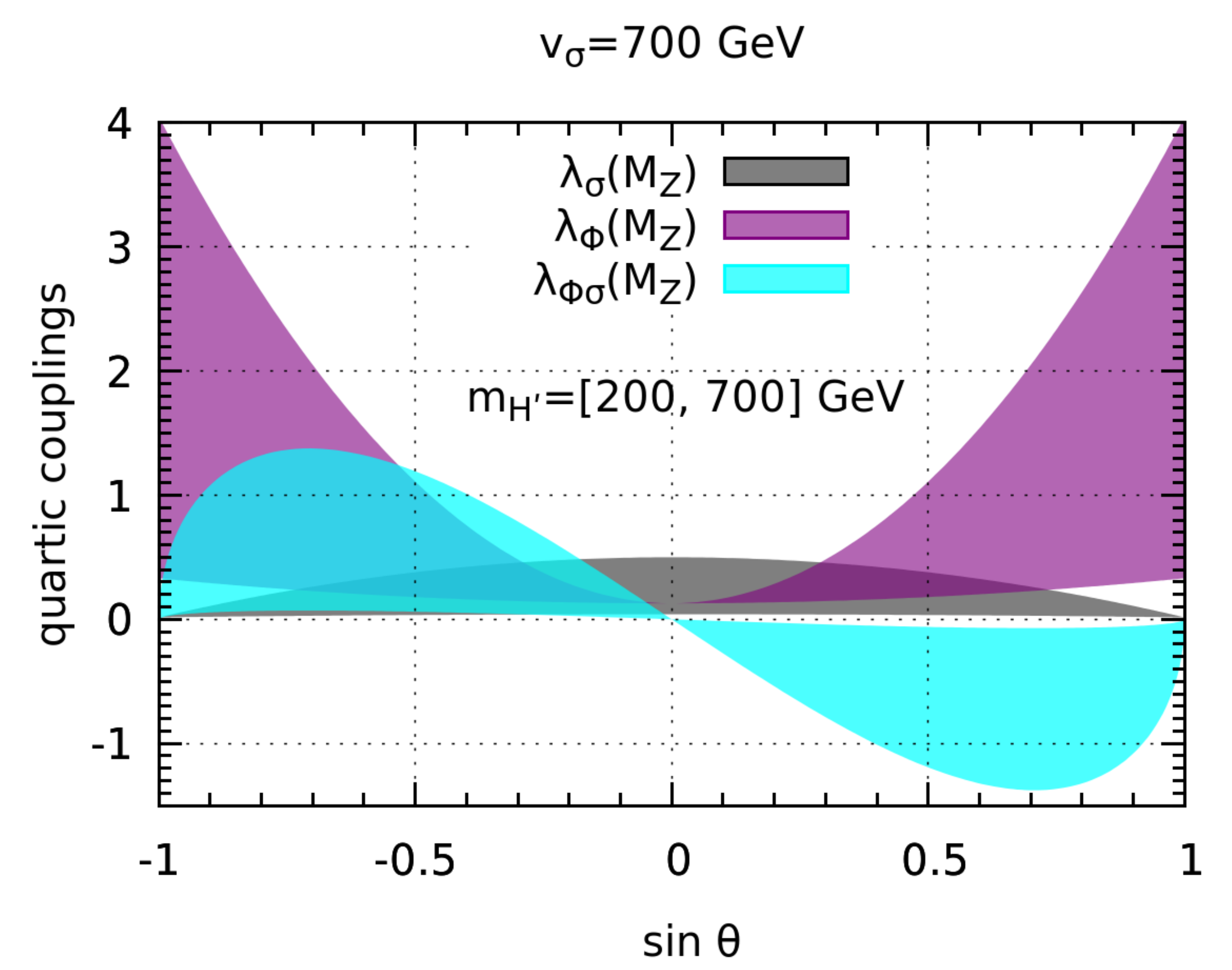}
\includegraphics[width=0.49\textwidth]{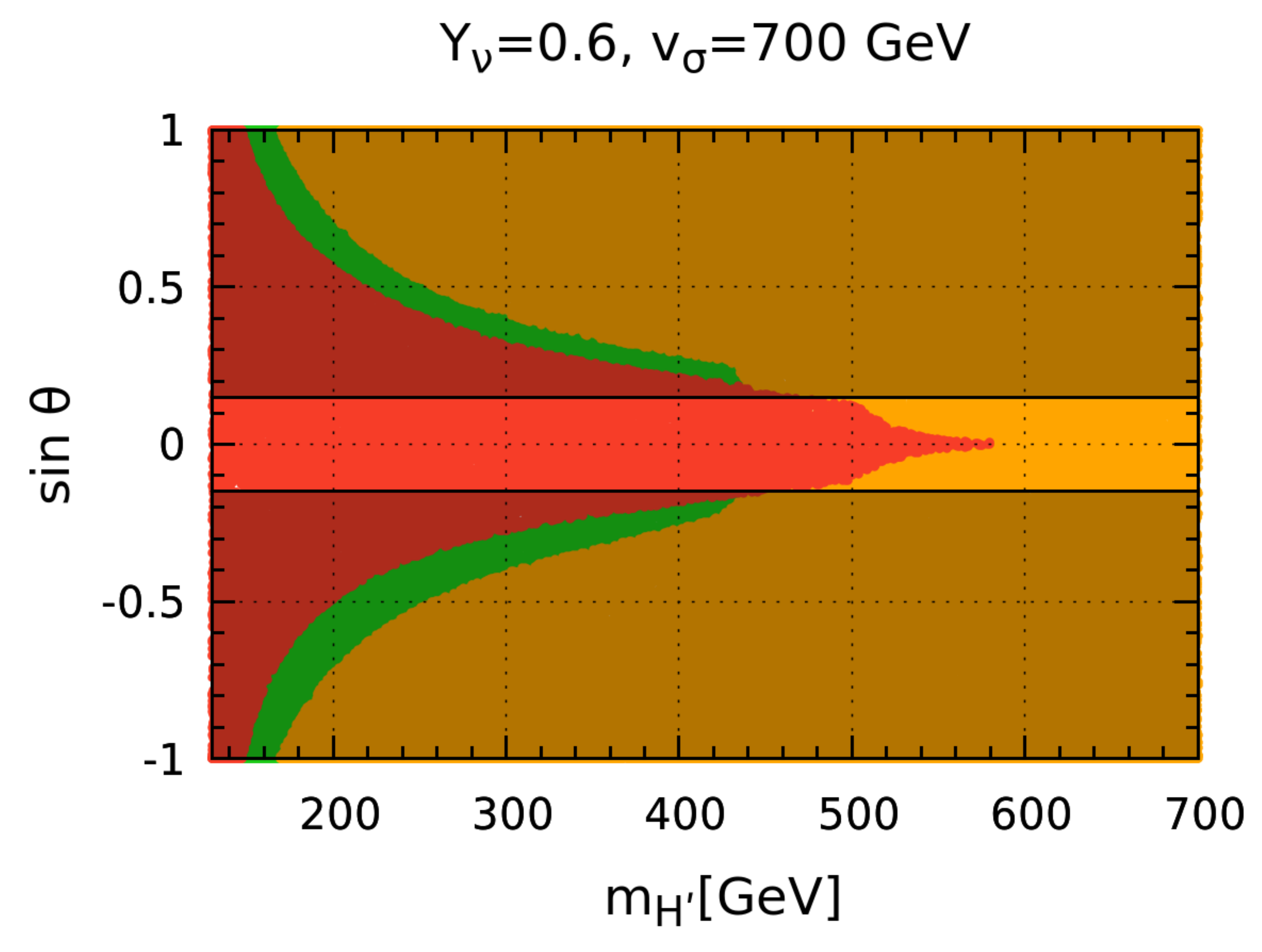}
\caption{\footnotesize{\textbf{Upper left panel:} Values of $m_{H'}$ and mixing angle $\theta$ leading to a stable vacuum with all couplings perturbative~(green),
    non-perturbative couplings at some energy scale~(brown), or an unstable potential~(red). The region between the black lines is ruled out by LEP-LHC constraints.
We have fixed the heavy neutrino mass scale $\Lambda=10$ TeV and $v_\sigma=700$ GeV and imposed the stability-perturbativity constraints up to $M_P$. 
\textbf{Upper right panel:} Values of $\lambda_\Phi$, $\lambda_\sigma$ and $\lambda_{\Phi \sigma}$ at the electroweak scale as a function of $\sin\theta$.
The bands correspond to variation of lighter Higgs mass $m_{H_2}$ in the range of 130 GeV to 700 GeV. 
\textbf{Bottom panel:} same as upper left panel but with $Y_\nu=0.6$.}}
\label{stability for case 2 with 700 GeV}
\end{figure}

\begin{figure}[h]
\centering
\includegraphics[width=0.49\textwidth]{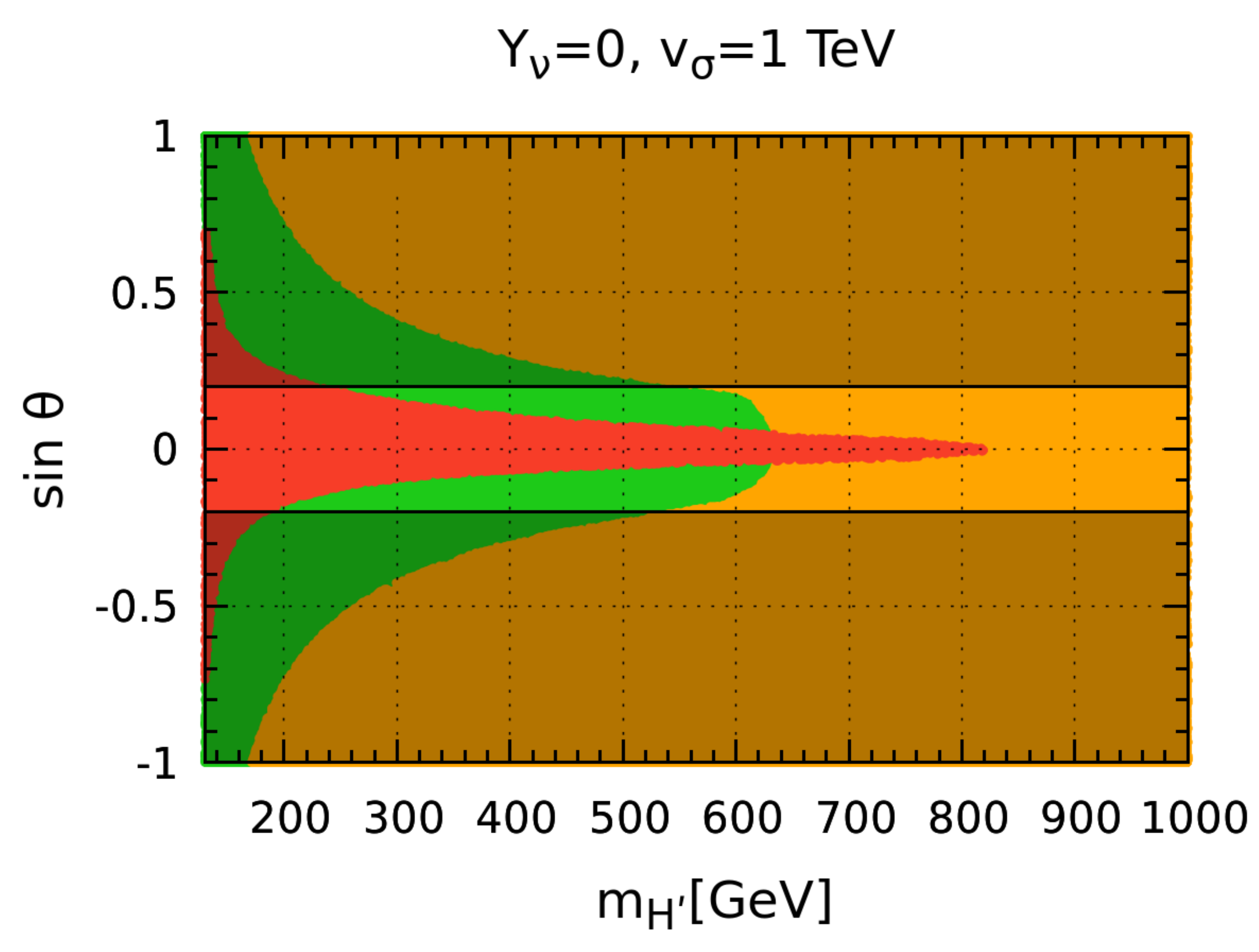}
\includegraphics[width=0.49\textwidth]{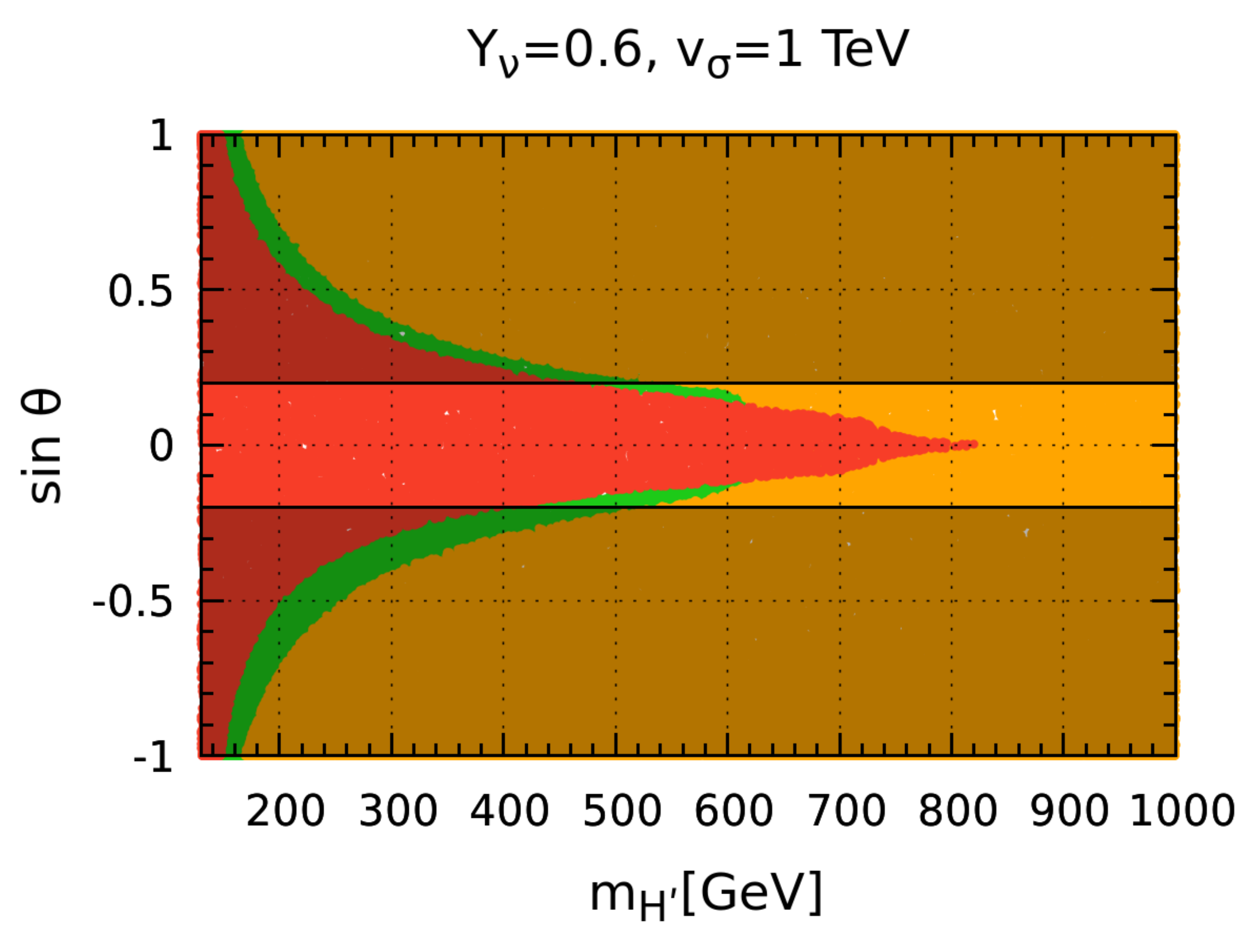}
\caption{\footnotesize{Values of $m_{H'}$ and mixing angle $\theta$ leading to a bounded-from-below and perturbative vacuum~(green), non-perturbative couplings at some energy scale~(brown),
    or unstable vacuum~(red). We have fixed the neutrino scale $\Lambda=10$ TeV and $v_\sigma=1$ TeV, taking $Y_\nu=0$ and 0.6 for the left and right panels, respectively.
    Stability-perturbativity constraints are imposed up to the Planck scale.}}
\label{stability for case 2 with 1 TeV}
\end{figure}

\begin{figure}[h]
\centering
\includegraphics[width=0.49\textwidth]{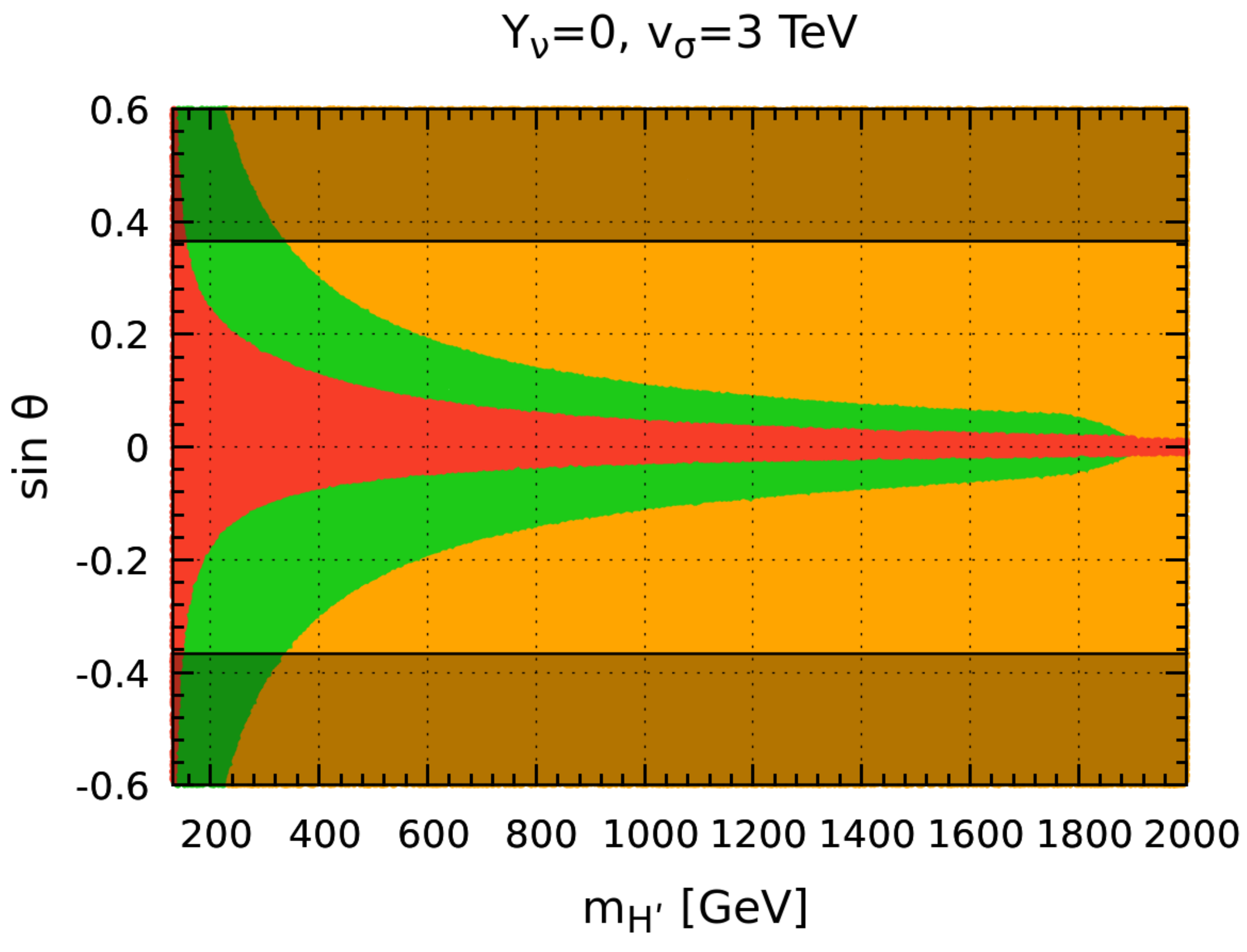}
\includegraphics[width=0.49\textwidth]{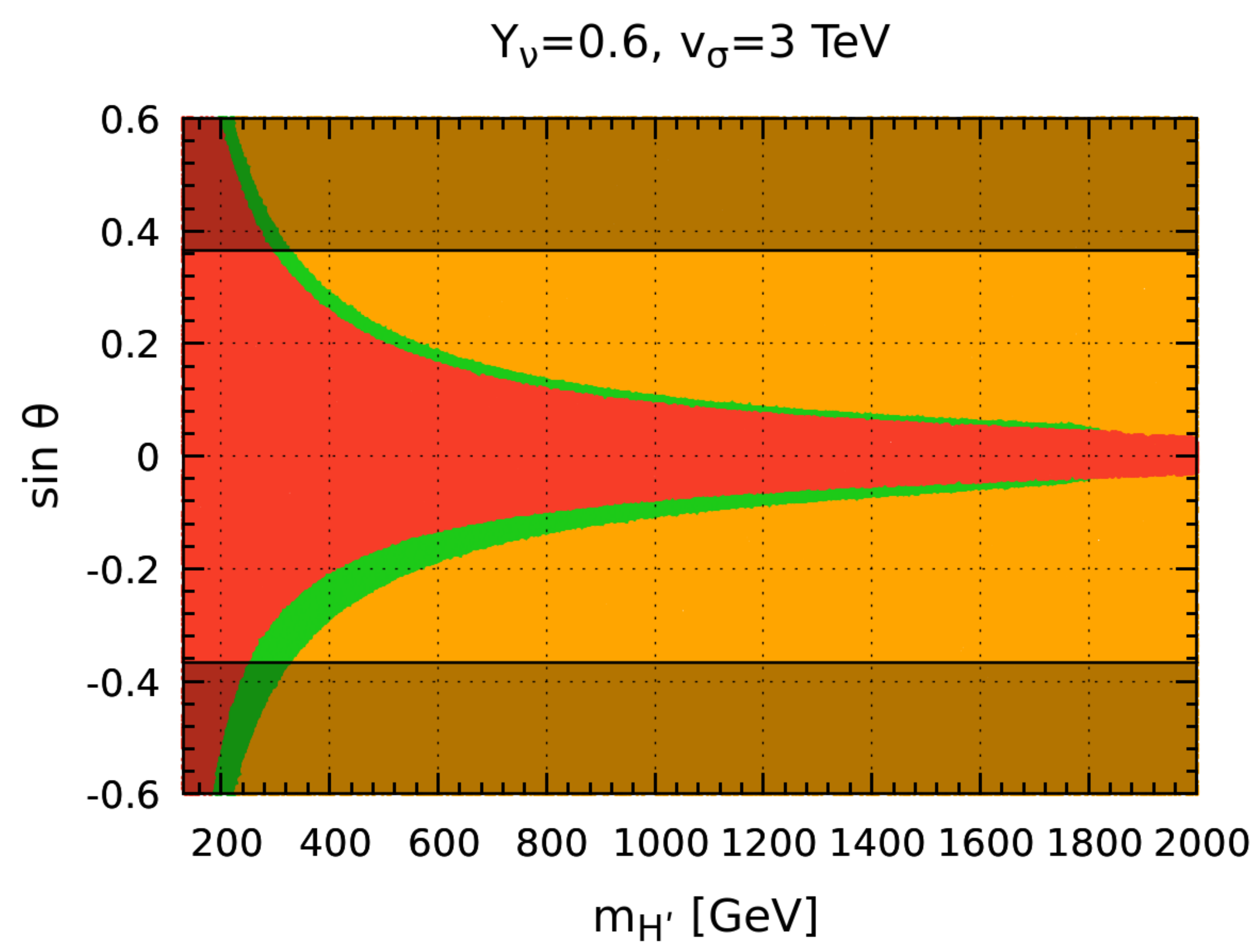}
\caption{\footnotesize{Same as in Fig.~\ref{stability for case 2 with 1 TeV} but for $v_\sigma=3$ TeV.}}
\label{stability for case 2 with 3 TeV}
\end{figure}

Let us now discuss the stability-perturbativity implications for Case II when the lighter CP even scalar $H_1 = H_{125}$ while the heavier scalar $m_{H_2} > 130$ GeV.
We will confront the electroweak vacuum consistency requirements in the majoron low-scale seesaw with the LEP-LHC constraints. 
For this case we choose three benchmarks for the dynamical lepton number breaking scale $v_\sigma$, namely $v_\sigma=700$ GeV, $1$ TeV and $3$ TeV.
The results for these are shown in Fig.~\ref{stability for case 2 with 700 GeV}, \ref{stability for case 2 with 1 TeV} and \ref{stability for case 2 with 3 TeV}, respectively. 

We start with the results for $v_\sigma=700$ GeV. The color codes in Figs.~\ref{stability for case 2 with 700 GeV} are same as in Fig.~\ref{stability2} and the neutrino scale fixed at $\Lambda=10$ TeV. 
Unlike the previous section~\ref{sec:case1}, in the $Y_\nu = 0$ case (upper left panel) we do have a region (green) where the vacuum is stable and all couplings are perturbative up to the Planck scale.
However, LEP-LHC constraints (black shaded region) coming from Eqs.~(\ref{eq:muf}) and (\ref{eq:inv}) rule out large part of the green space. 
Even then some green region remains which satisfies all the constraints. 
%
 
The different regions of the upper left panel of Fig.~\ref{stability for case 2 with 700 GeV} can be understood from the corresponding upper right panel. 
From Fig.~\ref{stability for case 2 with 700 GeV} (see also Eq.~\eqref{definition lambda phi} and \eqref{definition lambda sigma}) one can see that for $|\sin\theta|\sim 1$ and small $m_{H'} \lsim 200$ GeV,
one finds that $\lambda_\Phi(M_Z)$, $\lambda_\sigma(M_Z)$ are not too large (top-right panel), in contrast to the case of large $H'$ masses (bottom-left). 
On the other hand $\lambda_{\Phi\sigma}(M_Z)$ is large enough to counter the negative contribution of top-Yukawa coupling in the evolution of $\lambda_{\Phi}$~(see Eq.~\ref{one-loop-lambdaphi}). 
Hence, we obtain the green region. 

On the other hand for $|\sin\theta|\sim\mathcal{O}(1)$ and large $m_{H'}> 180$ GeV, using Eq.~\eqref{definition lambda phi} one finds that $\lambda_\Phi(M_Z)$ is too large.
From Eq.~(\ref{one-loop-lambdaphi}) we see that for large $\lambda_\Phi(M_Z)$ the one-loop running of $\lambda_\Phi$ can be approximated as $\beta_{\lambda_\Phi}\approx 24\lambda_\Phi^2$.
Following Eq.~(\ref{qua}) we find that this actually becomes non-perturbative well below the Planck scale. 
For $|\sin\theta|\sim 0$ and $m_{H'}<550$ GeV, $\lambda_\Phi(M_Z)\sim\mathcal{O}(0.1)$ and $\lambda_{\Phi\sigma}(M_Z)$ is not large enough to counter the negative top-Yukawa coupling contribution.
As a result, it leads to an unstable vacuum, as indicated in red.  
For large $m_{H'}$ and $|\sin\theta|\sim 0$, $\lambda_\sigma(M_Z)$ is large but $\lambda_{\Phi\sigma}(M_Z)$ is small. Hence, the one-loop $\lambda_\sigma$ evolution can be approximated as
$\beta_{\lambda_\sigma}\approx 20\lambda_\sigma^2$~(see Eq.~\ref{lambdasigma-running}) and following Eq.~(\ref{qua}) the coupling becomes non-perturbative before reaching the Planck scale.

Keeping all other parameters the same, we now switch on the neutrino Yukawa coupling $Y_\nu = 0.6$ to analyze its effect~\footnote{
  Note that we restrict the Yukawa coupling $Y_\nu\leq 0.6$, as with $Y_\nu\geq 0.7$ and $\Lambda=10$ TeV, we get either unstable vacuum or non-perturbative dynamics
  for the entire range of $m_{H'}$ and $\sin\theta$. }, see the lower panel in Fig.~\ref{stability for case 2 with 700 GeV}. 
As expected, upon switching on the neutrino Yukawa coupling, the vacuum will be unstable over a larger parameter space. 
One sees from the lower panel in Fig.~\ref{stability for case 2 with 700 GeV}, that the region with unstable vacuum increases appreciably.
Correspondingly, the green region where the vacuum is stable and all couplings are perturbative decreases in size. 
After imposing the LEP-LHC constraints (region between the black lines), we find that no viable allowed region remains, as all the green region falls within the collider-forbidden region. 
Thus, stability-perturbativity in conjunction with LEP-LHC constraints, completely rule out this benchmark.\\[-.3cm]

The main conclusions drawn for the case of $v_\sigma = 700$ GeV hold for larger $v_\sigma$ values.  
In Figs.~\ref{stability for case 2 with 1 TeV} and \ref{stability for case 2 with 3 TeV}, we have shown the results for $v_\sigma=1$ TeV and $3$ TeV, respectively. 


The different regions in these figures can be explained in a way similar to the case of $v_\sigma=700$ GeV. 
The most important change now is that the collider constraints from Eqs.~(\ref{eq:muf}) and (\ref{eq:inv}) get relaxed with increasing value of $v_\sigma$. 
Indeed, from Table.~\ref{tab:3} and Fig.~\ref{present limit for case 2}, one sees that, the larger the $v_\sigma$, the more relaxed are the LHC constraints.
The fact that these constraints are weaker for $v_\sigma = 3$ TeV than for $v_\sigma = 1$ TeV explains why the green region in Fig.~\ref{stability for case 2 with 3 TeV}
allowed by these constraints is much larger than shown in Fig.~\ref{stability for case 2 with 3 TeV}.
Notice that, for such relatively large $v_\sigma$, we obtain a stable vacuum consistent with LHC constraints even for non-zero Yukawa couplings. 

\begin{figure}[h]
\centering
\includegraphics[width=0.49\textwidth]{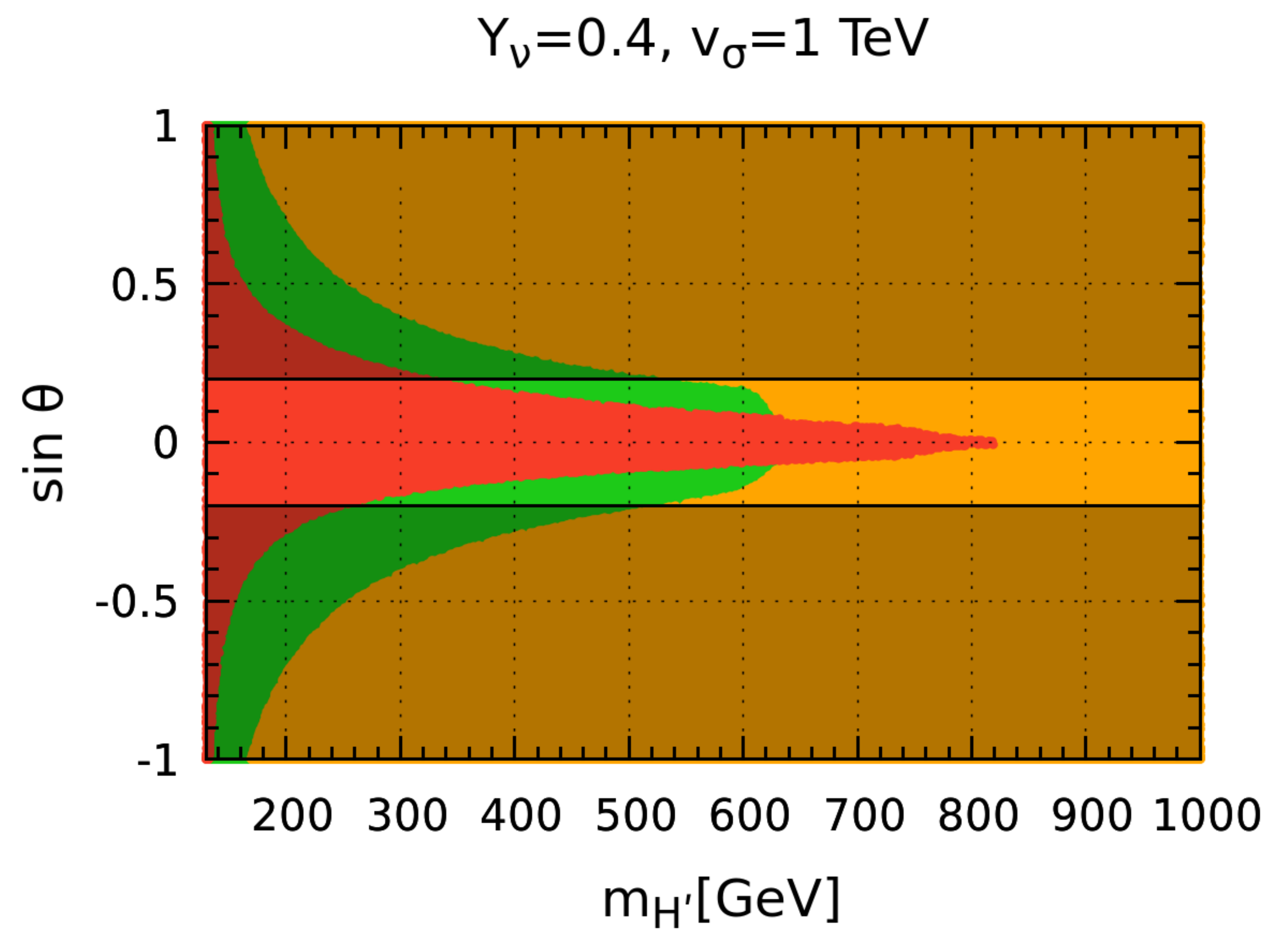}
\includegraphics[width=0.49\textwidth]{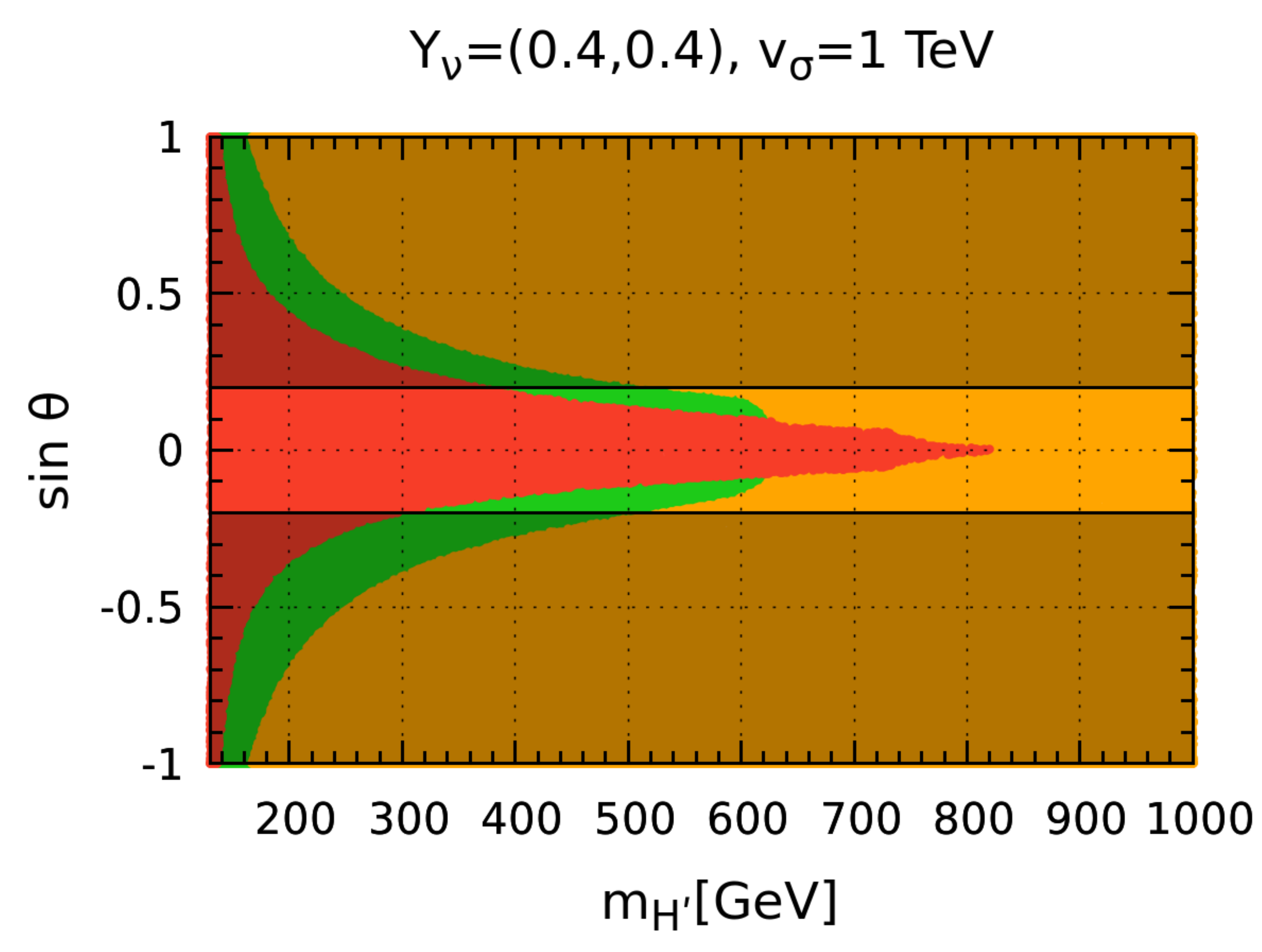}
\includegraphics[width=0.49\textwidth]{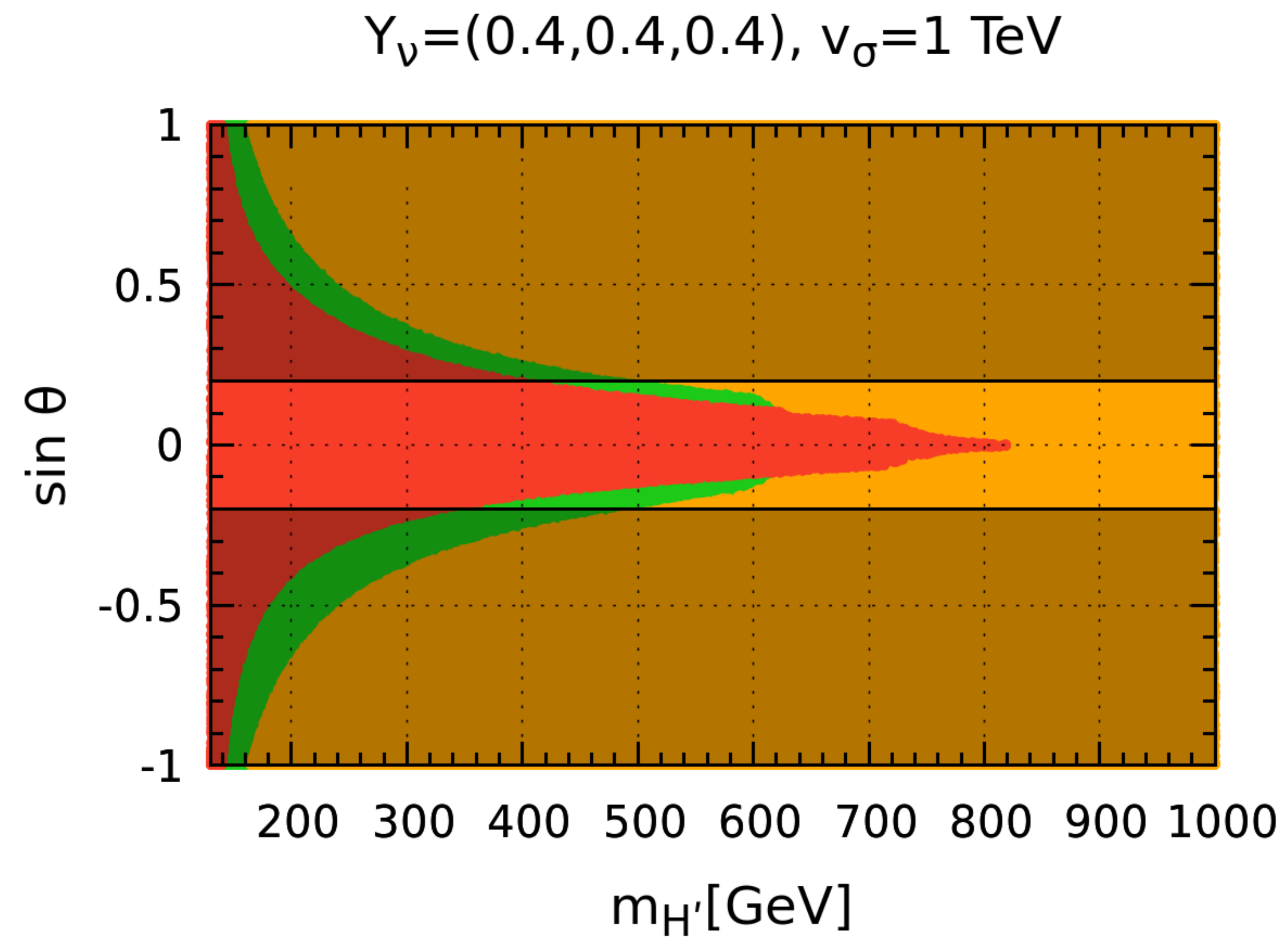}
\caption{\footnotesize{The left figure of upper panel is for the reference (3,1,1) majoron inverse seesaw with $Y_\nu = 0.4$.
    The right upper panel is for (3,2,2) case with $Y_\nu =\text{Diag}(0.4,0.4)$.
    The lower panel is for (3,3,3) case with $Y_\nu =\text{Diag}(0.4, 0.4, 0.4)$.
    We have fixed $v_\sigma=1$ TeV, the neutrino mass scale $\Lambda=10$ TeV and have imposed the stability-perturbativity constraints up to Planck scale.
    The color code is same as in Fig~\ref{stability for case 2 with 1 TeV}.}}
\label{Three generation with 1 TeV}
\end{figure}
\begin{figure}[h]
\centering
\includegraphics[width=0.49\textwidth]{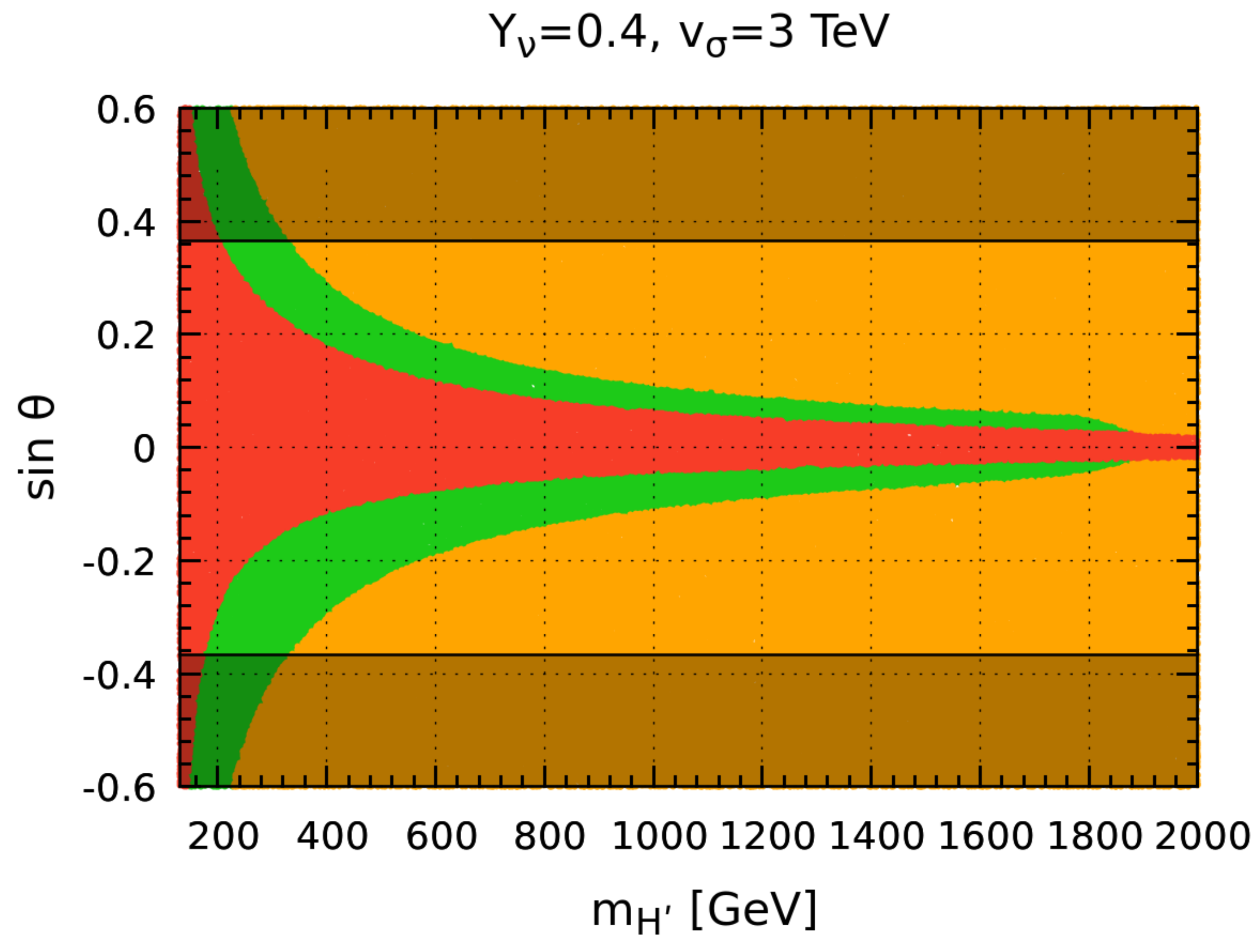}
\includegraphics[width=0.49\textwidth]{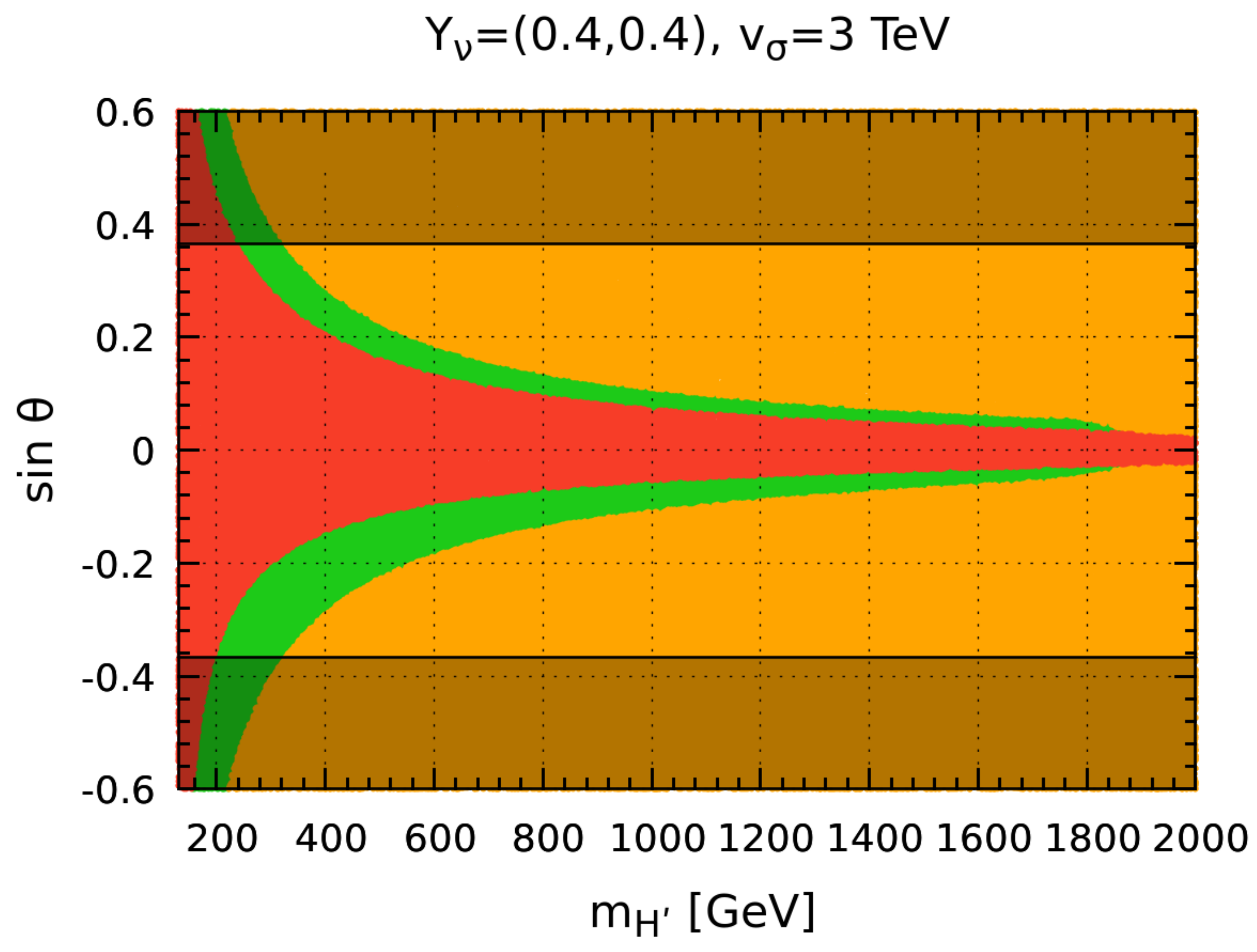}
\includegraphics[width=0.49\textwidth]{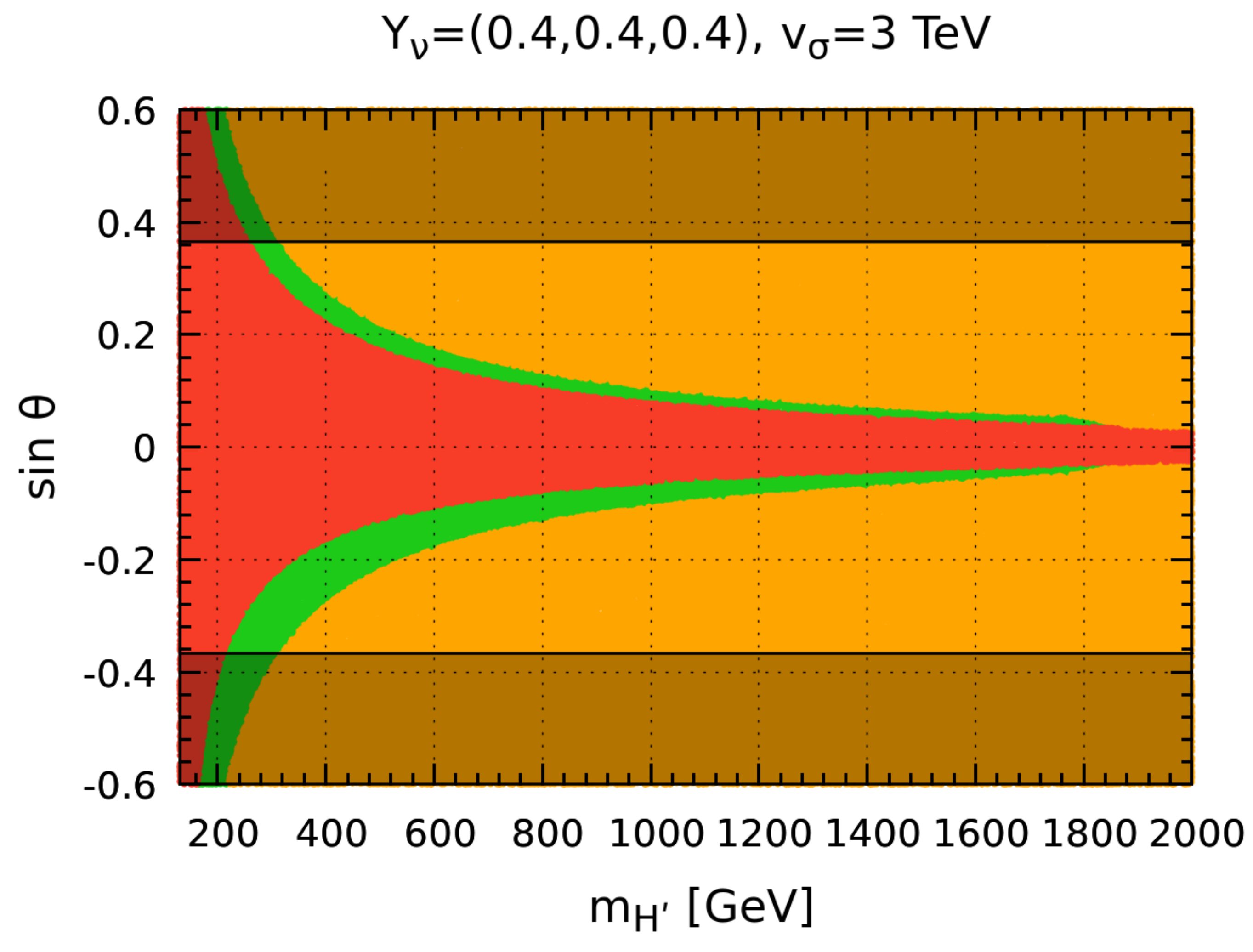}
\caption{\footnotesize{Same as Fig.~\ref{Three generation with 1 TeV}, but for $v_\sigma=3$ TeV. }}
\label{Three generation with 3 TeV}
\end{figure}

We remark that the above analysis has been performed for the missing partner $(3,1,1)$ inverse seesaw mechamism, where there is only one Yukawa coupling.  
For higher $(3,n,n)$ inverse seesaw schemes with $n \geq 2$, the results will be similar, but more Yukawa couplings means that larger regions will be ruled out by the stability-perturbativity requirement.
We have also considered these cases in detail.
In Fig.~\ref{Three generation with 1 TeV} and \ref{Three generation with 3 TeV} we compare the stability properties of the $(3,1,1)$ and (3,2,2) missing partner seesaw with the standard sequential
(3,3,3) inverse seesaw.

We have taken the Yukawa coupling $|Y_\nu| = 0.4$ for the (3,1,1) case, while for the $(3, n, n)$ with $n \geq 2$ and we assumed $Y_\nu^{ii} = 0.4$ and the off-diagonal entries as $Y_\nu^{ij} = 0$. 
One sees from these figures that when the dynamical lepton number breaking scale $v_\sigma$ is high i.e. $v_\sigma \gsim 1$ TeV, thanks to the relaxed LHC constraints (Fig.~\ref{present limit for case 2}),  the Higgs vacuum can be still kept stable up to the Planck scale, even for appreciable Yukawa couplings. 
Of course, the presence of additional fermions means that the maximum values of $Y_\nu^{ii}$, for which Higgs vacuum stability can be achieved up to $M_P$, is somewhat reduced compared to the Yukawa coupling $Y_\nu=0.6$ considered in Figs.~\ref{stability for case 2 with 700 GeV}, \ref{stability for case 2 with 1 TeV} and \ref{stability for case 2 with 3 TeV}. 
However, we can still have regions consistent with LHC measurements and vacuum stability-perturbativity for the case $n\geq 2$ even for moderate Yukawa coupling values.\\[-.3cm] 

To sum up, in contrast to Case I, for Case II we can have regions consistent with LHC constraints and vacuum stability-perturbativity all the way up to the Planck scale even for relatively large Yukawas.
This is possible provided the scale of dynamical lepton number breaking $v_\sigma$, is sufficiently high, i.e. $v_\sigma \gsim 1$ TeV. 
Finally, if one demands the consistency of the vacuum only up to an energy scale far below the Planck scale, then the green stability-perturbativity regions will enlarge considerably.

\section{Summary and outlook} 
\label{Conclusion}

We have examined the dynamical inverse seesaw mechanism as a simple benchmark for electroweak breaking and Higgs boson physics.
We first briefly summarized the issue of vacuum stability in the context of inverse seesaw mechanism with explicit lepton number violation and compared with the SM, Fig.~\ref{fig:RG-SM-Running}.
The addition of fermion singlets ($\nu^c$ and $S$) has a destabilizing effect on the running of the Higgs quartic coupling $\lambda$ and we found that with sizeable Yukawa coupling $Y_\nu$ the quartic
coupling $\lambda$ becomes negative before the \sm instability scale $\mu\sim 10^{10}$ GeV, Fig.~\ref{fig:Three-species}. 
Further we have examined the issue of vacuum stability in the simplest dynamical realization of this scenario within context of the \SM gauge theory. 
The Higgs sector is the simplest generalization of that of the SM, adding a singlet $\sigma$ whose vacuum expectation value $v_\sigma$ drives the spontaneous violation of lepton number
and neutrino mass generation. 
It has two characteristic features:
i) the existence of a massless Nambu-Goldstone boson, $J$, dubbed majoron, and
ii) two $CP$-even neutral Higgs $H_1$ and $H_2$ given as a simple admixture of the doublet and singlet scalar given by $\sin\theta$.
This leads to potentially large Higgs decays into the invisible majorons, in~Eq.~(\ref{eq:inv}), modifying also the rates for \sm decay channels in a simple manner, Fig.~\ref{muf range}.
In Fig.~\ref{present limit for case 2} we presented the regions of $|\sin\theta|$ versus $v_\sigma$ allowed by the current LHC limits on the invisible Higgs decay in~Eq.~(\ref{eq:inv}) (magenta) and 
    the signal strength parameter $\mu_f$ in Eq.~(\ref{eq:muf})~(gray), see also Fig.~\ref{fig:invisible-vs-vev1}.
We have examined the implications of existing Higgs measurements at LEP-II and LHC,   
 considering two cases:       Case I) where the heaviest scalar $H_2 = H_{125}$ is the Higgs boson with $m_{H_2}=125$ GeV, with $15\,\text{GeV}\leq m_{H_1}\leq 115\,\text{GeV}$, and
                              Case II) where the lightest scalar is the $H_1 = H_{125}$ Higgs with $m_{H_1}=125$ GeV, and $m_{H_2}\geq 130$ GeV.
                              For case I), both the LHC and LEP-II measurements are applicable, while for case II) only LHC measurements are relevant.
The resulting regions excluded by these experiments are shown in Figs.~\ref{fig:LEPandBR}, \ref{muWW and LEP}, \ref{muZZ vs Invisible} and \ref{Invisible vs Invisible}. 
    
Moreover, the stability properties of the electroweak vacuum can be substatially improved, see Figs.~\ref{fig:both-contributions-to-lambda} and \ref{fig:Three-species1TeV}. 
We further study how the vacuum stability can further restrict the parameter space, as illustrated in Figs.~\ref{stability}, \ref{lambda plot}, \ref{stability2}, \ref{stability for case 2 with 700 GeV},
\ref{stability for case 2 with 1 TeV}, \ref{stability for case 2 with 3 TeV}, \ref{Three generation with 1 TeV} and \ref{Three generation with 3 TeV}.
The take-home message implied by our findings is that, with appreciable mixing angle between $CP$-even neutral Higgs $H_1$ and $H_2$, we can have stable vacuum even for relatively large Yukawa coupling.
Therefore further experimental improvements, either at the LHC or at the future FCC~\cite{Abada:2019ono,Abada:2019lih}, will further restrict the allowed regions.
Needless to say, an open conflict between stability-perturbativity and experiment could mean the breakdown of the theory, and the presence of new physics below the Planck scale.
We are still far from this long term goal and, in the meatime, our dynamical inverse seesaw picture offers a very simple benchmark for electroweak breaking
and precision Higgs boson studies at upcoming collider facilities.

\begin{acknowledgments}
  The work of S.M. and J.V. are supported by the Spanish grant FPA2017-85216-P (AEI/FEDER, UE), PROMETEO/2018/165 (Generalitat Valenciana).
  J.C.R is supported in part by the Portuguese Funda\c{c}\~{a}o para a Ciencia e Tecnologia (FCT) under contracts UIDB/00777/2020, UIDP/00777/2020, CERN/FIS-PAR/0004/2017, and PTDC/FISPAR/29436/2017.
  RS is supported by the Government of India, SERB Startup Grant SRG/2020/002303.
\end{acknowledgments}
\appendix
\label{Appendix}
\section{RGEs: Inverse seesaw with majoron}
\label{app:inverse-seesaw-majoron}
In our work we have used the package SARAH~\cite{Staub:2015kfa} to perform the renormalization group analysis of the dynamical inverse seesaw model.
The $\beta$ function of a given parameter $c$ is given by,
\begin{align*}
 \mu\frac{dc}{d\mu}\equiv\beta_{c}=\frac{1}{16\pi^{2}}\beta_{c}^{(1)}+\frac{1}{(16\pi^{2})^{2}}\beta_{c}^{(2)} \, .
\end{align*}
where $\mu$ is the running scale and $\beta_{c}^{(1)}$, $\beta_{c}^{(2)}$ are the one-loop and two-loop renormalization group terms.

In the presence of the majoron the one- and two-loop renormalization group corrections for the quartic scalar couplings are modified to

\subsection{Quartic scalar couplings}
 \underline{\bf{One Loop}}
{\allowdisplaybreaks  \begin{align} 
\beta_{\lambda_\Phi}^{(1)} & =  
+\lambda_{\Phi\sigma}^{2} + 24 \lambda_\Phi^{2} 
- \frac{9}{5} g_{1}^{2} \lambda_\Phi - 9 g_{2}^{2} \lambda_\Phi  + 12 \lambda_\Phi y_t^2 +4 \lambda_\Phi \mbox{Tr}\Big({Y_\nu  Y_{\nu}^{\dagger}}\Big)   \nonumber \\ 
& - 6  y_t^4  - 2 \mbox{Tr}\Big({Y_\nu  Y_{\nu}^{\dagger}  Y_\nu  Y_{\nu}^{\dagger}}\Big) 
 +\frac{27}{200} g_{1}^{4} +\frac{9}{20} g_{1}^{2} g_{2}^{2} +\frac{9}{8} g_{2}^{4}
 \end{align}
 \underline{\bf{Two Loop}}
  \begin{align}
\beta_{\lambda_\Phi}^{(2)} & =  
-\frac{3411}{2000} g_{1}^{6} -\frac{1677}{400} g_{1}^{4} g_{2}^{2} -\frac{289}{80} g_{1}^{2} g_{2}^{4} +\frac{305}{16} g_{2}^{6} -4 \lambda_{\Phi\sigma}^{3} +\frac{1887}{200} g_{1}^{4} \lambda_\Phi +\frac{117}{20} g_{1}^{2} g_{2}^{2} \lambda_\Phi -\frac{73}{8} g_{2}^{4} \lambda_\Phi \\ \nonumber
&-10 \lambda_{\Phi\sigma}^{2} \lambda_\Phi 
 +\frac{108}{5} g_{1}^{2} \lambda_\Phi^{2} +108 g_{2}^{2} \lambda_\Phi^{2} -312 \lambda_\Phi^{3} 
  -4 \lambda_{\Phi\sigma}^{2} \mbox{Tr}\Big({Y_S  Y_S^*}\Big) 
 -\frac{171}{100} g_{1}^{4} y_t^2 +\frac{63}{10} g_{1}^{2} g_{2}^{2} y_t^2 \\ \nonumber
&-\frac{9}{4} g_{2}^{4} y_t^2 +\frac{17}{2} g_{1}^{2} \lambda_\Phi y_t^2 
 +\frac{45}{2} g_{2}^{2} \lambda_\Phi y_t^2 +80 g_{3}^{2} \lambda_\Phi y_t^2 -144 \lambda_\Phi^{2} y_t^2 -\frac{9}{100} g_{1}^{4} \mbox{Tr}\Big({Y_\nu  Y_{\nu}^{\dagger}}\Big) \nonumber \\ 
 &-\frac{3}{10} g_{1}^{2} g_{2}^{2} \mbox{Tr}\Big({Y_\nu  Y_{\nu}^{\dagger}}\Big) -\frac{3}{4} g_{2}^{4} \mbox{Tr}\Big({Y_\nu  Y_{\nu}^{\dagger}}\Big) +\frac{3}{2} g_{1}^{2} \lambda_\Phi \mbox{Tr}\Big({Y_\nu  Y_{\nu}^{\dagger}}\Big) +\frac{15}{2} g_{2}^{2} \lambda_\Phi \mbox{Tr}\Big({Y_\nu  Y_{\nu}^{\dagger}}\Big) 
 -48 \lambda_\Phi^{2} \mbox{Tr}\Big({Y_\nu  Y_{\nu}^{\dagger}}\Big) \nonumber \\ 
 &-\frac{8}{5} g_{1}^{2} y_t^4 -32 g_{3}^{2} y_t^4 -3 \lambda_\Phi y_t^4 - \lambda_\Phi \mbox{Tr}\Big({Y_\nu  Y_{\nu}^{\dagger}  Y_\nu  Y_{\nu}^{\dagger}}\Big) 
  +30 y_t^6 +10 \mbox{Tr}\Big({Y_\nu  Y_{\nu}^{\dagger}  Y_\nu  Y_{\nu}^{\dagger}  Y_\nu  Y_{\nu}^{\dagger}}\Big) 
 \end{align}
 \underline{\bf{One Loop}}
  \begin{align}  
\beta_{\lambda_{\Phi\sigma}}^{(1)} & =  
\frac{1}{10} \lambda_{\Phi\sigma} \Big( + 40 \lambda_{\Phi\sigma} + 80 \lambda_{\sigma} + 120 \lambda_\Phi 
+ 40 \mbox{Tr}\Big({Y_S  Y_S^*}\Big) + 60 y_t^2 + 20 \mbox{Tr}\Big({Y_\nu  Y_{\nu}^{\dagger}}\Big) 
 - 9 g_{1}^{2} - 45 g_{2}^{2} \Big)
  \end{align}
 \underline{\bf{Two Loop}}
  \begin{align}  
\beta_{\lambda_{\Phi\sigma}}^{(2)} & =  
+\frac{1671}{400} g_{1}^{4} \lambda_{\Phi\sigma} +\frac{9}{8} g_{1}^{2} g_{2}^{2} \lambda_{\Phi\sigma} -\frac{145}{16} g_{2}^{4} \lambda_{\Phi\sigma} +\frac{3}{5} g_{1}^{2} \lambda_{\Phi\sigma}^{2} +3 g_{2}^{2} \lambda_{\Phi\sigma}^{2} -11 \lambda_{\Phi\sigma}^{3} -48 \lambda_{\Phi\sigma}^{2} \lambda_{\sigma} \nonumber \\
&-40 \lambda_{\Phi\sigma} \lambda_{\sigma}^{2} 
 +\frac{72}{5} g_{1}^{2} \lambda_{\Phi\sigma} \lambda_\Phi +72 g_{2}^{2} \lambda_{\Phi\sigma} \lambda_\Phi -72 \lambda_{\Phi\sigma}^{2} \lambda -60 \lambda_{\Phi\sigma} \lambda_\Phi^{2} -8 \lambda_{\Phi\sigma}^{2} \mbox{Tr}\Big({Y_S  Y_S^*}\Big) \nonumber \\
& -32 \lambda_{\Phi\sigma} \lambda_{\sigma} \mbox{Tr}\Big({Y_S  Y_S^*}\Big) 
 +\frac{17}{4} g_{1}^{2} \lambda_{\Phi\sigma} y_t^2 +\frac{45}{4} g_{2}^{2} \lambda_{\Phi\sigma} y_t^2 +40 g_{3}^{2} \lambda_{\Phi\sigma} y_t^2 -12 \lambda_{\Phi\sigma}^{2} y_t^2 \nonumber \\ 
 &-72 \lambda_{\Phi\sigma} \lambda_\Phi y_t^2 +\frac{3}{4} g_{1}^{2} \lambda_{\Phi\sigma} \mbox{Tr}\Big({Y_\nu  Y_{\nu}^{\dagger}}\Big) +\frac{15}{4} g_{2}^{2} \lambda_{\Phi\sigma} \mbox{Tr}\Big({Y_\nu  Y_{\nu}^{\dagger}}\Big) -4 \lambda_{\Phi\sigma}^{2} \mbox{Tr}\Big({Y_\nu  Y_{\nu}^{\dagger}}\Big) \nonumber \\ 
 &-24 \lambda_{\Phi\sigma} \lambda_\Phi \mbox{Tr}\Big({Y_\nu  Y_{\nu}^{\dagger}}\Big)  
  -24 \lambda_{\Phi\sigma} \mbox{Tr}\Big({Y_S  Y_S^*  Y_S  Y_S^*}\Big) -\frac{27}{2} \lambda_{\Phi\sigma} y_t^4 -\frac{9}{2} \lambda_{\Phi\sigma} \mbox{Tr}\Big({Y_\nu  Y_{\nu}^{\dagger}  Y_\nu  Y_{\nu}^{\dagger}}\Big) 
   \end{align}
 \underline{\bf{One Loop}}
  \begin{align}  
\beta_{\lambda_{\sigma}}^{(1)} & =  
2 \Big(10 \lambda_{\sigma}^{2}  + \lambda_{\Phi\sigma}^{2} 
+ 4 \lambda_{\sigma} \mbox{Tr}\Big({Y_S  Y_S^*}\Big)  
- 8 \mbox{Tr}\Big({Y_S  Y_S^*  Y_S  Y_S^*}\Big)   \Big)
\label{lambdasigma-oneloop}
 \end{align}
 \underline{\bf{Two Loop}}
  \begin{align}  
\beta_{\lambda_{\sigma}}^{(2)} & =  
+\frac{12}{5} g_{1}^{2} \lambda_{\Phi\sigma}^{2} +12 g_{2}^{2} \lambda_{\Phi\sigma}^{2} -8 \lambda_{\Phi\sigma}^{3} -20 \lambda_{\Phi\sigma}^{2} \lambda_{\sigma} -240 \lambda_{\sigma}^{3} 
 -80 \lambda_{\sigma}^{2} \mbox{Tr}\Big({Y_S  Y_S^*}\Big) -12 \lambda_{\Phi\sigma}^{2} y_t^2 \nonumber \\
& -4 \lambda_{\Phi\sigma}^{2} \mbox{Tr}\Big({Y_\nu  Y_{\nu}^{\dagger}}\Big) +16 \lambda_{\sigma} \mbox{Tr}\Big({Y_S  Y_S^*  Y_S  Y_S^*}\Big) +256 \mbox{Tr}\Big({Y_S  Y_S^*  Y_S  Y_S^*  Y_S  Y_S^*}\Big) 
\end{align}}
\subsection{Yukawa Couplings}
Likewise, in the presence of the majoron the one- and two-loop renormalization group corrections for the Yukawa couplings in the inverse seesaw model are modified to\\
 \underline{\bf{One Loop}}
{\allowdisplaybreaks  \begin{align} 
\beta_{Y_\nu}^{(1)} & =  
\frac{3}{2} {Y_\nu  Y_{\nu}^{\dagger}  Y_\nu}  
+ Y_\nu \Big(3 y_t^2 + \mbox{Tr}\Big({Y_\nu  Y_{\nu}^{\dagger}}\Big)
- \frac{9}{20} g_{1}^{2}  -\frac{9}{4} g_{2}^{2}   \Big)
  \end{align}
 \underline{\bf{Two Loop}}
  \begin{align}   
\beta_{Y_\nu}^{(2)} & =  
\frac{1}{80} \Big(279 g_{1}^{2} {Y_\nu  Y_{\nu}^{\dagger}  Y_\nu} +675 g_{2}^{2} {Y_\nu  Y_{\nu}^{\dagger}  Y_\nu} -960 \lambda_\Phi {Y_\nu  Y_{\nu}^{\dagger}  Y_\nu}  
  +120 {Y_\nu  Y_{\nu}^{\dagger}  Y_\nu  Y_{\nu}^{\dagger}  Y_\nu} -540 {Y_\nu  Y_{\nu}^{\dagger}  Y_\nu} y_t^2 \nonumber \\
&-180 {Y_\nu  Y_{\nu}^{\dagger}  Y_\nu} \mbox{Tr}\Big({Y_\nu  Y_{\nu}^{\dagger}}\Big)  
 +2 Y_\nu \Big(21 g_{1}^{4} -54 g_{1}^{2} g_{2}^{2} -230 g_{2}^{4} +20 \lambda_{\Phi\sigma}^{2} +240 \lambda_\Phi^{2}  
  +85 g_{1}^{2} y_t^2 \nonumber \\
&+225 g_{2}^{2} y_t^2 +800 g_{3}^{2} y_t^2 
 +15 g_{1}^{2} \mbox{Tr}\Big({Y_\nu  Y_{\nu}^{\dagger}}\Big) +75 g_{2}^{2} \mbox{Tr}\Big({Y_\nu  Y_{\nu}^{\dagger}}\Big)    
   -270 y_t^4 -90 \mbox{Tr}\Big({Y_\nu  Y_{\nu}^{\dagger}  Y_\nu  Y_{\nu}^{\dagger}}\Big) \Big)\Big)
    \end{align}
 \underline{\bf{One Loop}}
  \begin{align}  
\beta_{y_t}^{(1)} & =  
\frac{3}{2} y_t^3  + y_t \Big( 3 y_t^2 + \mbox{Tr}\Big({Y_\nu  Y_{\nu}^{\dagger}}\Big)
- 8 g_{3}^{2}  -\frac{17}{20} g_{1}^{2}  -\frac{9}{4} g_{2}^{2}   \Big)
 \end{align}
 \underline{\bf{Two Loop}}
  \begin{align}  
\beta_{y_t}^{(2)} & =  
+\frac{1}{80} \Big(120 y_t^5 + y_t^3 \Big(1280 g_{3}^{2}    -180 \mbox{Tr}\Big({Y_\nu  Y_{\nu}^{\dagger}}\Big)  + 223 g_{1}^{2}    -540 y_t^2  + 675 g_{2}^{2}  -960 \lambda_\Phi \Big)\nonumber \\ 
 &+y_t \Big(\frac{1187}{600} g_{1}^{4} -\frac{9}{20} g_{1}^{2} g_{2}^{2} -\frac{23}{4} g_{2}^{4} +\frac{19}{15} g_{1}^{2} g_{3}^{2} +9 g_{2}^{2} g_{3}^{2} -108 g_{3}^{4} +\frac{1}{2} \lambda_{\Phi\sigma}^{2} +6 \lambda_\Phi^{2}  +\frac{17}{8} g_{1}^{2} y_t^2 \nonumber  \\
& +\frac{45}{8} g_{2}^{2} y_t^2
 +20 g_{3}^{2} y_t^2 +\frac{3}{8} g_{1}^{2} \mbox{Tr}\Big({Y_\nu  Y_{\nu}^{\dagger}}\Big) +\frac{15}{8} g_{2}^{2} \mbox{Tr}\Big({Y_\nu  Y_{\nu}^{\dagger}}\Big)  
    -\frac{27}{4} y_t^4 -\frac{9}{4} \mbox{Tr}\Big({Y_\nu  Y_{\nu}^{\dagger}  Y_\nu  Y_{\nu}^{\dagger}}\Big) \Big)
     \end{align}
 \underline{\bf{One Loop}}
  \begin{align}  
\beta_{Y_S}^{(1)} & =  2 Y_S \mbox{Tr}\Big({Y_S  Y_S^*}\Big)  + 4 {Y_S  Y_S^*  Y_S} 
 \end{align}
 \underline{\bf{Two Loop}}
  \begin{align}  
\beta_{Y_S}^{(2)} & =  
28 {Y_S  Y_S^*  Y_S  Y_S^*  Y_S}  -4 {Y_S  Y_S^*  Y_S} \Big(3 \mbox{Tr}\Big({Y_S  Y_S^*}\Big)  + 8 \lambda_{\sigma} \Big) + Y_S \Big(-12 \mbox{Tr}\Big({Y_S  Y_S^*  Y_S  Y_S^*}\Big) \nonumber \\
& + 4 \lambda_{\sigma}^{2}  + \lambda_{\Phi\sigma}^{2}\Big)
\end{align}} 
\section{Some comments on the RGEs}
\label{app:few discussions of RGEs}
\subsection{Landau Pole}
\begin{itemize}
\item For any coupling $c$, if $\beta_c=Ac^2$ one has
\begin{align}
\mu\frac{dc(\mu)}{d\mu}=Ac^2(\mu) \Rightarrow
\int_{M_Z}^\mu\frac{dc(\mu)}{Ac^2(\mu)}=\int_{M_Z}^\mu \frac{d\mu}{\mu} \nonumber \\
\Rightarrow c(\mu)=\frac{c(M_Z)}{1-A c(M_Z)\text{log}\frac{\mu}{M_Z}}
\label{qua}
\end{align}
so one has an analytical solution.
Depending on the values of $A$ and $c(M_Z)$ Eq.~\ref{qua} has a singularity, i.e. the Landau pole singularity, at a scale: $\mu_L=M_Ze^{\frac{1}{A c(M_Z)}}$.
The presence of this Landau pole indicates that the coupling $c(\mu)$ grows strong at large renormalisation scale $\mu$. We see that larger the $c(M_Z)$ smaller the Landau scale $\mu_L$. \\

We see from Eq.~\ref{lambdasigma-running} that if $Y_S\approx 0$, one-loop evolution of $\lambda_\sigma$ can be approximated as
\begin{align}
\beta_{\lambda_{\sigma}}^{(1)} & \approx 
2 \Big(10 \lambda_{\sigma}^{2}  + \lambda_{\Phi\sigma}^{2}  \Big)
\end{align}
Hence we see that when $\lambda_{\Phi\sigma}$ can be neglected, $\lambda_\sigma(\mu)$ follows Eq.~\ref{qua}.

\item Eq.\ref{qua} can be generalized in the following way:
\begin{align}
\mu\frac{dc(\mu)}{d\mu}=A c^n(\mu) \Rightarrow  c(\mu)=\frac{c(M_Z)}{\Big(1-(n-1)A c(M_Z)^{(n-1)}\text{log}\frac{\mu}{M_Z}\Big)^{\frac{1}{n-1}}}
\label{general Landau}
\end{align}
From this we see that for $n>1$ we can have a Landau pole.
\end{itemize}

\subsection{Continuous growth}
If $\beta_c=A c^n$ with $n\leq 1$, we see from Eq.~\ref{general Landau} that in this case we will not have pole but $c(\mu)$ can grows continuosly. For example with $n=\frac{1}{2}$,
\begin{align}
c(\mu)=c(M_Z)\Big(1+\frac{A}{2\sqrt{c(M_Z)}}\text{log}\frac{\mu}{M_Z}\Big)^2,
\end{align}
and with $n=0$,
\begin{align}
\mu\frac{d c(\mu)}{d\mu}=A \Rightarrow c(\mu)=c(M_Z)+A\text{log}\frac{\mu}{M_Z}
\label{qua1}
\end{align}
Although running of $c(\mu)$ will not encounter any Landau pole but it can go to non-perturbative region with relatively large values of $c(M_Z)$ and $A$.\\

\subsection{Saturation}
%
If $\beta_c$ has a zero at the finite value $c(\mu_*)$ then the growth of $c$ will be saturated at $c(\mu_*)$ for $\mu\to\infty$.
To illustrate this let us consider the following form of $\beta_c$, $\beta_c=(A-B c(\mu))$. This can have a zero at $c(\mu_*)=\frac{A}{B}$.
\begin{align}
\mu\frac{d c(\mu)}{d\mu}=(A-B c(\mu)) \Rightarrow \int_{\mu_*}^{\mu} \frac{d c(\mu)}{\Big(A-B c(\mu)\Big)}=\int_{\mu_*}^{\mu}\frac{d\mu}{\mu} \nonumber \\
\Rightarrow c(\mu)=\frac{1}{B\mu^B}\Big(-A\mu_*^B+B\mu_*^B c(\mu_*)+A\mu^B\Big)
\end{align}
Now using $c(\mu^*)=\frac{A}{B}$ in the above equation, we find $c(\mu)=\frac{A}{B}$ for any $\mu$. Note that this will happen for any $\beta_c$ which has a zero at the finite value $c(\mu_*)$.\\

One loop and two-loop RGEs for the quartic couplings are in general very complex.
However we find numerically that, depending on the starting values of the quartic couplings, we can have zeros in the $\beta$ function at some scale $\mu_*$ so that the subsequent evolution of the corresponding coupling saturates. 


\bibliographystyle{utphys}
\bibliography{bibliography} 

\providecommand{\href}[2]{#2}\begingroup\raggedright\begin{thebibliography}{10}

\bibitem{Aad:2012tfa}
{\bfseries ATLAS} Collaboration, G.~Aad {\em et~al.}, ``{Observation of a new
  particle in the search for the Standard Model Higgs boson with the ATLAS
  detector at the LHC},''
  \href{http://dx.doi.org/10.1016/j.physletb.2012.08.020}{{\em Phys. Lett.}
  {\bfseries B716} (2012) 1--29},
\href{http://arxiv.org/abs/1207.7214}{{\ttfamily arXiv:1207.7214 [hep-ex]}}.

\bibitem{Chatrchyan:2012ufa}
{\bfseries CMS} Collaboration, S.~Chatrchyan {\em et~al.}, ``{Observation of a
  New Boson at a Mass of 125 GeV with the CMS Experiment at the LHC},''
  \href{http://dx.doi.org/10.1016/j.physletb.2012.08.021}{{\em Phys. Lett.}
  {\bfseries B716} (2012) 30--61},
  \href{http://arxiv.org/abs/1207.7235}{{\ttfamily arXiv:1207.7235 [hep-ex]}}.

\bibitem{Abada:2019ono}
{\bfseries FCC} Collaboration, A.~Abada {\em et~al.}, ``{HE-LHC: The
  High-Energy Large Hadron Collider},''
  \href{http://dx.doi.org/10.1140/epjst/e2019-900088-6}{{\em Eur. Phys. J. ST}
  {\bfseries 228} no.~5, (2019) 1109--1382}.

\bibitem{Abada:2019lih}
{\bfseries FCC} Collaboration, A.~Abada {\em et~al.}, ``{FCC Physics
  Opportunities},''
  \href{http://dx.doi.org/10.1140/epjc/s10052-019-6904-3}{{\em Eur. Phys. J.}
  {\bfseries C79} no.~6, (2019) 474}.

\bibitem{Kajita:2016cak}
T.~Kajita, ``{Nobel Lecture: Discovery of atmospheric neutrino oscillations},''
  \href{http://dx.doi.org/10.1103/RevModPhys.88.030501}{{\em Rev.Mod.Phys.}
  {\bfseries 88} (2016) 030501}.

\bibitem{McDonald:2016ixn}
A.~B. McDonald, ``{Nobel Lecture: The Sudbury Neutrino Observatory: Observation
  of flavor change for solar neutrinos},''
  \href{http://dx.doi.org/10.1103/RevModPhys.88.030502}{{\em Rev.Mod.Phys.}
  {\bfseries 88} (2016) 030502}.

\bibitem{deSalas:2020pgw}
P.~de~Salas {\em et~al.}, ``{2020 Global reassessment of the neutrino
  oscillation picture},'' \href{http://dx.doi.org/10.1007/JHEP02(2021)071}{{\em
  JHEP} {\bfseries 02} (2021) 144},
  \href{http://arxiv.org/abs/2006.11237}{{\ttfamily arXiv:2006.11237
  [hep-ph]}}.

\bibitem{Schechter:1980gr}
J.~Schechter and J.~W.~F. Valle, ``{Neutrino Masses in SU(2) x U(1)
  Theories},'' \href{http://dx.doi.org/10.1103/PhysRevD.22.2227}{{\em
  Phys.Rev.} {\bfseries D22} (1980) 2227}.

\bibitem{Chikashige:1980ui}
Y.~Chikashige, R.~N. Mohapatra, and R.~Peccei, ``{Are There Real Goldstone
  Bosons Associated with Broken Lepton Number?},''
  \href{http://dx.doi.org/10.1016/0370-2693(81)90011-3}{{\em Phys.Lett.}
  {\bfseries B98} (1981) 265}.

\bibitem{Schechter:1981cv}
J.~Schechter and J.~W.~F. Valle, ``Neutrino decay and spontaneous violation of
  lepton number,'' \href{http://dx.doi.org/10.1103/PhysRevD.25.774}{{\em Phys.
  Rev. D} {\bfseries 25} (Feb, 1982) 774--783}.
  \url{https://link.aps.org/doi/10.1103/PhysRevD.25.774}.

\bibitem{Mohapatra:1986bd}
R.~Mohapatra and J.~W.~F. Valle,
  \href{http://dx.doi.org/10.1103/PhysRevD.34.1642}{``{Neutrino Mass and Baryon
  Number Nonconservation in Superstring Models},''} vol.~D34, p.~1642.
\newblock 1986.

\bibitem{Khan:2012zw}
S.~Khan, S.~Goswami, and S.~Roy, ``{Vacuum Stability constraints on the minimal
  singlet TeV Seesaw Model},''
  \href{http://dx.doi.org/10.1103/PhysRevD.89.073021}{{\em Phys.Rev.}
  {\bfseries D89} (2014) 073021},
  \href{http://arxiv.org/abs/1212.3694}{{\ttfamily arXiv:1212.3694 [hep-ph]}}.

\bibitem{Rodejohann:2012px}
W.~Rodejohann and H.~Zhang, ``{Impact of massive neutrinos on the Higgs
  self-coupling and electroweak vacuum stability},''
  \href{http://dx.doi.org/10.1007/JHEP06(2012)022}{{\em JHEP} {\bfseries 1206}
  (2012) 022}, \href{http://arxiv.org/abs/1203.3825}{{\ttfamily arXiv:1203.3825
  [hep-ph]}}.

\bibitem{Bonilla:2015kna}
C.~Bonilla, R.~M. Fonseca, and J.~W.~F. Valle, ``{Vacuum stability with
  spontaneous violation of lepton number},''
  \href{http://dx.doi.org/10.1016/j.physletb.2016.03.037}{{\em Phys. Lett.}
  {\bfseries B756} (2016) 345--349},
  \href{http://arxiv.org/abs/1506.04031}{{\ttfamily arXiv:1506.04031
  [hep-ph]}}.

\bibitem{Rose:2015fua}
L.~Delle~Rose, C.~Marzo, and A.~Urbano, ``{On the stability of the electroweak
  vacuum in the presence of low-scale seesaw models},''
  \href{http://dx.doi.org/10.1007/JHEP12(2015)050}{{\em JHEP} {\bfseries 1512}
  (2015) 050}, \href{http://arxiv.org/abs/1506.03360}{{\ttfamily
  arXiv:1506.03360 [hep-ph]}}.

\bibitem{Lindner:2015qva}
M.~Lindner, H.~H. Patel, and B.~Radov{\v{c}}i{\'c}, ``{Electroweak Absolute,
  Meta-, and Thermal Stability in Neutrino Mass Models},''
  \href{http://dx.doi.org/10.1103/PhysRevD.93.073005}{{\em Phys.Rev.}
  {\bfseries D93} (2016) 073005},
  \href{http://arxiv.org/abs/1511.06215}{{\ttfamily arXiv:1511.06215
  [hep-ph]}}.

\bibitem{Ng:2015eia}
J.~Ng and A.~de~la Puente, ``{Electroweak Vacuum Stability and the Seesaw
  Mechanism Revisited},''
  \href{http://dx.doi.org/10.1140/epjc/s10052-016-3981-4}{{\em Eur.Phys.J.}
  {\bfseries C76} (2016) 122},
  \href{http://arxiv.org/abs/1510.00742}{{\ttfamily arXiv:1510.00742
  [hep-ph]}}.

\bibitem{Bambhaniya:2016rbb}
G.~Bambhaniya, P.~Bhupal~Dev, S.~Goswami, S.~Khan, and W.~Rodejohann,
  ``{Naturalness, Vacuum Stability and Leptogenesis in the Minimal Seesaw
  Model},'' \href{http://dx.doi.org/10.1103/PhysRevD.95.095016}{{\em Phys.Rev.}
  {\bfseries D95} (2017) 095016},
  \href{http://arxiv.org/abs/1611.03827}{{\ttfamily arXiv:1611.03827
  [hep-ph]}}.

\bibitem{Garg:2017iva}
I.~Garg, S.~Goswami, K.~Vishnudath, and N.~Khan, ``{Electroweak vacuum
  stability in presence of singlet scalar dark matter in TeV scale seesaw
  models},'' \href{http://dx.doi.org/10.1103/PhysRevD.96.055020}{{\em
  Phys.Rev.} {\bfseries D96} (2017) 055020},
  \href{http://arxiv.org/abs/1706.08851}{{\ttfamily arXiv:1706.08851
  [hep-ph]}}.

\bibitem{Mandal:2019ndp}
S.~Mandal, R.~Srivastava, and J.~W.~F. Valle, ``{Consistency of the dynamical
  high-scale type-I seesaw mechanism},''
  \href{http://dx.doi.org/10.1103/PhysRevD.101.115030}{{\em Phys. Rev. D}
  {\bfseries 101} no.~11, (2020) 115030},
  \href{http://arxiv.org/abs/1903.03631}{{\ttfamily arXiv:1903.03631
  [hep-ph]}}.

\bibitem{Mandal:2020lhl}
S.~Mandal, R.~Srivastava, and J.~W. Valle, ``{Electroweak symmetry breaking in
  the inverse seesaw mechanism},''
  \href{http://arxiv.org/abs/2009.10116}{{\ttfamily arXiv:2009.10116
  [hep-ph]}}.

\bibitem{GonzalezGarcia:1988rw}
M.~Gonzalez-Garcia and J.~W.~F. Valle, ``{Fast Decaying Neutrinos and
  Observable Flavor Violation in a New Class of Majoron Models},''
  \href{http://dx.doi.org/10.1016/0370-2693(89)91131-3}{{\em Phys.Lett.}
  {\bfseries B216} (1989) 360--366}.

\bibitem{Joshipura:1992hp}
A.~S. Joshipura and J.~W.~F. Valle, ``Invisible higgs decays and neutrino
  physics,''
  \href{http://dx.doi.org/https://doi.org/10.1016/0550-3213(93)90337-O}{{\em
  Nuclear Physics B} {\bfseries 397} no.~1, (1993) 105 -- 122}.
  \url{http://www.sciencedirect.com/science/article/pii/055032139390337O}.

\bibitem{TheATLASandCMSCollaborations:2015bln}
{\bfseries ATLAS, CMS} Collaboration, G.~Aad {\em et~al.}, ``{Measurements of
  the Higgs boson production and decay rates and constraints on its couplings
  from a combined ATLAS and CMS analysis of the LHC pp collision data at $
  \sqrt{s}=7 $ and 8 TeV},''
  \href{http://dx.doi.org/10.1007/JHEP08(2016)045}{{\em JHEP} {\bfseries 08}
  (2016) 045}, \href{http://arxiv.org/abs/1606.02266}{{\ttfamily
  arXiv:1606.02266 [hep-ex]}}.

\bibitem{Aad:2019mbh}
{\bfseries ATLAS} Collaboration, G.~Aad {\em et~al.}, ``{Combined measurements
  of Higgs boson production and decay using up to $80$ fb$^{-1}$ of
  proton-proton collision data at $\sqrt{s}=$ 13 TeV collected with the ATLAS
  experiment},'' \href{http://dx.doi.org/10.1103/PhysRevD.101.012002}{{\em
  Phys.Rev.} {\bfseries D101} (2020) 012002},
  \href{http://arxiv.org/abs/1909.02845}{{\ttfamily arXiv:1909.02845
  [hep-ex]}}.

\bibitem{Bonilla:2015uwa}
C.~Bonilla, J.~W.~F. Valle, and J.~C. Romao, ``{Neutrino mass and invisible
  Higgs decays at the LHC},''
  \href{http://dx.doi.org/10.1103/PhysRevD.91.113015}{{\em Phys. Rev.}
  {\bfseries D91} no.~11, (2015) 113015},
  \href{http://arxiv.org/abs/1502.01649}{{\ttfamily arXiv:1502.01649
  [hep-ph]}}.

\bibitem{Bonilla:2015jdf}
C.~Bonilla, J.~C. Romao, and J.~W.~F. Valle, ``{Electroweak breaking and
  neutrino mass: invisible Higgs decays at the LHC (type II seesaw)},''
  \href{http://dx.doi.org/10.1088/1367-2630/18/3/033033}{{\em New J. Phys.}
  {\bfseries 18} no.~3, (2016) 033033},
  \href{http://arxiv.org/abs/1511.07351}{{\ttfamily arXiv:1511.07351
  [hep-ph]}}.

\bibitem{Fontes:2019uld}
D.~Fontes, J.~C. Romao, and J.~W. Valle, ``{Electroweak Breaking and Higgs
  Boson Profile in the Simplest Linear Seesaw Model},''
  \href{http://dx.doi.org/10.1007/JHEP10(2019)245}{{\em JHEP} {\bfseries 1910}
  (2019) 245}, \href{http://arxiv.org/abs/1908.09587}{{\ttfamily
  arXiv:1908.09587 [hep-ph]}}.

\bibitem{Sirunyan:2018owy}
{\bfseries CMS} Collaboration, A.~M. Sirunyan {\em et~al.}, ``{Search for
  invisible decays of a Higgs boson produced through vector boson fusion in
  proton-proton collisions at $\sqrt{s} =$ 13 TeV},''
\href{http://arxiv.org/abs/1809.05937}{{\ttfamily arXiv:1809.05937 [hep-ex]}}.

\bibitem{Aaboud:2019rtt}
{\bfseries ATLAS} Collaboration, M.~Aaboud {\em et~al.}, ``{Combination of
  searches for invisible Higgs boson decays with the ATLAS experiment},'' {\em
  Submitted to: Phys. Rev. Lett.} (2019) ,
\href{http://arxiv.org/abs/1904.05105}{{\ttfamily arXiv:1904.05105 [hep-ex]}}.

\bibitem{Buttazzo:2013uya}
D.~Buttazzo {\em et~al.}, ``{Investigating the near-criticality of the Higgs
  boson},'' \href{http://dx.doi.org/10.1007/JHEP12(2013)089}{{\em JHEP}
  {\bfseries 1312} (2013) 089},
  \href{http://arxiv.org/abs/1307.3536}{{\ttfamily arXiv:1307.3536 [hep-ph]}}.

\bibitem{Degrassi:2012ry}
G.~Degrassi {\em et~al.}, ``{Higgs mass and vacuum stability in the Standard
  Model at NNLO},'' \href{http://dx.doi.org/10.1007/JHEP08(2012)098}{{\em JHEP}
  {\bfseries 1208} (2012) 098},
  \href{http://arxiv.org/abs/1205.6497}{{\ttfamily arXiv:1205.6497 [hep-ph]}}.

\bibitem{Alekhin:2012py}
S.~Alekhin, A.~Djouadi, and S.~Moch, ``{The top quark and Higgs boson masses
  and the stability of the electroweak vacuum},''
  \href{http://dx.doi.org/10.1016/j.physletb.2012.08.024}{{\em Phys.Lett.}
  {\bfseries B716} (2012) 214--219},
  \href{http://arxiv.org/abs/1207.0980}{{\ttfamily arXiv:1207.0980 [hep-ph]}}.

\bibitem{CentellesChulia:2020dfh}
S.~Centelles~Chuli{\'a}, R.~Srivastava, and A.~Vicente, ``{The Inverse Seesaw
  Family: Dirac And Majorana},''
  \href{http://arxiv.org/abs/2011.06609}{{\ttfamily arXiv:2011.06609
  [hep-ph]}}.

\bibitem{Casas:1999cd}
J.~Casas, V.~Di~Clemente, A.~Ibarra, and M.~Quiros, ``{Massive neutrinos and
  the Higgs mass window},''
  \href{http://dx.doi.org/10.1103/PhysRevD.62.053005}{{\em Phys.Rev.}
  {\bfseries D62} (2000) 053005}.

\bibitem{Romao:1992zx}
J.~Romao, F.~de~Campos, and J.~W.~F. Valle, ``{New Higgs signatures in
  supersymmetry with spontaneous broken R parity},''
  \href{http://dx.doi.org/10.1016/0370-2693(92)91183-A}{{\em Phys.Lett.}
  {\bfseries B292} (1992) 329--336},
  \href{http://arxiv.org/abs/hep-ph/9207269}{{\ttfamily arXiv:hep-ph/9207269
  [hep-ph]}}.

\bibitem{LopezFernandez:1993tk}
A.~Lopez-Fernandez {\em et~al.}, ``{Model independent Higgs boson mass limits
  at LEP},'' \href{http://dx.doi.org/10.1016/0370-2693(93)90518-M}{{\em
  Phys.Lett.} {\bfseries B312} (1993) 240--246}.

\bibitem{DeCampos:1994fi}
F.~De~Campos {\em et~al.}, ``{Limits on associated production of visibly and
  invisibly decaying Higgs bosons from Z decays},''
  \href{http://dx.doi.org/10.1016/0370-2693(94)90557-6}{{\em Phys.Lett.}
  {\bfseries B336} (1994) 446--456},
  \href{http://arxiv.org/abs/hep-ph/9407328}{{\ttfamily arXiv:hep-ph/9407328
  [hep-ph]}}.

\bibitem{deCampos:1996bg}
F.~de~Campos {\em et~al.}, ``{Searching for invisibly decaying Higgs bosons at
  LEP-2},'' \href{http://dx.doi.org/10.1103/PhysRevD.55.1316}{{\em Phys.Rev.}
  {\bfseries D55} (1997) 1316--1325},
  \href{http://arxiv.org/abs/hep-ph/9601269}{{\ttfamily arXiv:hep-ph/9601269
  [hep-ph]}}.

\bibitem{Abdallah:2004wy}
{\bfseries DELPHI} Collaboration, J.~Abdallah {\em et~al.}, ``{Searches for
  neutral higgs bosons in extended models},''
  \href{http://dx.doi.org/10.1140/epjc/s2004-02011-4}{{\em Eur.Phys.J.}
  {\bfseries C38} (2004) 1--28}.

\bibitem{Sirunyan:2018koj}
{\bfseries CMS} Collaboration, A.~M. Sirunyan {\em et~al.}, ``{Combined
  measurements of Higgs boson couplings in proton{\textendash}proton collisions
  at $\sqrt{s}=13\,\text {Te}\text {V} $},''
  \href{http://dx.doi.org/10.1140/epjc/s10052-019-6909-y}{{\em Eur.Phys.J.}
  {\bfseries C79} (2019) 421},
  \href{http://arxiv.org/abs/1809.10733}{{\ttfamily arXiv:1809.10733
  [hep-ex]}}.

\bibitem{Romao:1992dc}
J.~Romao {\em et~al.}, ``{Detection of intermediate mass Higgs bosons from
  spontaneously broken R-parity supersymmetry},''
  \href{http://dx.doi.org/10.1142/S0217732394000642}{{\em Mod.Phys.Lett.}
  {\bfseries A9} (1994) 817--828},
  \href{http://arxiv.org/abs/hep-ph/9211258}{{\ttfamily arXiv:hep-ph/9211258
  [hep-ph]}}.

\bibitem{Berezhiani:1992cd}
Z.~Berezhiani, A.~Smirnov, and J.~W.~F. Valle, ``{Observable Majoron emission
  in neutrinoless double beta decay},''
  \href{http://dx.doi.org/10.1016/0370-2693(92)90126-O}{{\em Phys.Lett.}
  {\bfseries B291} (1992) 99--105}.

\bibitem{Staub:2015kfa}
F.~Staub, ``{Exploring new models in all detail with SARAH},''
  \href{http://dx.doi.org/10.1155/2015/840780}{{\em Adv.High Energy Phys.}
  {\bfseries 2015} (2015) 840780},
  \href{http://arxiv.org/abs/1503.04200}{{\ttfamily arXiv:1503.04200
  [hep-ph]}}.

\end{thebibliography}\endgroup
\end{document}